\documentclass[aps,preprint,showpacs,superscriptaddress,groupedaddress,amsmath]{revtex4}

\usepackage{amsmath,amsfonts,graphicx,bm,amssymb,array}
\usepackage[usenames]{color}
\usepackage{array}
\usepackage{float}
\usepackage{sidecap}
\usepackage{graphicx}
\usepackage{bm}
\usepackage{axodraw4j}
\usepackage{subfigure}

\def\etmiss{E\!\!\!\!\slash_{T}}

\def\pslash{\not{\hbox{\kern-4pt $p$}}}
\def\qslash{\not{\hbox{\kern-4pt $q$}}}
\def\lv{\not{\hbox{\kern-4pt $L$}}}
\def\lsim{\mathrel{\raise.3ex\hbox{$<$\kern-.75em\lower1ex\hbox{$\sim$}}}}
\def\gsim{\mathrel{\raise.3ex\hbox{$>$\kern-.75em\lower1ex\hbox{$\sim$}}}}
\def\ifmath#1{\relax\ifmmode #1\else $#1$\fi}

\def\Wp{W'}
\def\ttbar{t\bar{t}}

\definecolor{DarkRed}{rgb}{0.55,0.00,0.00}

\newcommand{\nc}{\newcommand}
\nc{\postscript}[2]{\setlength{\epsfxsize}{#2\hsize}\centerline{\epsfbox{#1}}}
\nc{\beq}{\begin{equation}} \nc{\eeq}{\end{equation}}
\nc{\bea}{\begin{eqnarray}} \nc{\eea}{\end{eqnarray}}
\nc{\baa}{\begin{array}} \nc{\eaa}{\end{array}}
\nc{\bit}{\begin{itemize}} \nc{\eit}{\end{itemize}}
\nc{\ben}{\begin{enumerate}} \nc{\een}{\end{enumerate}}
\nc{\bce}{\begin{center}} \nc{\ece}{\end{center}}
\nc{\non}{\nonumber}
\nc{\eqns}[1]{\begin{eqnarray} #1 \end{eqnarray}}
\nc{\braket}[1]{\left( #1 \right)}

\newcommand{\Lagr}{L}

\usepackage{booktabs}
\usepackage[latin1]{inputenc}

\begin{document}

\baselineskip=17pt

\thispagestyle{empty}
\vspace{20pt}
\font\cmss=cmss10 \font\cmsss=cmss10 at 7pt

\hfill

\begin{center}
{\Large \textbf
{
Single Top Production as a Probe of Heavy Resonances
}}

\vspace{15pt}

{\large
Elizabeth Drueke$^{a}$, Joseph Nutter$^{a}$, Reinhard Schwienhorst$^{a}$, Natascia Vignaroli$^{a}$, Devin G.\! E.\! Walker$^{b}$, Jiang-Hao Yu$^{c}$
}

\vspace{15pt}

$^{a}$\textit{Department of Physics and Astronomy, Michigan State University, East Lansing 48824, U.S.A.,} \\
$^{b}$\textit{SLAC National Accelerator Laboratory, 2575 Sand Hill Road, Menlo Park, CA 94025, U.S.A.,} \\
$^{c}$\textit{Theory Group, Department of Physics and Texas Cosmology Center, The University of Texas at Austin,  Austin, TX 78712 U.S.A.} \\
\vspace{10pt}
\end{center}


\begin{center}
\vspace{15pt}
\textbf{Abstract}

The single top-quark final state provides sensitivity to new heavy resonances produced in proton-proton collisions at the Large Hadron Collider. Particularly, the single top plus quark final state appears in models with heavy charged bosons or scalars, or in models with flavor-changing neutral currents involving the top quark. The cross sections and final state kinematics distinguish such models from each other and from standard model backgrounds. Several models of resonances decaying to a single top-quark final state are presented and their phenomenology is discussed.
\end{center}

\noindent Preprint UTTG-21-14\\
  
\vfill\eject
\noindent


\section{Introduction}
\label{sec:intro}

The CERN Large Hadron Collider (LHC) in Geneva, Switzerland has, since its first run in September 2008, been vital in probing new physics at high energies. Several compelling new physics scenarios predict extended gauge sectors with new heavy resonances which couple strongly to the third generation of quarks. Such theories include top-color models~\cite{Hill:2002ap}, extra-dimensional theories~\cite{Agashe:2003zs}, and composite Higgs models~\cite{Agashe:2004rs}. The top-quark production channels thus offer a very promising way to probe new physics at the LHC. In particular, single top production is studied. Several types of heavy resonances could indeed be observed in the single top final state. In this analysis, $W$-prime bosons and several types of colored resonances -- scalar color octets, Kaluza-Klein gluons ($KKg$), and color-triplet scalars -- will be considered. $W$-prime bosons are predicted in many Beyond-the-SM (BSM) theories ($G(221)$ models~\cite{Hsieh:2010zr,Chivukula:2006cg}, Randall-Sundrum (RS)~\cite{Randall:1999ee} theories, and composite Higgs models~\cite{Agashe:2004rs}, among many others) and have a relevant decay branching-ratio into the $tb$ mode, which this analysis is focused on, in many motivated scenarios. Scalar and vector color octets are analyzed in the $tc$ flavor violating decay. Such a channel offers the possibility of testing the flavor structure of many compelling BSM theories such as RS, composite Higgs, and top-coloron models~\cite{Hill:2002ap,Aquino:2006vp,Perelstein:2005ka,Delaunay:2010dw}. Color-triplet scalars could appear from R-Parity violating supersymmetric theories~\cite{Barbier:2004ez}, $E_6$ grand unification models~\cite{Hewett:1988xc}, and as ``excited quarks'' in some types of composite models~\cite{Baur:1987ga}. A scalar color-triplet in the $tb$ decay mode is considered. In this paper, search strategies for this wide class of heavy resonances at the LHC are presented. A significant part of the analysis is devoted to the study of the most suitable kinematic variables to distinguish the different types of resonances from the backgrounds and from each other. 

Several different models of resonances in the single top plus jet final state have been proposed, though most have not yet been studied experimentally. Phenomenological analyses of the single top plus jet final state at the LHC have focused on $\Wp$ production~\cite{Heikinheimo:2014tba}, including at next-to-leading order (NLO) in QCD~\cite{Sullivan:2002jt,Ayazi:2010jd, Duffty:2012rf}, quark excitations~\cite{Hassanain:2009at,Shiu:2007tn}, charged Higgs~\cite{Dittmaier:2007uw,Hashemi:2013raa}, as well as on searches in the $tW$ final state~\cite{Nutter:2012an}. Colored resonances have previously been considered in the form of a top pion~\cite{Tait:2000sh} and colored diquarks~\cite{Karabacak:2012rn}.

The benchmark models are presented in Sec.~\ref{sec:theory}. Section~\ref{sec:analysis} presents the event selection. The separation of the signals from the backgrounds is discussed for the 8~TeV LHC and a resonance mass of 750~GeV in Sec.~\ref{sec:lowmass}, while the different signals are separated from each other at the 14~TeV and a resonance mass of 3000~GeV in Sec.~\ref{sec:highmass}. Our conclusions are in Sec.~\ref{sec:concl}.


\section{The Benchmark Models}
\label{sec:theory}
In this study, the focus is on several types of heavy resonances which can contribute to the production of single top quarks at the LHC, specifically scalar color octets, vector color octets (Kaluza-Klein gluons), scalar color-triplets and $\Wp$~bosons. All events are generated with MadGraph5~\cite{Alwall:2011uj}. The detailed specifications for the generations can be found in Sec.~\ref{sec:eventgen}. 

\subsection{A scalar and a vector/KKg color octet from an extended color gauge sector }
\label{subsec:coloron}
Colored vector bosons from new strong dynamics, or $KKg$'s in a dual 5D picture, have been searched for mainly in the $\ttbar$ channel ~\cite{Lillie:2007yh, Agashe:2006hk}. In this paper, the single top channel through the flavor violating $KKg \to tc$ decay is analyzed. This channel could give insight into the flavor structure of many compelling BSM theories such as RS~\cite{Randall:1999ee}, composite Higgs \cite{Agashe:2004rs}, and top-coloron models~\cite{Hill:2002ap}. The $KKg$ has been studied by ATLAS~\cite{ATLAS:2012aja} and CMS~\cite{CMS:2012zdh} in the $\ttbar$ mode, and limits were estimated for 14~TeV running in the $KKg \to tc$ mode during the Snowmass 2013 conference~\cite{Drueke:2013wsa}. 

The benchmark adopted here is the simple renormalizable model of an extended color gauge sector introduced in~\cite{Chivukula:2013kw}, which realizes next-to-minimal flavor violation (NMFV)~\cite{Agashe:2005hk,Barbieri:2012bh}. In this model, the third generation quarks couple differently than the light quarks under an extended $SU(3)_1 \times SU(3)_2$ color gauge group. The mixing between light and third generation quarks is induced by the interactions of all three generation quarks with a set of new heavy vector-like quarks. The model reproduces the CKM mixing and generates flavor-changing neutral currents (FCNCs) from non-standard interactions. Due to the specific structure of the model, dangerous FCNCs are naturally suppressed and a large portion of the model parameter space is allowed by the data on meson mixing processes and on $b\to s \gamma$~\cite{Chivukula:2013kw}. 

The model has the color gauge structure $SU(3)_1 \times SU(3)_2$. The extended color symmetry is broken down to $SU(3)_C$ by the (diagonal) expectation value, $\langle \Phi \rangle \propto u \cdot {\cal I}$, of a scalar field $\Phi$, which transforms as a ({$\bf 3, \bar{3} $}) under $SU(3)_1 \times SU(3)_2$. It is assumed that color gauge breaking occurs at a scale much higher than the electroweak scale, $u \gg v$. 

Breaking the color symmetry induces a mixing between the $SU(3)_1$ and the $SU(3)_2$ gauge fields $A^{1}_{\mu}$, $A^{2}_{\mu}$, which is diagonalized by a field rotation determined by

\begin{equation}\label{eq:ctomega}
\cot\omega=\frac{g_1}{g_2} \qquad g_s = g_1 \sin\omega = g_2 \cos\omega \ ,
\end{equation} 

\noindent where $g_s$ is the QCD strong coupling and $g_1$, $g_2$ are the $SU(3)_1$ and $SU(3)_2$ gauge couplings, respectively. The mixing diagonalization reveals two color vector boson mass eigenstates: the mass-less SM gluon and a new massive color-octet vector boson $G^{*}$ given by

\begin{equation}\label{eq:Gstar}
G^{*}_{\mu}= \cos\omega A^{1}_{\mu} - \sin\omega A^{2}_{\mu} \qquad M_{G^{*}}=\frac{g_s u}{\sin\omega \cos\omega} \ .
\end{equation} 

In the NMFV model, the third generation quarks couple differently than the light quarks under the extended color group. $q_L=(t_L, b_L)$, $t_R$, and $b_R$, as well as a new weak-doublet of vector-like quarks, transform as ({$\bf 3, 1$}) under $SU(3)_1 \times SU(3)_2$ while the light generation quarks are charged under $SU(3)_2$ and transform as ({$\bf 1, 3$}). The $G^{*}$ interactions with the color currents associated with $SU(3)_1$ and $SU(3)_2$ are given by
 
\begin{equation}\label{eq:Gstar-current}
g_s \left(\cot\omega J^{\mu}_1 - \tan\omega J^{\mu}_2 \right)G^{*}_{\mu} \ .
\end{equation}

\subsubsection{The vector color octet ($KKg$) ${\bf G^{*}}$ }
\label{subsubsec:kkg}
\begin{figure}[!h!tbp]
  \centering
    \includegraphics[width=0.48\textwidth]{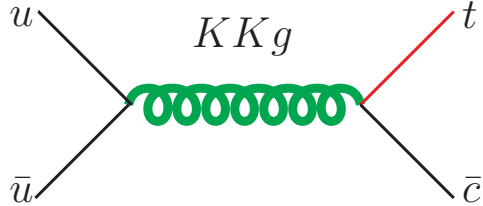}
  \caption{Example Feynman diagram for $KKg$ ($G^{*}$) production with decay to $tc$.}
      \label{fig:KKgFeyn}
\end{figure}

The color vector boson from an extended color group, Kaluza-Klein gluon or $G^{*}$, can be produced at the LHC by quark-antiquark fusion (Fig.~\ref{fig:KKgFeyn}) determined by the $G^{*}$ coupling to light quarks $g_s \tan\omega$ (Eq.~(\ref{eq:Gstar-current})). Gluon-gluon fusion production is forbidden at tree level by $SU(3)_C$ gauge invariance. 

The $G^{*}$ decay widths are:

\begin{align}
\begin{split}
& \Gamma[G^{*} \to \ttbar] = \frac{g^2_s}{24\pi} M_{G^{*}}\cot^2\omega \sqrt{1-4 \frac{m^2_t}{M^2_{G^{*}}}} \left( 1+2\frac{m^2_t}{M^2_{G^{*}}} \right) \,,\\
& \Gamma[G^{*} \to b\bar b] = \frac{g^2_s}{24\pi} M_{G^{*}}\cot^2\omega \,,\\
& \Gamma[G^{*} \to j j] = \frac{g^2_s}{6\pi} M_{G^{*}}\tan^2\omega \,. \\
\end{split}
\end{align}

\noindent Additionally, the NMFV flavor structure of the model generates a $G^{*} \to tc$ flavor violating decay with rate

\begin{equation}
\Gamma[G^{*} \to t_L \bar c_L]=\Gamma[G^{*} \to c_L \bar t_L]\simeq \left(V_{cb}\right)^2 \frac{g^2_s}{48\pi} M_{G^{*}} \left( \cot\omega+\tan\omega \right)^2 \ ,
\end{equation}

\noindent where $V_{cb}=0.0415$ is the CKM matrix element. \footnote{$G^{*}$ FCNCs are induced by the mixing among left-handed quarks generated by the exchange of heavy vector-like quarks. This mixing is controlled by the $3\times 3$ matrices $U_L$ and $D_L$ in the up- and down-quark sectors, respectively. In particular, the $G^{*} \to tc$ flavor violating decay is controlled by the $(U_L)_{23}$ element. The CKM mixing matrix is given by $V_{CKM}=U^{\dagger}_L D_L$. At first order in the mixing parameters, $(U_L)_{23}\equiv V_{cb} - (D_L)_{23}$. The non-diagonal elements of $D_L$ are strongly constrained by the data on $b\to s\gamma$~\cite{Chivukula:2013kw}. $(D_L)_{23}$ is thus forced to be small and, as a consequence, $(U_L)_{23}\simeq V_{cb}$. } 

A complete analysis of the limits on the $G^{*}$ coupling and mass from direct LHC searches and from flavor observables has been performed in~\cite{Chivukula:2013kw}. The LHC searches for di-jet resonances~\cite{CMS:2012nba,ATLAS:2012qjz} place the strongest limits from collider searches \footnote{Weaker bounds come from the LHC searches in the $\ttbar$ final state~\cite{TheATLAScollaboration:2013kha}.}, excluding $G^{*}$ masses in the range 1000-4300~GeV for $\cot\omega=0.5$, when the $G^{*}$ couples more strongly to light quarks, and gradually reducing to the range 1000-2400~GeV for $\cot\omega=2.5$, when the $G^{*}$ couples more strongly to third-generation quarks. Outside the range $0.5 \leq \cot\omega \leq 2.5$, the $G^{*}$ resonance is broader, $\Gamma/M \gtrsim 0.2$, and cannot be excluded by current LHC searches. This analysis considers the $G^{*}$ contribution to single top production, $pp \to G^{*} \to tc$, for 
\begin{equation}
\cot\omega=2.6 \qquad \Gamma/M \simeq 0.25 \, .
\end{equation} 

\noindent For this $\cot\omega$ value, the strongest limits come from the data on $K$-meson mixing, which can still allow for $G^{*}$ above $\sim$500~GeV for reasonable values of the quark-mixing parameters. \footnote{The mass limits on $G^{*}$ from the contribution to $\text{Im} [C^{1}_K]$ depend on the value of the $\alpha_2$ coupling between the second-generation quarks and the new vector-like quarks. The CKM matrix is correctly reproduced for $\alpha_2=\mathcal{O}(\lambda^2) $, where $\lambda$ is the Wolfenstein parameter. $\alpha_2=\lambda^2 /4$ gives a bound $M_{G^{*}}\gtrsim$ 500~GeV. Weaker bounds are allowed for $\alpha_2<\lambda^2 /4$.}

\subsubsection{The scalar color octet  (coloron)  ${\bf G_H}$}
\label{subsubsec:coloron}
The $SU(3)_1 \times SU(3)_2 \to SU(3)_C$ breaking induced by the expectation value of the ({$\bf 3,\bar{ 3}$}) scalar field $\Phi$ generates color-octet and color-singlet scalars, whose phenomenology have been discussed in~\cite{Chivukula:2013hga}. This paper analyzes the contribution to single top production of the color-octet scalar, which will be dubbed as the \textit{coloron}. 

The most general renormalizable potential for $\Phi$ is~\cite{Bai:2010dj}:

\begin{equation}
\label{eq:potential}
V(\Phi)=-m^2_{\Phi}\text{Tr}(\Phi\Phi^\dagger) -\mu (\text{det}\Phi+\text{H.c.})+\frac{\xi}{2}\left[ \text{Tr}(\Phi\Phi^\dagger) \right]^2+\frac{k}{2}\text{Tr}(\Phi\Phi^\dagger\Phi\Phi^\dagger) \ ,
\end{equation}

\noindent where 

\begin{equation}
\text{det} \Phi = \frac{1}{6}\epsilon^{ijk}\epsilon^{i'j'k'}\Phi_{ii'}\Phi_{jj'}\Phi_{kk'} \ ,
\end{equation}

\noindent and where, without loss of generality, one can choose $\mu >0$. Assuming $m^2_\Phi >0$, $\Phi$ acquires a (positive) diagonal expectation value:

\begin{equation}
\langle \Phi \rangle = u \cdot \mathcal{I} \,.
\end{equation}

\noindent The $\Phi$ expansion around the vacuum gives:

\begin{equation}\label{eq:Phi}
\Phi=u+\frac{1}{\sqrt{6}}\left(\phi_R+i\phi_I\right)+\left(G^a_H+iG^a_G\right)T^a \ ,
\end{equation}

\noindent where $\phi_R$ and $\phi_I$ are singlets under $SU(3)_C$ Additionally, $G^a_G$, $a=1,\dots,8$, are the Nambu-Goldstone bosons associated with the color-symmetry breaking,  which will be eaten by the $G^{*}$'s, and $G^a_H$ are color octets. This analysis focuses on the color-octet $G_H$. 

$G_H$ can be produced in pairs through its interactions with gluons:

\begin{equation}\label{eq:GH-gluons}
\frac{g^2_s}{2}f^{abc}f^{ade}G^b_{\mu}G^{\mu d}G^c_H G^e_H +g_s f^{abc} G^a_{\mu} G^b_H \partial^{\mu} G^c_H \ ,
\end{equation}

\noindent or it can be produced singly via gluon-gluon fusion. This occurs at one-loop order through the cubic interaction

\begin{equation}\label{eq:GH-cubic}
\frac{\mu}{6} d_{abc} G^a_H G^b_H G^c_H   \,,
\end{equation}

\noindent which arises from the $\mu(\det\Phi+\text{H.c.})$ term in the potential (\ref{eq:potential}); where $d_{abc}$ is the SU(3) totally symmetric tensor. The single production of $G_H$ can be described by the effective coupling

\begin{equation}\label{eq:single-eff}
-\frac{1}{4} C_{ggG} d_{abc} G^a_{\mu\nu} G^{\mu\nu b} G^c_H
\end{equation}

\noindent with

\begin{equation}\label{eq:cGHg}
C_{ggG}=\sqrt{\frac{1}{6}}\frac{\alpha_s}{\pi }\frac{\mu}{M^2_{G_H}}\left(\frac{\pi^2}{9}-1\right) \ .
\end{equation}

Note that single production is suppressed by a factor $(\pi^2/9 -1)^2$, which is an accidental suppression factor coming from the loop.
Above the threshold for decays into a single top quark, $G_H$ has two main decay modes: the decay into gluons (c.~f. Sec.~\ref{subsec:coloron}), which occurs at loop-level similar to single coloron production, and the flavor-violating decay into $tc$. The corresponding rates are:

\begin{align}
\begin{split}\label{eq:G-rates}
 & \Gamma \left[G_H \to (\bar{c}_L t_R +\bar{t}_R c_L )\right] =\left(V_{cb}\right)^2 \frac{M_{G_H}}{16 \pi} \frac{m^2_t}{u^2}\left(1-\frac{m^2_t}{M^2_{G_H}}\right)^2 \,, \\
& \Gamma \left[G_H \to gg \right]=\frac{5 \alpha^2_s}{1536 \pi^3}\frac{\mu^2}{M_{G_H}}\left(\frac{\pi^2}{9}-1\right)^2 \,.
\end{split}
\end{align}

\begin{figure}[!h!tbp]
  \centering
    \includegraphics[width=0.8\textwidth]{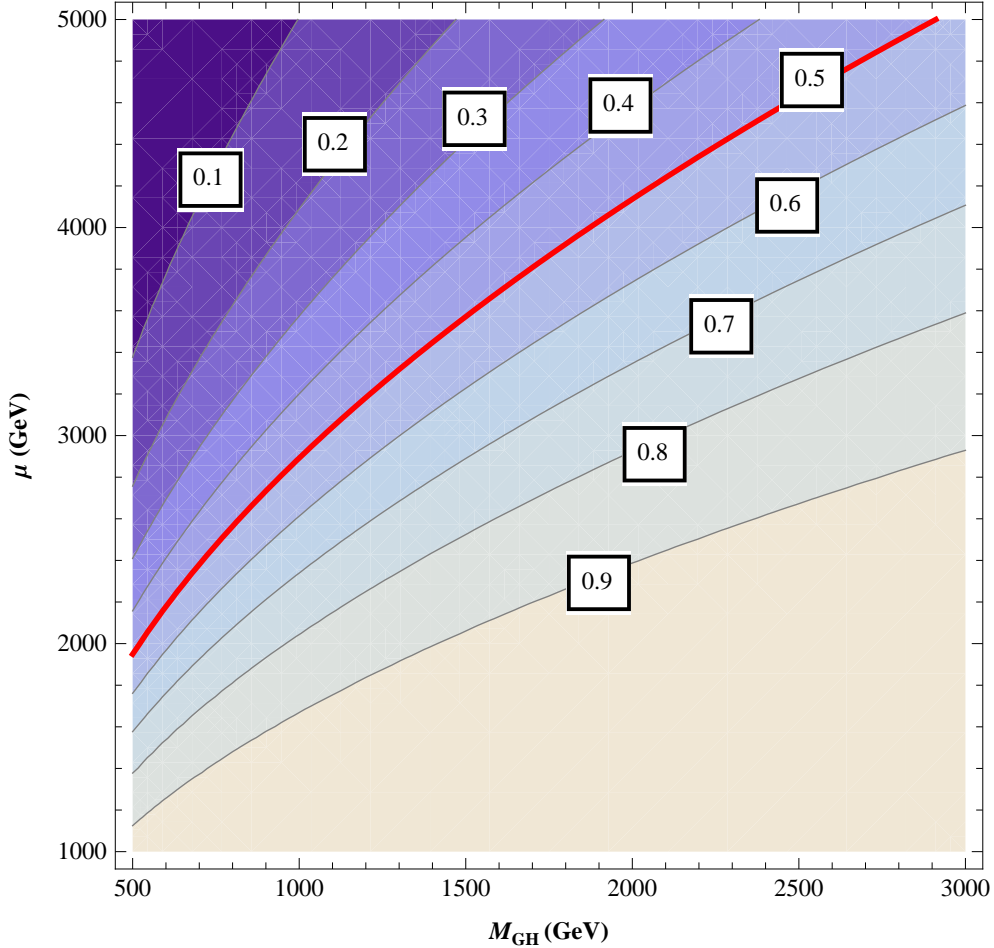}
  \caption{Contour plot of the $G_H \to tc$ branching ratio in the plane $(M_{G_H}, \mu)$. This analysis uses the values indicated by the red-thick curve, for which $G_H$ decays at the same rate into $tc$ and into $gg$. }
      \label{fig:GH-BR}
\end{figure}

\noindent We set $u=\mu$ (the stability of the potential in (\ref{eq:potential}) forbids $\mu>u$); and consider for simplicity the set of $(M_{G_H}, \mu)$ values in Fig.~\ref{fig:GH-BR} that give a 50\% $G_H$ decay into $tc$ and 50\% into $gg$. 
$G_H$ is a very narrow resonance, with a width of the order of 10$^{-4}$~GeV. 

\begin{figure}[!h!tbp]
  \centering
    \includegraphics[width=0.48\textwidth]{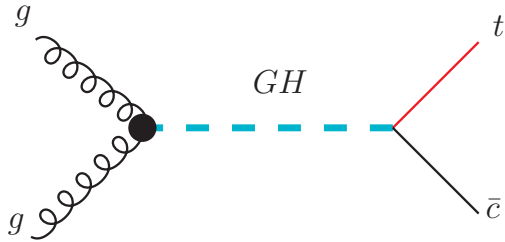}
  \caption{Feynman diagram for single coloron production with decay to $tc$.}
    \label{fig:ColoronFeyn}
\end{figure}

\begin{figure}[!h!tbp]
  \centering
  \subfigure[]{
    \includegraphics[width=0.48\textwidth]{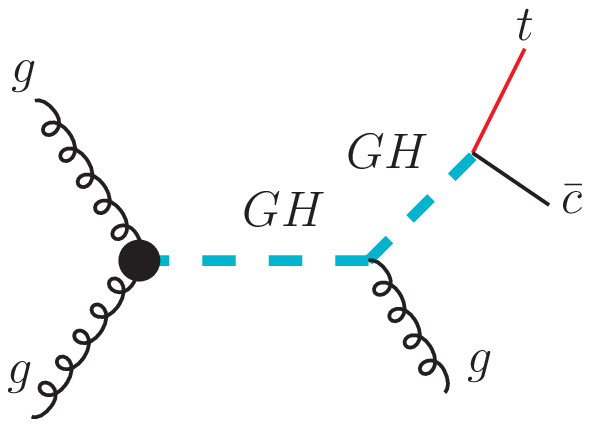}
    \label{fig:ColorongFeyna}
  }
  \subfigure[]{
    \includegraphics[width=0.48\textwidth]{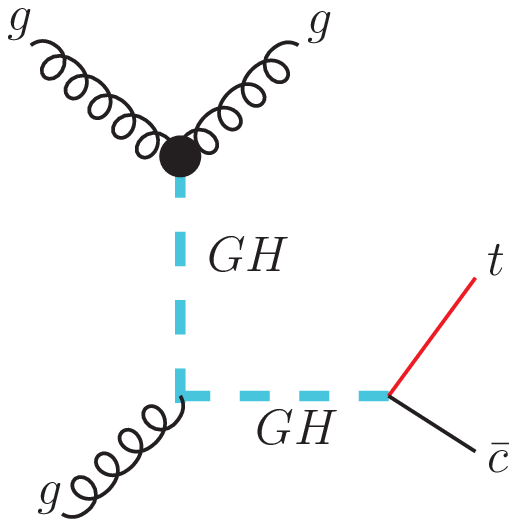}
    \label{fig:ColorongFeynb}
  }
  \caption{Feynman diagrams for single coloron production with decay to $tc$ in association with a gluon. The coloron couples to a gluon (a) in the final state and (b) in the initial state.}
\label{fig:ColorongFeyn}
\end{figure}

\begin{figure}[!h!tbp]
  \centering
  \subfigure[]{
    \includegraphics[width=0.48\textwidth]{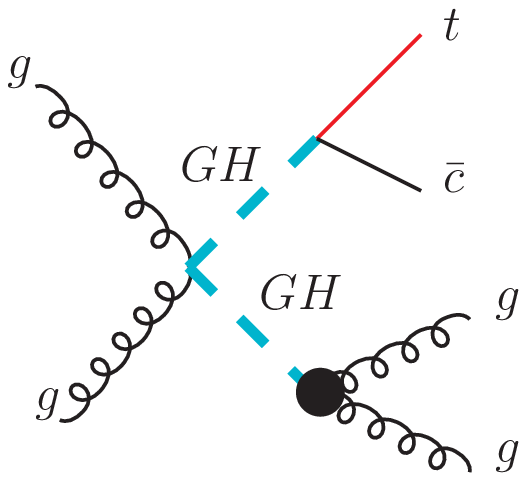}
    \label{fig:ColoronDoubleFeyn1}
  }
  \subfigure[]{
    \includegraphics[width=0.48\textwidth]{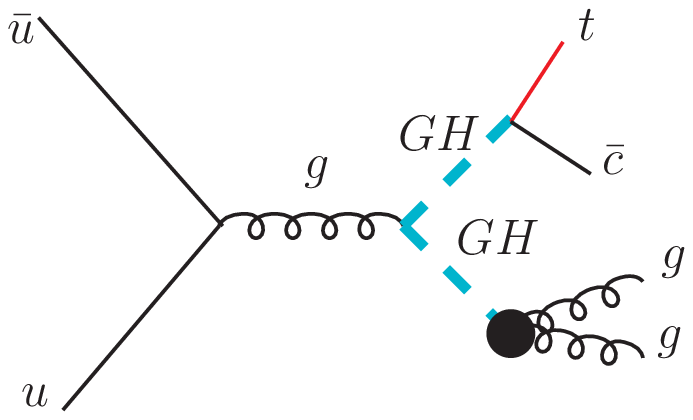}
    \label{fig:ColoronDoubleFeyn2}
  }
  \caption{Feynman diagrams for coloron pair production (a) via $gg$ and (b) via $q\bar{q}$ with decay to $tcgg$.}
    \label{fig:ColoronDoubleFeyn}
\end{figure}

The analysis is focused on the production of a single $G_H$ followed by the decay $G_H \to tc$, as shown in Fig.~\ref{fig:ColoronFeyn}. We note that since $G_H$ has color, the process of single coloron plus jet production includes contributions where the gluon is emitted by the coloron as shown in Fig.~\ref{fig:ColorongFeyna}, and where the coloron couples to an initial-state gluon as shown in Fig.~\ref{fig:ColorongFeynb}. Gluon emission from $G_H$ is described by the third term in Eq.~\ref{eq:GH-gluons}. These two modes are important for inclusive single coloron production. For the parton-level cuts given in Sec.~\ref{sec:eventgen}, these extra gluon processes contribute approximately 22\% of the tree-level cross section at the 8~TeV LHC and a resonance mass of 750~GeV, and approximately 29\% of the tree-level cross section at 14~TeV for a resonance mass of 3000~GeV.
Coloron pair production can also contribute as shown in Fig.~\ref{fig:ColoronDoubleFeyn}.

As discussed in~\cite{Chivukula:2013hga}, some limits on the $(M_{G_H}, \mu)$ parameter space, with $M_{G_H}\lesssim 450$~GeV, can be extracted from the CMS search for new physics in same-sign-dilepton channels~\cite{CMS:rxa} and from the ATLAS~\cite{ATLAS:2012ds} and CMS~\cite{Chatrchyan:2013izb} searches for pair-produced dijet resonances. No bounds are set on the $G_H$ parameter space for $M_{G_H}$ above 450~GeV. 
 
\subsection{Color-triplet}
\label{subsec:triplet}

\begin{figure}[!h!tbp]
  \centering
    \includegraphics[width=0.48\textwidth]{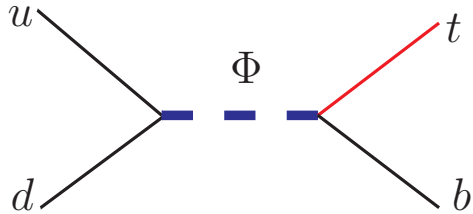}
  \caption{Feynman diagram for production and decay of a color-triplet $\Phi$.}
      \label{fig:TripletFeyn}
\end{figure}

The proton-proton initial state of the LHC makes it particularly sensitive to resonances that carry color. The color-triplet $\Phi$ is a heavy hadronic resonance with fractional electric charge, based on the model introduced in Ref.~\cite{Han:2010rf}. Figure~\ref{fig:TripletFeyn} shows the Feynman diagram for the production of a single color-triplet through colored charged particles, with decay to $tb$. At the LHC, quark-quark fusion with color structure $3\otimes 3$ produces such colored particles. According to the color decomposition $3 \otimes 3 = 6 \oplus \overline{3}$, where $3$, $\overline{3}$, and $6$ are the triplet, anti-triplet and  sextet  representations, the colored particles could be color sextet or anti-triplet, a so-called "diquark"~\cite{Karabacak:2012rn}. Since the LHC is a proton-proton machine, the production of the color-triplet receives an enhancement from the parton luminosity of the quark-quark initial state. Such color sextet and triplet particles are predicted in many new physics models, such as superstring inspired E6 grand unification models~\cite{Hewett:1988xc} and other kinds of diquark models.

The contributing quark-quark initial states are $QQ$, $QU$, $QD$, and $UD$, where $Q, U, D$ denote SM quark doublet, up-type singlet, and down-type singlet, respectively. The diquark particles could be the spin-$0$ scalars with $SU(3)\times SU(2)_L\times U(1)_Y$ quantum numbers

\bea
	\Phi \simeq (6 \oplus \overline{3}, 3, \frac13), \quad \Phi_U \simeq (6 \oplus \overline{3}, 1, \frac13),
\eea

\noindent and the spin-$1$ vectors

\bea
	V^\mu_U \simeq  (6 \oplus \overline{3}, 2, \frac56), \quad V^\mu_D \simeq  (6 \oplus \overline{3}, 2, -\frac16).
\eea

\noindent To produce the $tb$ final state, the charge of the colored particle needs to be $\frac13$. The gauge invariant Lagrangian can be written as~\cite{Han:2010rf}

\begin{eqnarray}
\mathcal{L}_{\rm diquark}
& = & K^j_{ab}\ \left[
\kappa_{\alpha\beta}\  \overline{Q^C_{\alpha a}}i\sigma_2 \Phi^{j} Q_{\beta b}   + \lambda_{\alpha\beta}\ \Phi_U \overline{D^C_{\alpha a}}U_{\beta b} \right.
\nonumber\\
&&
\left. +\lambda^U_{\alpha\beta}\ \overline{Q^C_{\alpha a}}i\sigma_2\gamma_\mu{V^{j}_U}^\mu U_{\beta b}
+\lambda^D_{\alpha\beta}\ \overline{Q^{C}_{\alpha a}}i\sigma_2\gamma_\mu{V^{j}_D}^\mu D_{\beta b} \right] 
+\rm{h.c.},
\label{diquark.EQ}
\end{eqnarray}

\noindent where $\Phi^j = {1\over 2}\sigma_{k} \Phi_{k}^{j}$ with the $SU(2)_{L}$ Pauli matrices $\sigma_{k}$ and color factor $K^j_{ab}$. The coupling to $QQ$ is given by $\kappa_{\alpha\beta}$, and the coupling to $U$ and $D$ by $\lambda_{\alpha\beta}$. Here $a, b$ are quark color indices, and $j$ the diquark color index with $j=1-N_D$, where $N_D$ is the dimension of the ($N_D=3$) anti-triplet or ($N_D=6$) sextet representation. $C$ denotes charge conjugation, and $\alpha,\beta$ are the fermion generation indices. After electroweak symmetry breaking, all of the SM fermions are in the mass eigenstates. The relevant couplings of the colored diquark to the top quark and the bottom quark are then given by

\begin{eqnarray}
\mathcal{L}_{qqD} &=&  K_{ab}^{j} \left[  \kappa^\prime_{\alpha\beta}    \Phi    
                                      \overline{u^c}_{\alpha a} P_\tau d_{\beta b}\right.\nonumber\\
									  && \left. 									  
  +  \lambda^\prime_{\alpha\beta} V_{D}^{j\mu} \overline{u^c}_{\alpha a} 
\gamma_{\mu}P_\tau d_{\beta b} \right]+ \mathrm{h.c.},\label{eq:Lagrn}
\end{eqnarray}

\noindent where $P_\tau = \frac{1\pm \gamma_5}{2}$ are the chiral projection operators. Assuming that the flavor-changing neutral coupling is small, the third-generation couplings are

\begin{eqnarray}
	\mathcal{L}_{\rm top} &=&  K_{ab}^{j} \Phi \overline{t^c}_\alpha P_\tau b_\beta + 
	K_{ab}^{j} V^\mu \overline{t^c}_\alpha \gamma_\mu P_\tau b_\beta  + h.c.
\end{eqnarray}

The decay width of the color-triplet to $tb$ is given by
\begin{equation}
	\Gamma (\Phi \to t\,b ) = \frac{g_{\Phi}^2}{8\pi}(1-x_t^2)^2 + {\mathcal O}(x_f\times x_b) + {\mathcal O}(x_b^2) \;,
\end{equation}
where $x_t=m_t/m_\Phi$ and $x_b=m_b/m_\Phi$ and the color-triplet coupling to $tb$ is given by $g_{\Phi}$.

\subsection{$\bf \Wp$}
\label{subsec:wp}
\begin{figure}[!t!tbp]
  \centering
    \includegraphics[width=0.48\textwidth]{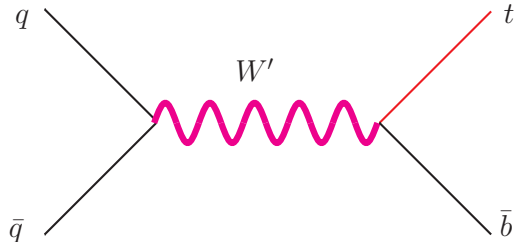}
  \caption[]{Feynman diagram of the $\Wp$ single top production.}
  \label{fig:WPrimeFeyn}
\end{figure}

Many BSM theories ($G(221)$ models \cite{Hsieh:2010zr,Chivukula:2006cg}, RS theories, and composite Higgs models among many others) consider an extended electroweak gauge symmetry and predict the existence of a heavy $W$-prime ($\Wp$) resonance. In many motivated scenarios, $\Wp$ has a relevant decay branching ratio to third generation quarks. This analysis focuses on the $\Wp \to tb$ channel (Fig.~\ref{fig:WPrimeFeyn}) in a general Lorentz invariant parametrization of the $\Wp$ interaction with fermions~\cite{Sullivan:2002jt, Duffty:2012rf}:
 
\begin{equation}\label{eq:L-Wprime}
  \Lagr =  \frac{g}{2\sqrt{2}}V^{'}_{i,j}\overline{f}_i\gamma_{\nu}(g^{'}_{R}(1 + \gamma^5)+g^{'}_{L}(1-\gamma^5))W^{'\nu}f_j+H.c. \ ,
\end{equation}

\noindent where $g=g_{SM}=e/\sin\theta_W$ is the SM weak coupling, $V^{'}_{i,j}$ corresponds to the CKM matrix in the case of $\Wp$ interactions with quarks and to the $\delta_{i,j}$ diagonal matrix for the $\Wp$ interactions with leptons, although in principle, $V^{'}_{i,j}$ could be different for left-handed and right-handed couplings of $\Wp$~\cite{Buras:2010pz}. The right-handed and left-handed gauge couplings to fermions are given by $g^{'}_{R}g$ and $g^{'}_{L}g$, respectively. Two coupling scenarios are considered without loss of generality:

\begin{align}
\begin{split}
 & g^{'}_{R}=1 \qquad g^{'}_{L}=0 \qquad \text{right-handed $\Wp$} \,, \\
 & g^{'}_{R}=0 \qquad g^{'}_{L}=1 \qquad \text{left-handed $\Wp$} \,.
\end{split}
\end{align}
\noindent
The direct coupling of $\Wp$ to SM gauge bosons is neglected. Interference with the SM $W$~boson is considered for the case of left-handed couplings~\cite{Boos:2006xe}. Moreover, it is demonstrated that contributions from the production of off-shell top quarks is relevant especially for $\Wp$~bosons with left-handed couplings; thus the process $\Wp \rightarrow W b \bar{b}$ is considered. The $\Wp$ is produced via the Drell-Yan process (Fig.~\ref{fig:WPrimeFeyn}). The left-handed (right-handed) $\Wp$ width-over-mass ratio is of the order of 0.08 $(g^{'}_L g)^2$ (0.08 $(g^{'}_R g)^2$) and the $\Wp \to tb$ branching ratio is about $25\%$ \cite{ Sullivan:2002jt, Sullivan}.  

ATLAS and CMS have recently performed searches for $W$-prime's in several decay channels, including $l\nu$~\cite{ATLAS-CONF-2014-017, CMS:2013rca}, hadronic decay~\cite{Aad:2014aqa,Chatrchyan:2013qha}, $WZ$~\cite{Aad:2014pha,Khachatryan:2014xja} and the $tb$ mode~\cite{Aad:2014xea,Aad:2014xra,Chatrchyan:2014koa}. The sensitivity at the 8~TeV LHC in the $tb$ decay mode approaches 2000~GeV for a SM-like $\Wp$ coupling to $tb$. However, these $\Wp$ searches focus on the high-mass reach for $\Wp$ and are not sensitive in the lower mass region where a $\Wp$ with smaller couplings might be hiding. The CMS lepton+jets analysis starts at $m_{\Wp}=800$~GeV, while the ATLAS single top analysis in the hadronic channel starts at $m_{\Wp}=1500$~GeV. Sensitivity at lower resonance masses is also provided by the Tevatron searches~\cite{Aaltonen:2009qu,Abazov:2011xs}.

The sensitivity of future hadron colliders to $\Wp \to tb$ was also explored for Snowmass~2013~\cite{Drueke:2013wsa}. Sensitivity up to several TeV can be achieved.

\subsection{Event generation}
\label{sec:eventgen}
All samples are generated at leading order (LO) using MadGraph5~\cite{Alwall:2011uj}, including both top and antitop production with leptonic (electron or muon) decays of the $W$~boson from the top-quark decay. The top-quark mass is set to 172.5~GeV~\cite{ATLAS:2014wva}, though off-shell top quarks are allowed in the event generation as discussed below. The parton distribution function (PDF) set CT10 is used~\cite{Lai:2010vv}, and the factorization and renormalization scales are set to the resonance particle mass unless otherwise indicated.

$KKg$ production uses the mixing parameter $\cot\omega = 2.6$, see Sec.~\ref{subsubsec:kkg}. For the coloron, only the tree-level diagram shown in Fig.~\ref{fig:ColoronFeyn} is included in the event generation in this parton-level analysis.
Coloron pair production is also considered since a mixed final state can mimic single coloron production. It produces the relevant final state if one coloron decays to $t\bar{c}$ or $\bar{t}c$ and the other to two gluons. All colorons have a branching ratio to $tc$ of 50\%. Color-triplet production and decay occurs with coupling $g_{\Phi}=0.3$, which corresponds to a branching fraction to $tb$ of roughly $1/3$. The same value is used for coupling to the first and third generations. In principle, the coupling is constrained by $D$~meson mixing for the second generation of quarks~\cite{Chen:2009xjb}. However, this analysis is not sensitive to second generation couplings.

\begin{figure}[!h!tbp]
  \centering
  \subfigure[]{
    \includegraphics[width=0.48\textwidth]{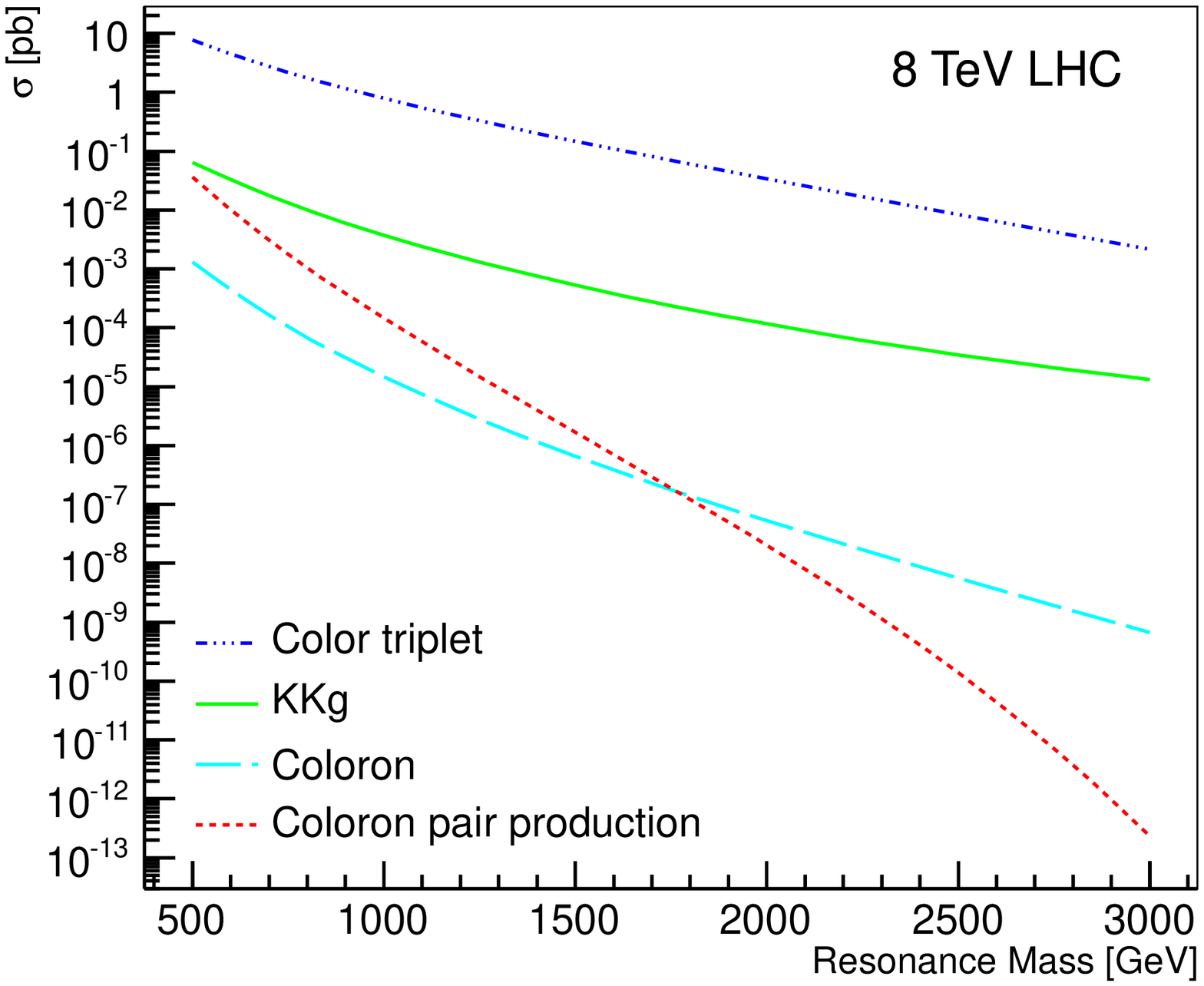}
    \label{fig:colcolkkga}
  }
  \subfigure[]{
    \includegraphics[width=0.48\textwidth]{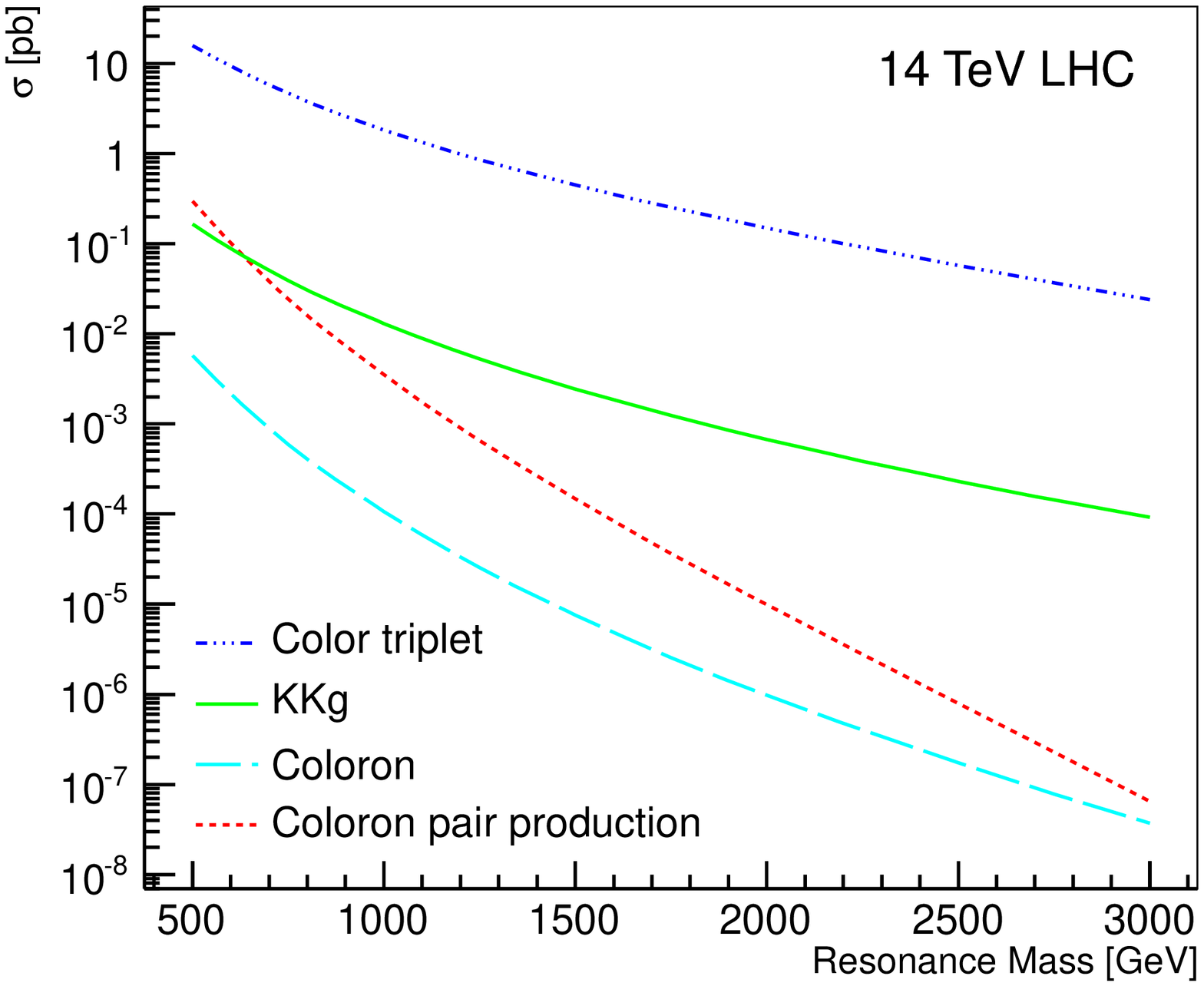}
    \label{fig:colcolkkgb}
  }
  \caption[]{Cross section as a function of resonance mass for $KKg$, color-triplet, single coloron~$\to tc$ and coloron pair~$\to tcgg$ production at a proton-proton collider at (a) 8~TeV and (b) 14~TeV.}
    \label{fig:colcolkkg}
\end{figure}

The cross sections for $KKg$ and color-triplet production, as well as single- and pair-production of colorons, are shown in Fig.~\ref{fig:colcolkkg} for two different proton-proton collider energies. The color-triplet cross section is largest since it benefits from the quark-quark initial state (with a small contribution from the antiquark-antiquark initial state). 
The $KKg$ initial state is $q\overline{q}$, hence the cross section is suppressed compared to color-triplet production. 

The initial state particles in single coloron production are both gluons, which result in a large cross section at low coloron mass. The cross section quickly decreases as the coloron mass increases. The initial state particles are similar in coloron pair production, although there is a $u \bar{u}$ initial state as well, see Fig.~\ref{fig:ColoronDoubleFeyn2}. An additional pattern visible in the coloron cross sections is that the pair production cross sections decrease more quickly than the single production one due to the limited available phase space. In general, the coloron cross sections are lower than those of the $KKg$ and color-triplet, reflecting the differences in coupling and initial states.

When increasing the collider energy from 8~TeV to 14~TeV, the color-triplet production only increases by a factor two, while the $KKg$ and single coloron production cross sections increase by about an order of magnitude, especially at higher masses. This is a result of gluons and anti-quarks in the initial state for $KKg$ and single coloron production. The coloron pair production increases even more due to the increase in available phase space.

\begin{figure}[!h!tbp]
  \centering
  \subfigure[]{
    \includegraphics[width=0.48\textwidth]{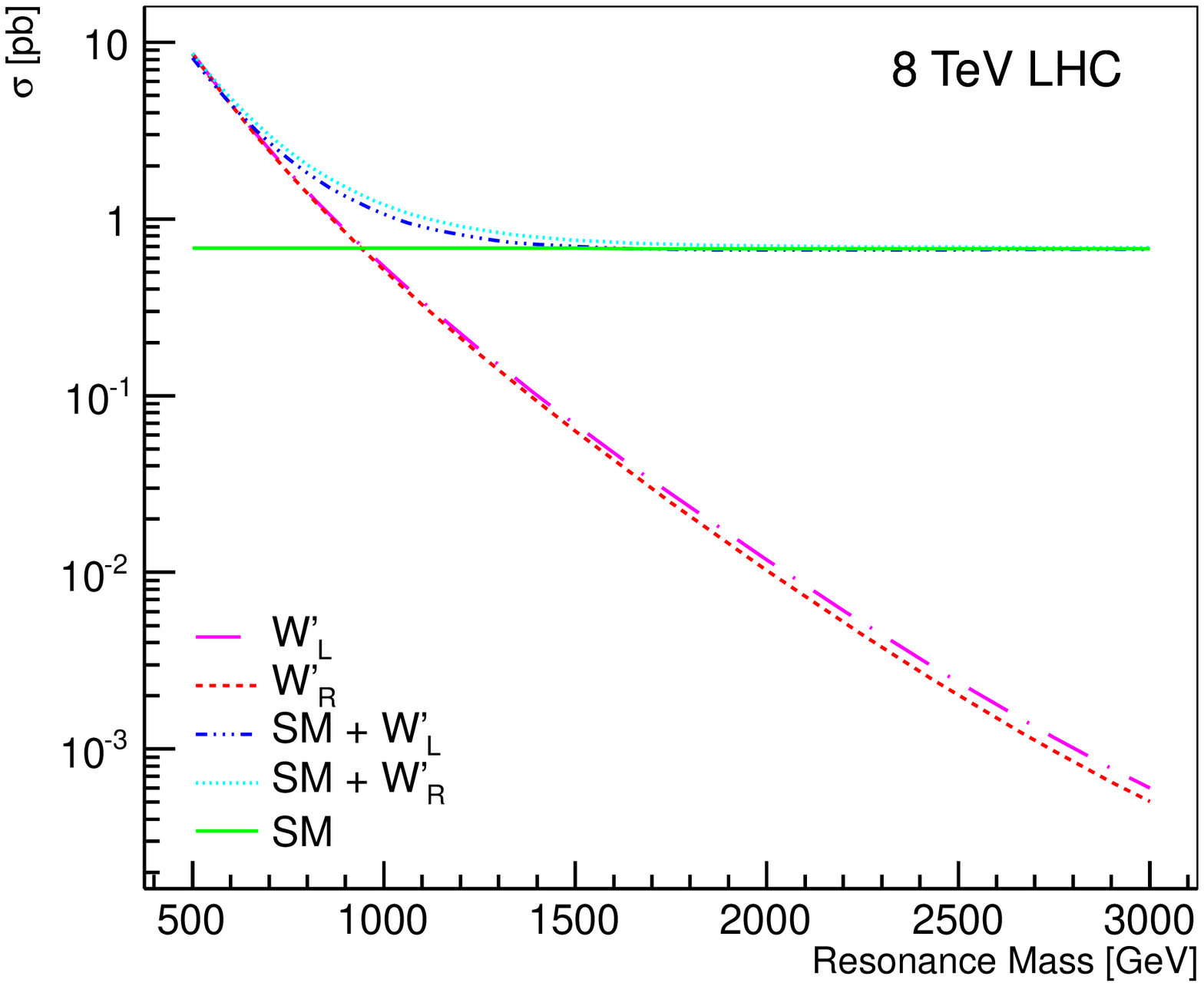}
    \label{fig:wpxseca}
  }
  \subfigure[]{
    \includegraphics[width=0.48\textwidth]{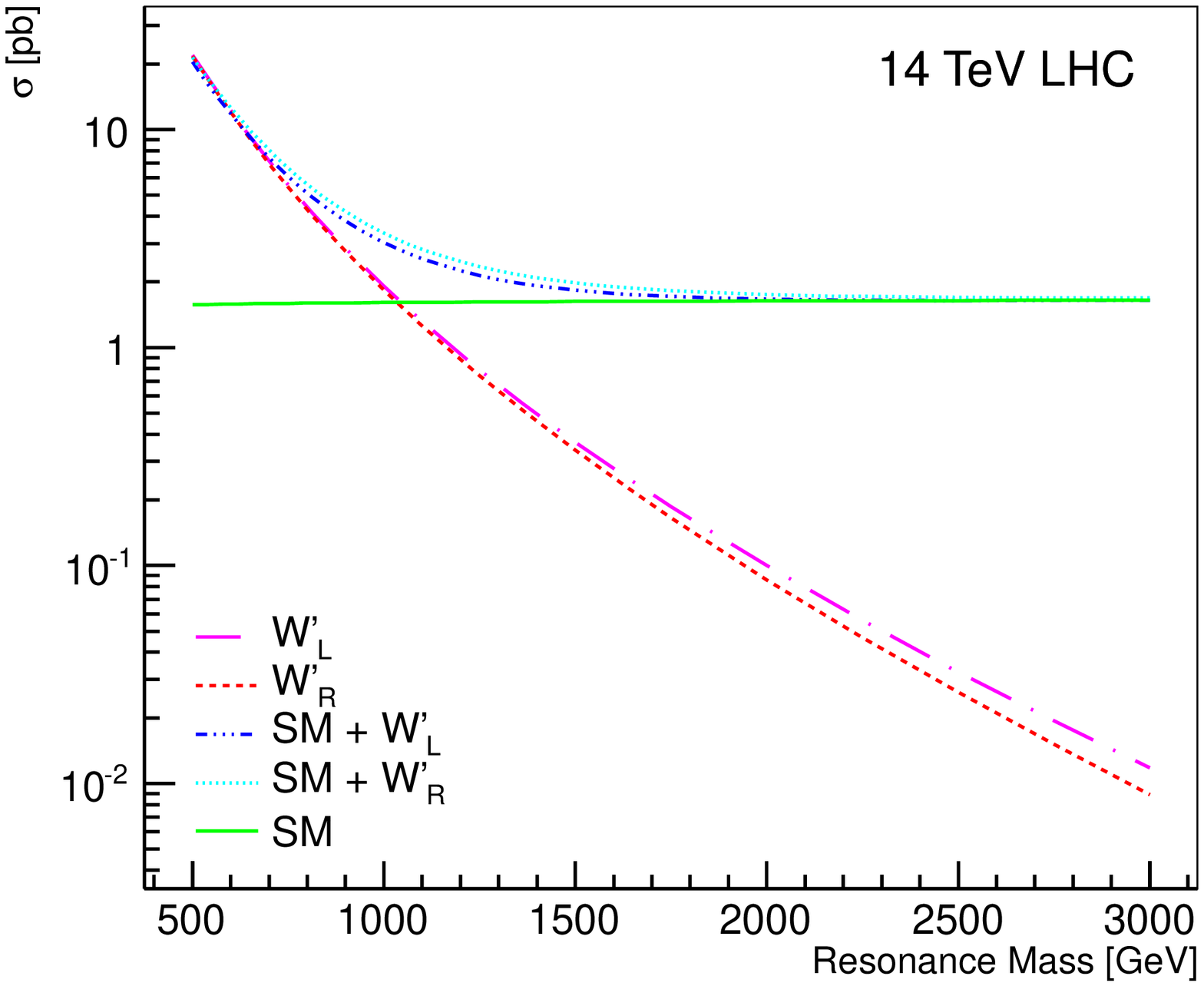}
    \label{fig:wpxsecb}
  }
  \caption[]{Cross section as a function of resonance mass for $\Wp$ production with different couplings and for $\Wp$ production together with SM $s$-channel production at a proton-proton collider at (a) 8~TeV and (b) 14~TeV.}
\label{fig:wpxsec}
\end{figure}

$\Wp$~boson samples are generated using the model described in Sec.~\ref{subsec:wp} for two $\Wp$ coupling scenarios: purely left-handed couplings ($\Wp_L$, with $g'_L=1$, $g'_R=0$) and purely right-handed couplings ($\Wp_R$, with $g'_L=0$, $g'_R=1$). The cross sections for $\Wp_L$ and $\Wp_R$ production are shown in Fig.~\ref{fig:wpxsec} for two different proton-proton collider energies.

\begin{figure}[!h!tbp]
  \centering
  \subfigure[]{
    \includegraphics[width=0.48\textwidth]{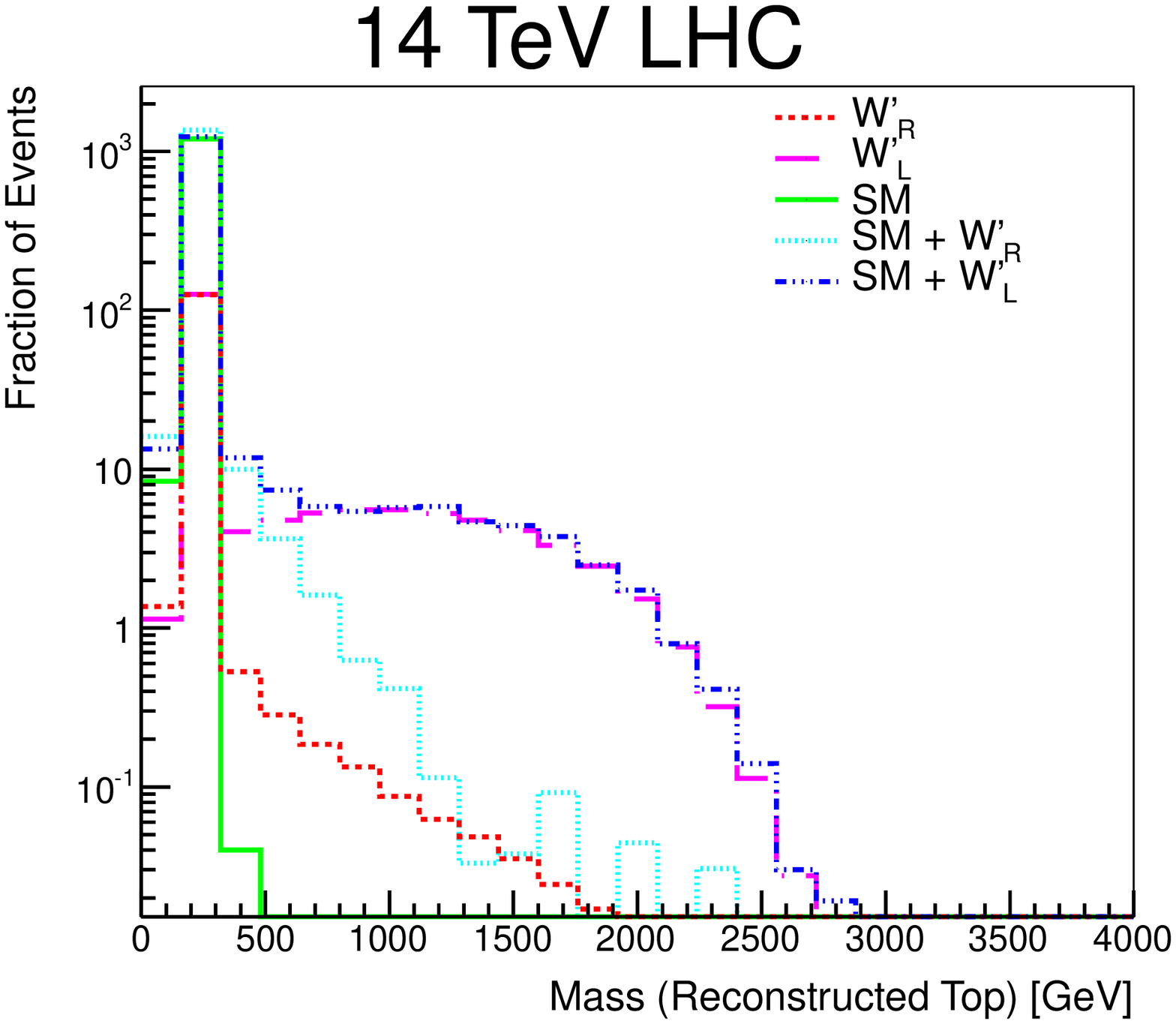}
    \label{fig:wpinttop}
  }
  \subfigure[]{
    \includegraphics[width=0.48\textwidth]{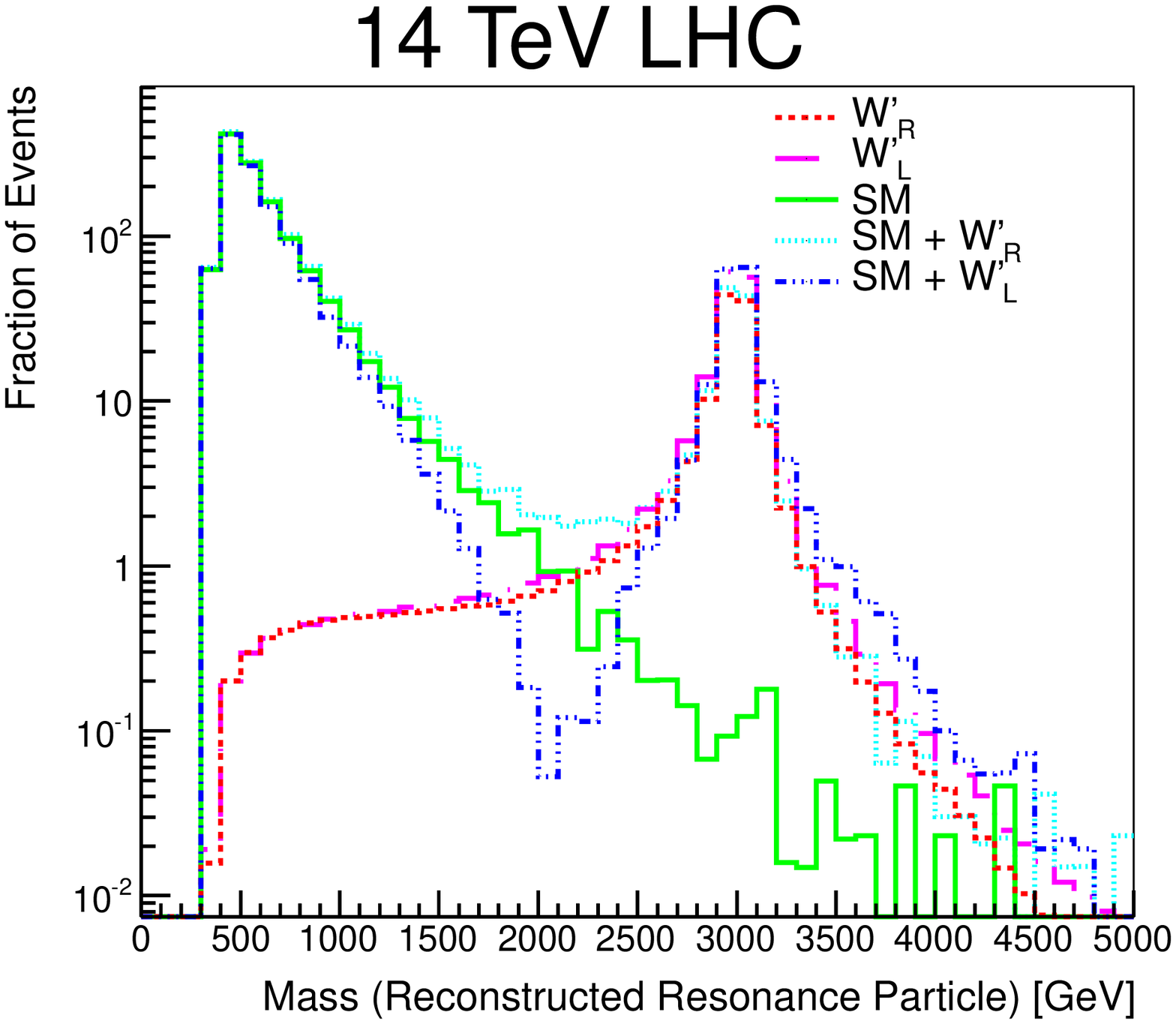}
    \label{fig:wpintpart}
  }
  \caption[]{Reconstruction of (a) the mass of the top quark and (b) the mass of the resonance particle for $\Wp$ signals with and without SM single top production, at a 14~TeV proton-proton collider. (Color online.) }
\label{fig:wpint}
\end{figure}

The left-handed and right-handed couplings of the $\Wp$ are principally the same, hence the production cross sections should be exactly the same. However, this is not the case. For the low $\Wp$~mass region in Fig.~\ref{fig:wpxsec}, the $\Wp_L$ and $\Wp_R$ cross sections are approximately the same, but they diverge for higher masses. The difference is as large as 20\% for a $\Wp$ mass of 3000~GeV. At such high resonance masses, the production of off-shell top quarks becomes relevant. A top quark produced from a $\Wp_R$ decay has to undergo a spin flip before it can decay via the left-handed weak interaction, hence the production of high-mass off-shell top quarks is suppressed in $\Wp_R$ production. This large off-shell top contribution for $\Wp_L$ production can also be seen in the parton-level invariant mass of the $Wb$ system shown in Fig.~\ref{fig:wpinttop}. Near 172.5~GeV, both $\Wp_L$ and $\Wp_R$ peak in the same location. When moving to higher masses, $\Wp_L$ production has a broad shoulder extending out to high virtual top-quark masses. Previous experimental analyses have not considered these off-shell top quarks that could potentially enhance the sensitivity to $\Wp_L$ production.

The production of $\Wp_L$ interferes with SM single top-quark production in the $s$-channel, hence the cross section for SM+$\Wp_L$ production is different from SM+$\Wp_R$ production at low $\Wp$ masses. This interference effect can also be seen as a dip in the parton-level distribution of the invariant mass of the $Wb\bar{b}$ system, shown in Fig.~\ref{fig:wpintpart}.

The largest backgrounds to the single top plus jet signature are from $W$~boson plus jet production, from SM single top production, from top-quark pair production, and from diboson production. Backgrounds are generated with MadGraph5~\cite{Alwall:2011uj}, using the generator given in Eq.~\ref{eq:MGcuts}. In particular the jet cuts are the same as those in the event selection in Eq.~\ref{eq:cuts}. $W$+jets events are generated as a $W$~boson in association with exactly two objects ($Wjj$), which can consist of light quarks, gluons, charm quarks or bottom quarks. Since this analysis is at the parton level, no parton showering is included and thus no matching procedure is required. The kinematic distributions we emphasize in this paper are not sensitive to the details of the modeling of low-$p_T$ jets in $W$+jets events.

The standard MadGraph5 generator cuts are used for quarks and gluons ($j$) and leptons ($l$): 
\begin{eqnarray}
\textrm{Quarks and gluons with } \qquad p_{T}^j&\geq&20\,{\rm GeV},\qquad
\left|\eta_{j}\right|\leq 5\,,\nonumber \\
\textrm{Lepton with } \qquad p_{T}^l&\geq&10\,{\rm GeV},\qquad
\left|\eta_{l}\right|\leq 2.5\,,\nonumber \\
j\textrm{-} j\textrm{ separation of } \qquad \Delta R &>& 0.4\,, \nonumber \\
j\textrm{-lepton separation of } \qquad \Delta R &>& 0.4\,. 
\label{eq:MGcuts}
\end{eqnarray}

Top pair production includes lepton+jets and dilepton decay modes (where lepton refers to an electron or muon). Single top production includes $t$-channel, $Wt$ and $s$-channel with decay to lepton+jets. Diboson production includes $WW$ and $WZ$, with one $W$~boson decaying to lepton/neutrino, and the other $W$~boson or the $Z$~boson decaying to jets.


Despite the simplistic modeling of the background processes, their event yield is in agreement within a factor of two of those in the ATLAS $W'\rightarrow t b$ search~\cite{Aad:2014xea}.
While we study these backgrounds and how to separate them from the signal only at the 8~TeV LHC, 
the relative importance of the different backgrounds is similar at the 14~TeV LHC, with the main change being that $W$+jets will be less important and that the top pair background will be more important.

\subsubsection{Cross section uncertainty}
\label{subsubsec:uncertainties}

We compute both the scale and the PDF uncertainties for the LO signal cross sections. To evaluate the scale uncertainty, we vary the factorization and renormalization scales up and down by a factor of two (together) from the default value, which is the resonance particle mass. The PDF uncertainty is evaluated by finding the largest variation among the set of CT10 uncertainty eigenvectors.
The resulting uncertainties are listed in Table~\ref{tab:unc8} for the 8~TeV LHC and~\ref{tab:unc14} for the 14~TeV LHC. 

\begin{table}[!h!tbp]
  \begin{center}
    \caption{Scale and PDF uncertainties for the different signals at a 8~TeV proton-proton collider.}
    \label{tab:unc8}
    \renewcommand{\arraystretch}{1.4}
    \begin{tabular}{|l|c|c|c|}                                                        
      \hline
       & \multicolumn{3}{c|}{8~TeV collider energy} \\
      \hline
      Signal & Cross Section & \multicolumn{2}{c|}{uncertainty [\%]} \\
      $m=750$~GeV&    [fb]       & Scale & PDF \\   
      \hline
      $\Wp_{R}$           & 1800     & 5.4 & 0.15 \\ 
      $\Wp_{L}$           & 1900     & 5.5 & 0.07 \\      
      $KKg$              & 13       & 24  & 0.27 \\ 
      Single coloron     & 0.10     & 34 & 0.08 \\
      Coloron pair       & 1.8      & 71 & 0.29 \\
      Color-triplet      & 2200     & 5.0 & 0.35 \\
     \hline
    \end{tabular}
  \end{center}
\end{table}

\begin{table}[!h!tbp]
  \begin{center}
    \caption{Scale and PDF uncertainties for the different signals at a 14~TeV proton-proton collider.}
    \label{tab:unc14}
    \renewcommand{\arraystretch}{1.4}
    \begin{tabular}{|l|c|c|c|}                                                        
      \hline
       & \multicolumn{3}{c|}{14~TeV collider energy} \\
      \hline
      Signal & Cross Section & \multicolumn{2}{c|}{uncertainty [\%]} \\
      $m=3000$~GeV&    [fb]       & Scale & PDF \\   
      \hline
      $\Wp_{R}$           & 8.9      & 9.9 & 0.09 \\ 
      $\Wp_{L}$           & 12       & 10 & 0.30 \\      
      $KKg$              & 0.09     & 23 & 0.21 \\ 
      Single coloron     & 3.74$\times 10^{-5}$ & 38 & 0.35 \\
      Coloron pair       & 6.5$\times 10^{-5}$ & 79 & 0.93 \\
      Color-triplet      & 24         & 9.6 & 0.15 \\
     \hline
    \end{tabular}
  \end{center}
\end{table}

The scale uncertainties for $\Wp$ and color-triplet double (from about 5\% to about 10\%) when going from 8~TeV to 14~TeV, while the uncertainties for $KKg$ and coloron remain large but unchanged.
The scale uncertainty of single coloron and coloron pair production are large due to the two initial state gluons. For single coloron production, the scale uncertainty is increased by the contribution of the additional gluon radiation diagrams from Fig.~\ref{fig:ColorongFeyn}. The additional gluon has a steeply falling $p_T$ distribution which makes it sensitive to the cuts from Eq.~\ref{eq:MGcuts}. The scale uncertainty for coloron pair production is even larger due to the production of two high-mass resonance particles. If a scale of twice the resonance mass is chosen, then the double coloron production scale uncertainty at 14~TeV is only 37\%. Coloron and $KKg$ production will all benefit from a NLO computation of the cross section~\cite{Chivukula:2011ng}.
Tables~\ref{tab:unc8} and~\ref{tab:unc14} show that the PDF uncertainties are negligible for the signals considered here, they are all less than 1\%. Other uncertainties (from top-quark mass or $\alpha_s$ or weak corrections) are also small.


\section{Analysis}
\label{sec:analysis}

We study single top plus jet resonances in two kinematic regions, each at a different collider energy. The low-mass analysis focuses on a resonance mass of 750~GeV at a collider energy of 8~TeV. Here, the individual signals are compared with the SM backgrounds. The high-mass analysis focuses on events with 3000~GeV resonance mass at a collider energy of 14~TeV, comparing kinematic distributions between different signals. 

The analysis is based on parton-level information from MadGraph~5. We select objects and apply cuts to model the ATLAS and CMS selection~\cite{Aad:2014xra,Chatrchyan:2014koa}. Jets are reconstructed from quarks and gluons using a cone algorithm with a cone size of $\Delta R=0.4$. Jets are required to have $p_{T}>25$~GeV and $|\eta|<2.5$. Random sampling is used to assign $b$-tag information to jets. A jet containing a $b$~quark is tagged with a probability of 80\%, a jet containing a $c$~ quark with a  probability of 10\%, and all other jets (containing light quarks and gluons) are tagged with a probability of 1\%. 

Leptons are required to be isolated, $\Delta R($lepton,~jet$)>0.2$. The missing transverse energy $\etmiss$ is calculated by adding the neutrino four-vector to the four-vectors of any jets with $p_T<25$~GeV or with $|\eta |>2.5$.

With these object definitions, events are required to pass a series of selection cuts: 
\begin{eqnarray}
\textrm{2 or 3 jets,} & \nonumber \\
\textrm{Leading jet:} & \qquad p_{T}^{j1}>150\;{\rm GeV}, \nonumber \\
\textrm{Second jet:} & \qquad p_{T}^{j2}>60\;{\rm GeV}, \nonumber \\
\textrm{At least one b-tagged jet, } & \nonumber \\
\textrm{Exactly one lepton (electron or muon): } & \qquad p_{T}^{\ell}>25\,{\rm GeV},\qquad
\left|\eta_{\ell}\right|<2.5,\nonumber \\
\textrm{Missing transverse energy:} & \qquad ~\etmiss >25~{\rm GeV}. \nonumber \\
\label{eq:cuts}
\end{eqnarray}

The acceptances for the different signals to pass these cuts are shown in Table~\ref{tab:acceptances} for the various signals. At the lower resonance mass of 750~GeV, the acceptance is 30\% to 43\%, with a much smaller acceptance for coloron pair production due to its larger number of jets. The acceptance goes up to around 50\% to 70\% for the higher resonance mass, and there is a larger variation in acceptance between the different signals due to their different couplings and initial states, which affect the event kinematics.

\begin{table}[!h!tbp]
  \begin{center}
    \caption{Acceptances for single top resonance signals to pass basic selection cuts (Eq.~\ref{eq:cuts}).}
    \label{tab:acceptances}
    \renewcommand{\arraystretch}{1.4}
    \begin{tabular}{|l|c|c|}
      \hline
      Resonance particle            & 8~TeV, & 14~TeV, \\
                                    & 750~GeV mass        & 3000~GeV mass \\
      \hline
      $\Wp_{R}$                      & 0.40 & 0.72 \\ 
      $\Wp_{L}$                      & 0.39 & 0.74 \\ 
      $KKg$                         & 0.30 & 0.52 \\ 
      Single coloron                & 0.39 & 0.68 \\ 
      Coloron pair                  & 0.07 & 0.05 \\ 
      Color-triplet                 & 0.43 & 0.82 \\ 
     \hline
    \end{tabular}
  \end{center}
\end{table}

\begin{figure}[!h!tbp]
  \centering
  \subfigure[]{
    \includegraphics[width=0.48\textwidth]{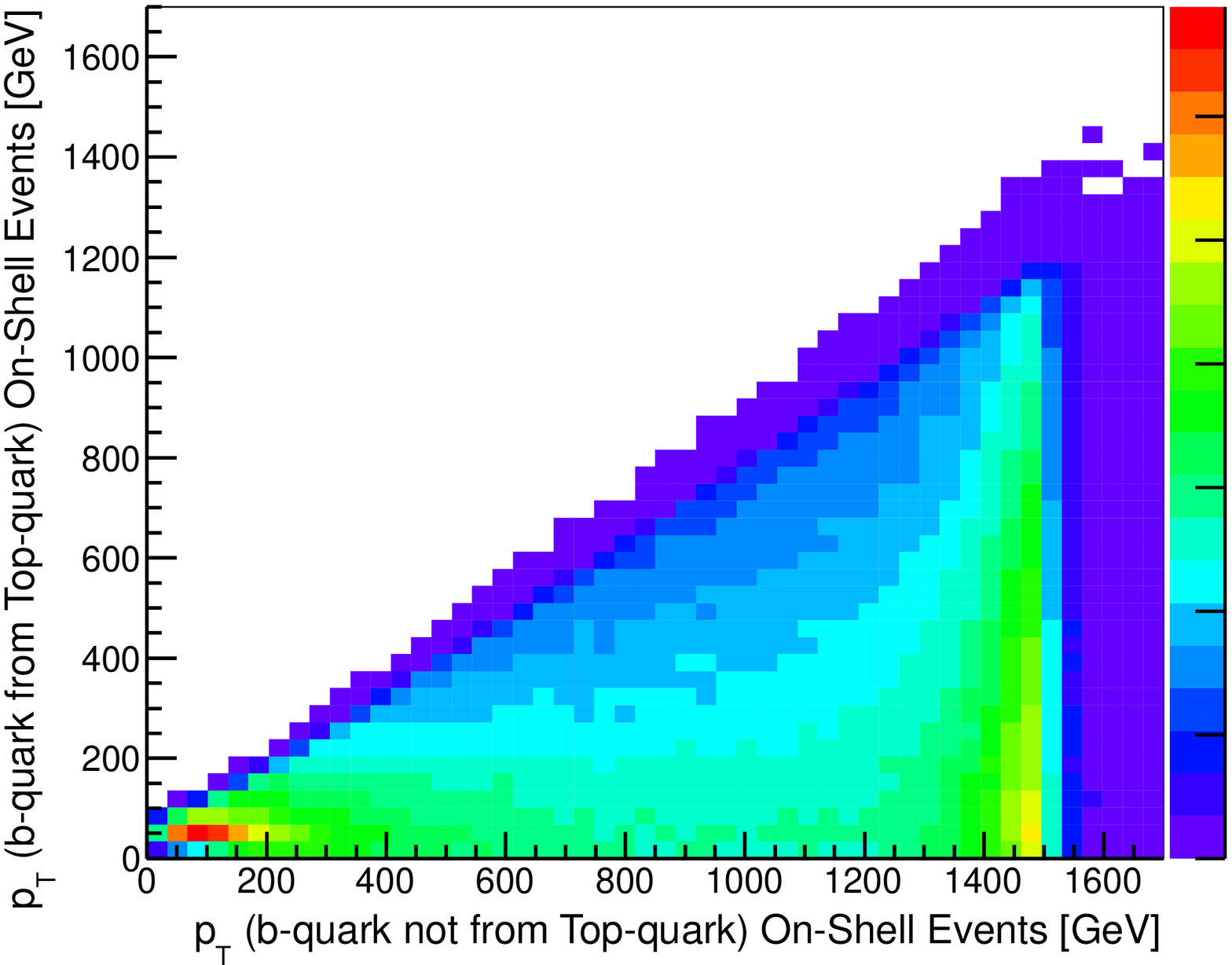}
    \label{fig:jet1vjet2wpr}
  }
  \subfigure[]{
    \includegraphics[width=0.48\textwidth]{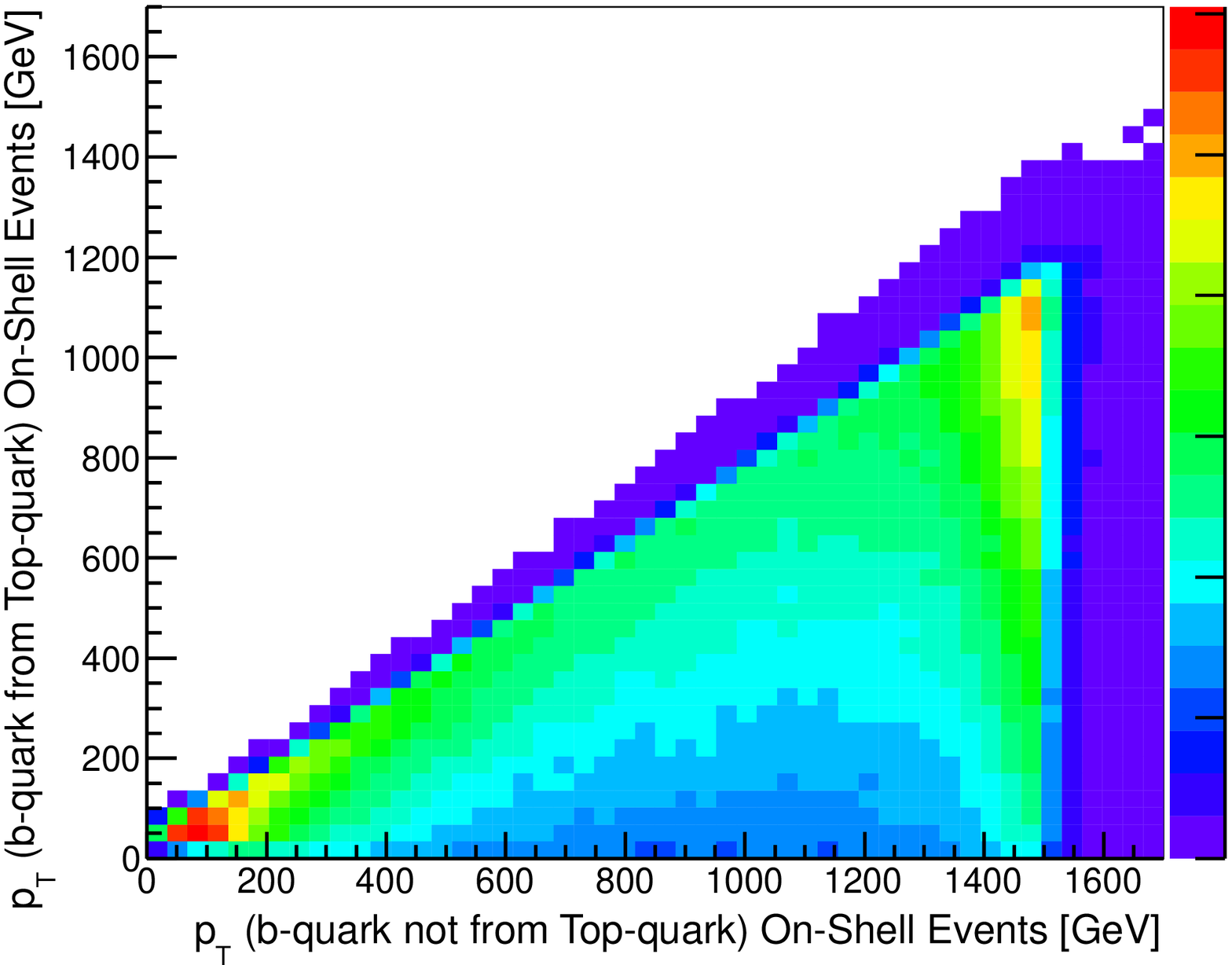}
    \label{fig:jet1vjet2wpl}
  }
  \caption{Distribution of the $p_T$ of the $b$~quark from the top-quark decay vs the $p_T$ of the quark produced together with the top quark in the event for $\Wp$ production at $m_{\Wp}=3000$~GeV at the 14~TeV LHC, for \subref{fig:jet1vjet2wpr} $\Wp_R$ and \subref{fig:jet1vjet2wpl} $\Wp_L$. The event count per bin follows rainbow colors. (Color online)}
\label{fig:jetvjetpt}
\end{figure}

The $W$~boson is reconstructed from the $x$, $y$ and $z$ components of the lepton and the $x$ and $y$ components of the $\etmiss$, using the parton-level longitudinal neutrino momentum for simplicity. This neutrino $p_Z$ is not available experimentally, where a $W$~boson mass constraint is commonly used. However, this choice does not have a big impact on our analysis which is not critically dependent on the longitudinal neutrino momentum. 
The top quark is reconstructed from the $W$~boson and the second jet. We always use the second jet in the event, regardless of $b$-tagging information. Figure~\ref{fig:jetvjetpt} shows the correlations of the jet transverse momenta for $\Wp$ signals with left-handed and right-handed couplings.

For resonance production at rest, the top quark is produced together with another quark, which for kinematic reasons will always have a momentum larger than the $b$~quark from the top decay. Only if the resonance particle itself receives a boost does this not have to be true. Figure~\ref{fig:jetvjetpt} reflects this relationship for high-$p_T$. Only at low transverse quark momenta does the $b$~quark from the top decay sometimes become the leading jet in the event. Most of these events are removed by the leading jet $p_T$ requirement, c.~f. Eq.~\ref{eq:cuts}. The effect of the $\Wp$ coupling can also be seen in Fig.~\ref{fig:jetvjetpt}. The $b$-quark from the top decay will move in the same direction as the top and thus get higher $p_T$ for a left-handed $\Wp$ decay. This also makes it more sensitive to the effects of limited experimental energy resolution and misreconstruction. The other signals have different couplings and land somewhere between $\Wp_R$ and $\Wp_L$. The $KKg$ is more similar to $\Wp_L$ (though it has some mis-identification as will be discussed in Sec.~\ref{sec:highmass}), while the coloron is more similar to $\Wp_R$. Note that if the top quark is allowed to go off-shell, then this relationship no longer holds, and for events in the high-top-mass tail of Fig.~\ref{fig:wpinttop}, there is no longer any correlation, and the $b$~quark from the top-quark decay is the leading jet half of the time. 

The resonance particle mass is then reconstructed by adding the top quark and the other jets in the event. The same algorithms to reconstruct the $W$~boson, top quark and resonance particle are applied at 8~TeV and at 14~TeV.

We use the kinematic properties of the lepton and jets together with these reconstructed objects to determine variables that separate the signals from the background at the 8~TeV LHC in Sec.~\ref{sec:lowmass}, for events passing the selection cuts from Eq.~\ref{eq:cuts}. We will discuss single $b$-tags and double $b$-tags separately as appropriate. We apply the same selection cuts in the 14~TeV discussion comparing different signal kinematics.


\section{Low-mass analysis}
\label{sec:lowmass}

In the low-mass analysis, a 750~GeV resonance particle mass is studied at a collider energy of 8~TeV. The interest of the LHC searches is understandably in higher mass sensitivity~\cite{Aad:2014xra}. However, as Fig.~\ref{fig:wpxseca} shows, processes other than $\Wp$ or color-triplets have smaller cross sections and existing analyses do not rule them out (and so far do not even consider them). Moreover, it is quite possible that a $\Wp$ exists at low resonance mass, albeit with couplings below the current limits. 

Here we explore kinematic variables that distinguish the various single top resonance signals from the SM backgrounds. Experiments have so far only probed for $\Wp$ production, and we will demonstrate that the kinematic properties of other signals can be very different from $\Wp$. Multivariate analyses as employed by ATLAS~\cite{ATLAS-CONF-2013-050} and CMS~\cite{Chatrchyan:2014koa} may not be sensitive to these signals since they are trained for specific signal kinematics.
The goal of this section is to find variables with strong discriminating power against background in favor of the various signals. Signals and backgrounds are normalized to an integrated luminosity of 20~fb$^{-1}$ with cross sections according to Sec.~\ref{sec:eventgen}. Individual MadGraph event weights are included and the normalization for each simulated event is given by
\begin{eqnarray}
\rm \frac{\textrm{Event weight} \cdot \textrm{Cross section} \cdot \textrm{Luminosity}}{\textrm{Total weight}} \,.
\label{eq:lowmassweighting}
\end{eqnarray}
We note that coloron pair production is not included in this section.

\begin{figure}[H]
  \centering
  \subfigure{
    \includegraphics[width=0.48\textwidth]{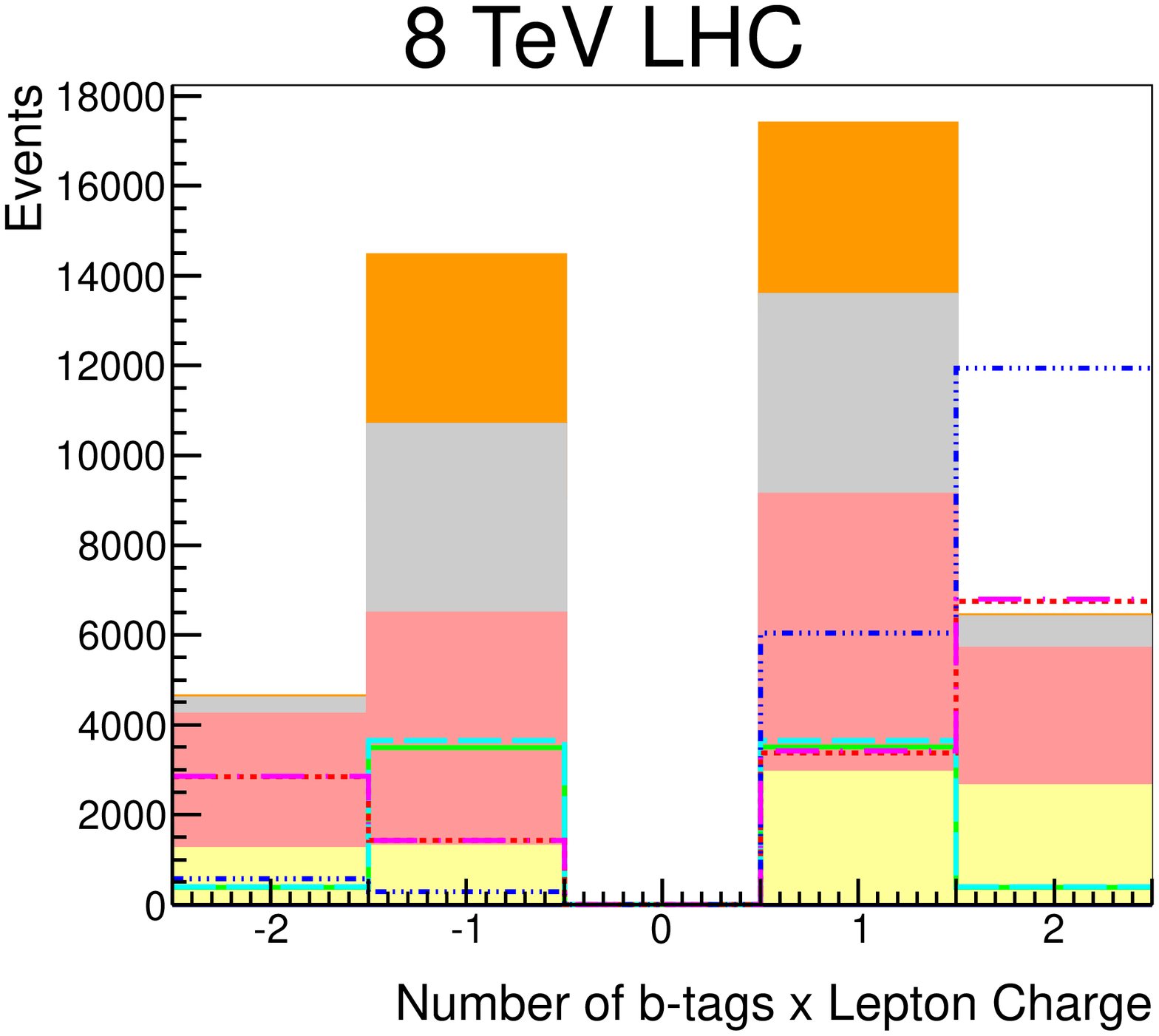}
  }
  \subfigure{
    \includegraphics[width=0.25\textwidth]{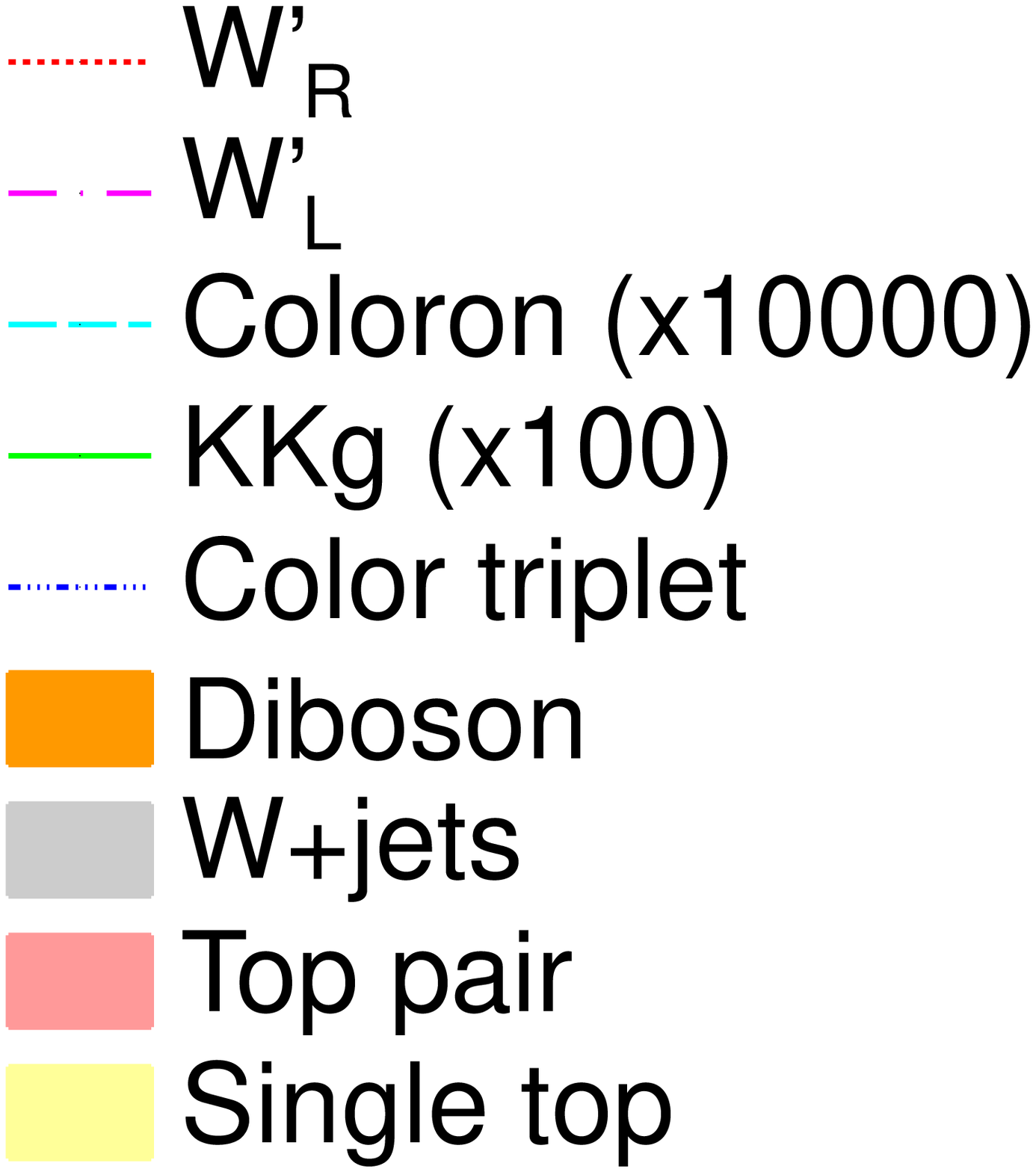}
  }
  \caption[]{Number of $b$-tagged jets multiplied by the charge of the lepton, in 20~fb$^{-1}$ at the 8~TeV LHC, for signal and background processes. (Color online.) }
  \label{fig:lowntagsandlepcharge}
\end{figure}

One of the most differentiating features in the signal and background events is the $b$-tag information. And since the LHC is a proton-proton collider, lepton charge (from the decay of the top quark) not only differentiates the different signals from each other (based on initial state quarks or gluons), but it also separates signals from backgrounds. We multiply the number of $b$-tagged jets by the lepton charge to fully explore this information and show the result in Fig.~\ref{fig:lowntagsandlepcharge}. 

It is clear that background dominates in the 1-tag region,  while the $\Wp$ and color-triplet signals will dominate in the 2-tag region. Only $KKg$ production mainly occupies the 1-tag region, as expected since there is only one $b$-quark in $KKg$ events. Color-triplet production also stands out in that it only occupies the positive lepton charge region due to the unique initial state. We will show kinematic distributions for both lepton charges and 1-tag and 2-tag events combined, except as noted otherwise.

\subsection{Object properties}
\label{subsec:lowobjectproperties}

\begin{figure}[!h!tbp]
  \centering
  \subfigure[]{
    \includegraphics[width=0.48\textwidth]{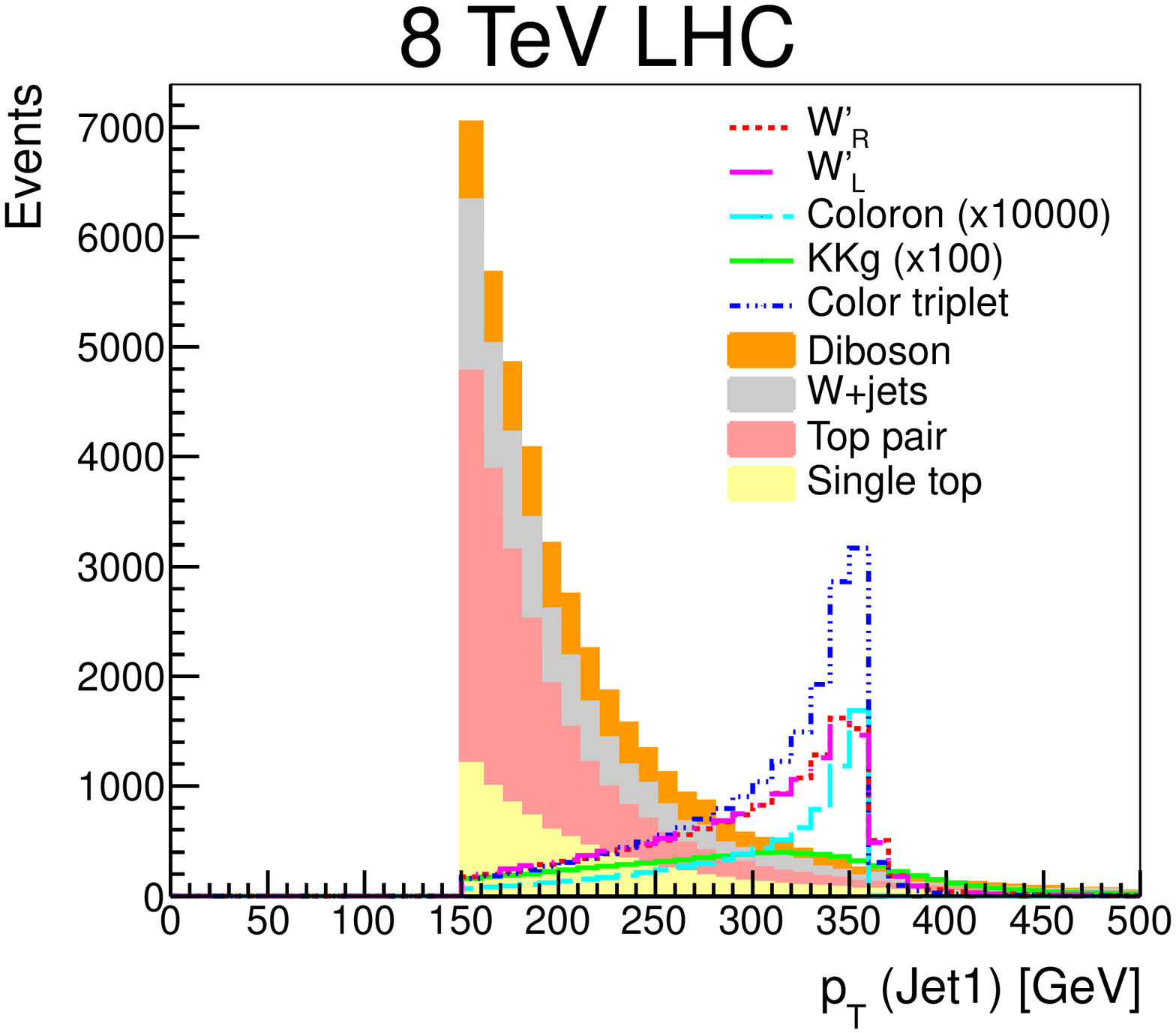}
    \label{fig:lowjet1pt}
  }
  \subfigure[]{
    \includegraphics[width=0.48\textwidth]{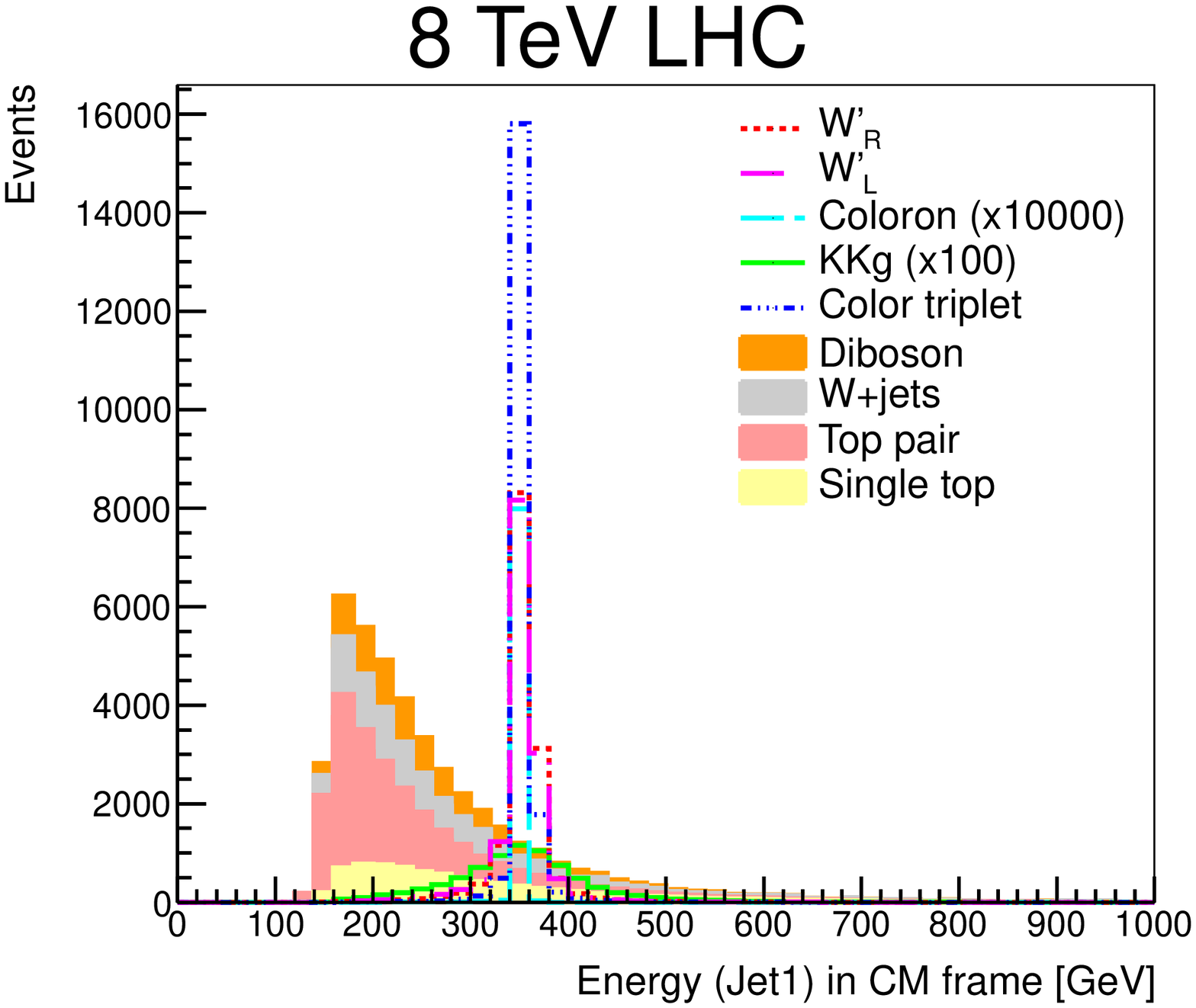}
    \label{fig:lowjet1booste}
  }
  \caption{The \subref{fig:lowjet1pt} $p_{T}$ and \subref{fig:lowjet1booste} energy in the CM frame of the leading jet in 20~fb$^{-1}$ at the 8~TeV LHC, for signal and background processes. (Color online.) }
\label{fig:lowjet1}
\end{figure}

We explore kinematic distributions in three categories: object properties, event kinematics and angular correlations. Here, we examine the basic kinematic distributions of jets, the lepton and $\etmiss$. The $p_T$ distribution of the leading jet is shown in Fig.~\ref{fig:lowjet1}.

Since the second jet reconstructs the top quark, the leading jet is consequently always the one produced together with the top quark. Its $p_T$ and energy distributions are shown in Fig.~\ref{fig:lowjet1}, where the energy is shown for the leading jet boosted into the center-of-mass (CM) frame. The signal $p_T$ distributions clearly shows the cut-off at half of the signal mass, while the background is smoothly falling in this region. By boosting the leading jet into the CM frame, its energy distribution peaks at exactly half of the resonance mass. This variable is therefore a powerful discriminator, equivalent in importance to the reconstructed resonance mass itself.

\begin{figure}[!h!tbp]
  \centering
  \subfigure[]{
    \includegraphics[width=0.48\textwidth]{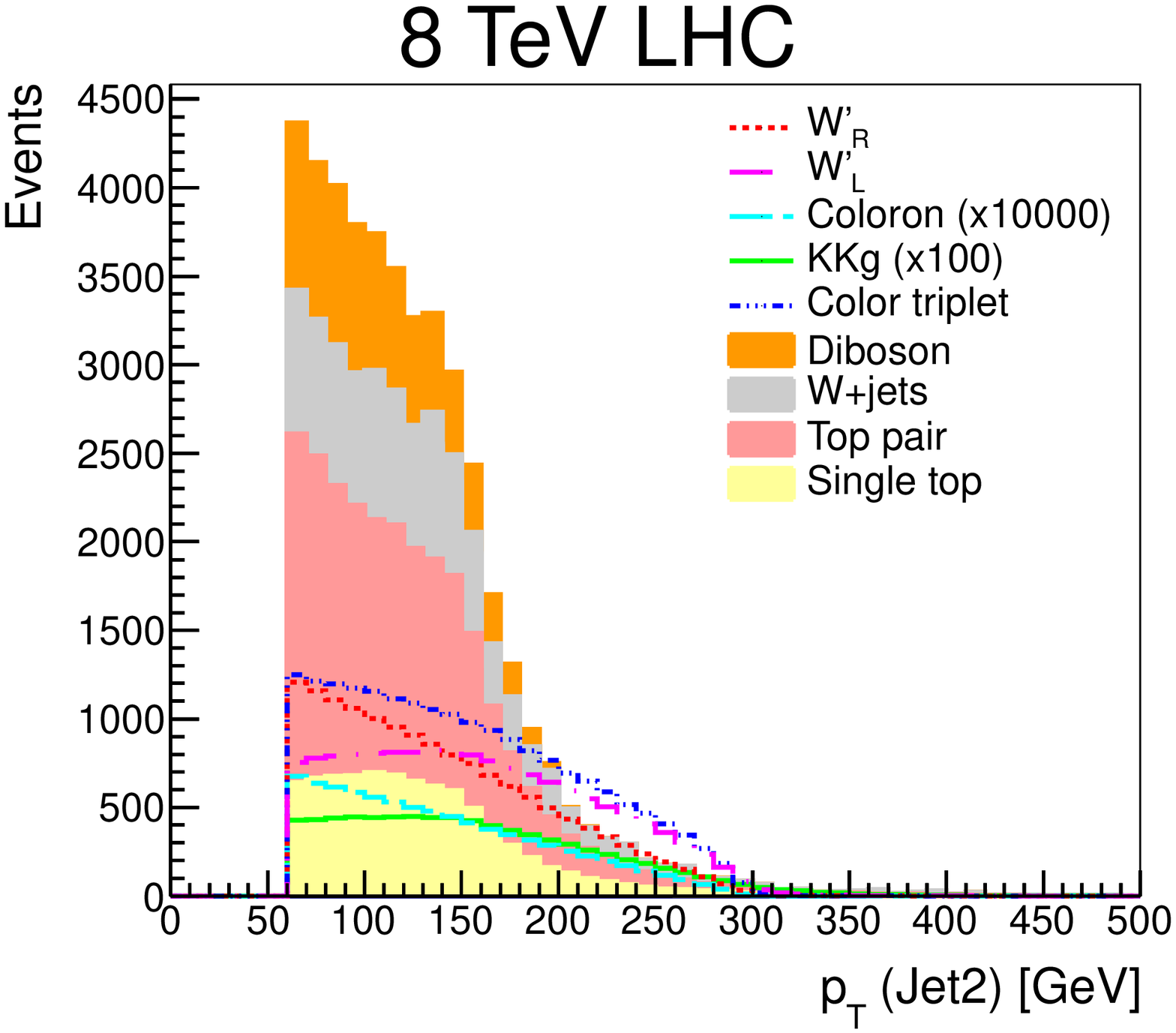}
    \label{fig:lowjet2pt}
  }
  \subfigure[]{
    \includegraphics[width=0.48\textwidth]{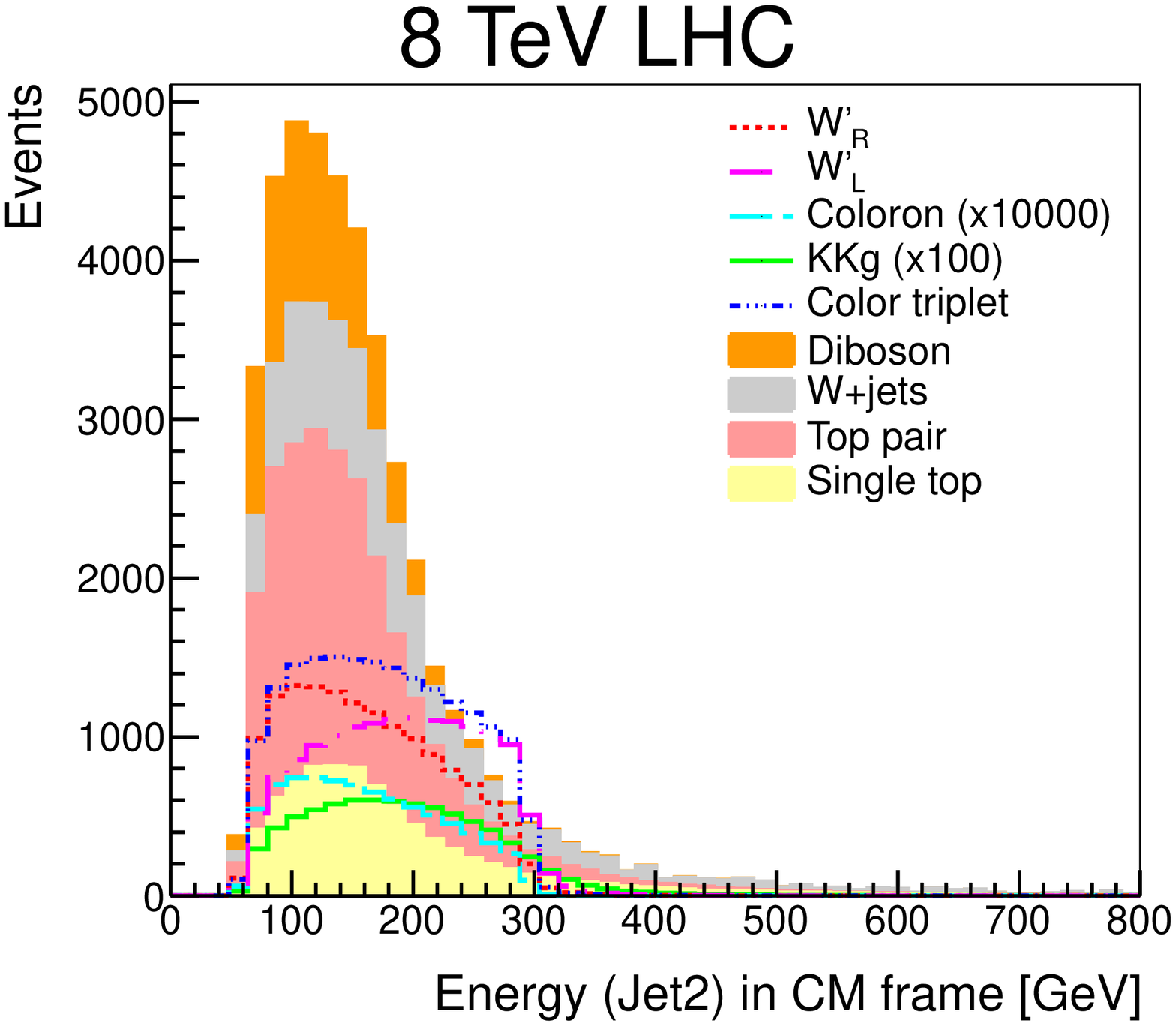}
    \label{fig:lowjet2booste}
  }
  \caption{The \subref{fig:lowjet2pt} $p_{T}$ and \subref{fig:lowjet2booste} energy in the CM frame of the second jet in 20~fb$^{-1}$ at the 8~TeV LHC, for signal and background processes. (Color online.)}
    \label{fig:lowjet2}
\end{figure}

In the distribution of the second jet $p_{T}$, shown in Fig.~\ref{fig:lowjet2pt}, there is no peak structure visible for the signal anymore because this is the jet from the top-quark decay. Nevertheless, the background and signals have different distributions, with the signal extending out to higher $p_T$. 

The energy of the second jet (also boosted into the CM frame) is shown in Fig.~\ref{fig:lowjet2booste}. Here, the signal has a smooth distribution rather than a peak structure, and it cuts off at just below 300~GeV. This is a feature of the phase space of the top-quark decay. The top quark has an energy of 375~GeV and decays into a $W$~boson (which takes at least 80~GeV) and a $b$~quark. The backgrounds have no such constraint and extend to much higher energies.

\begin{figure}[!h!tbp]
  \centering
  \subfigure[]{
    \includegraphics[width=0.48\textwidth]{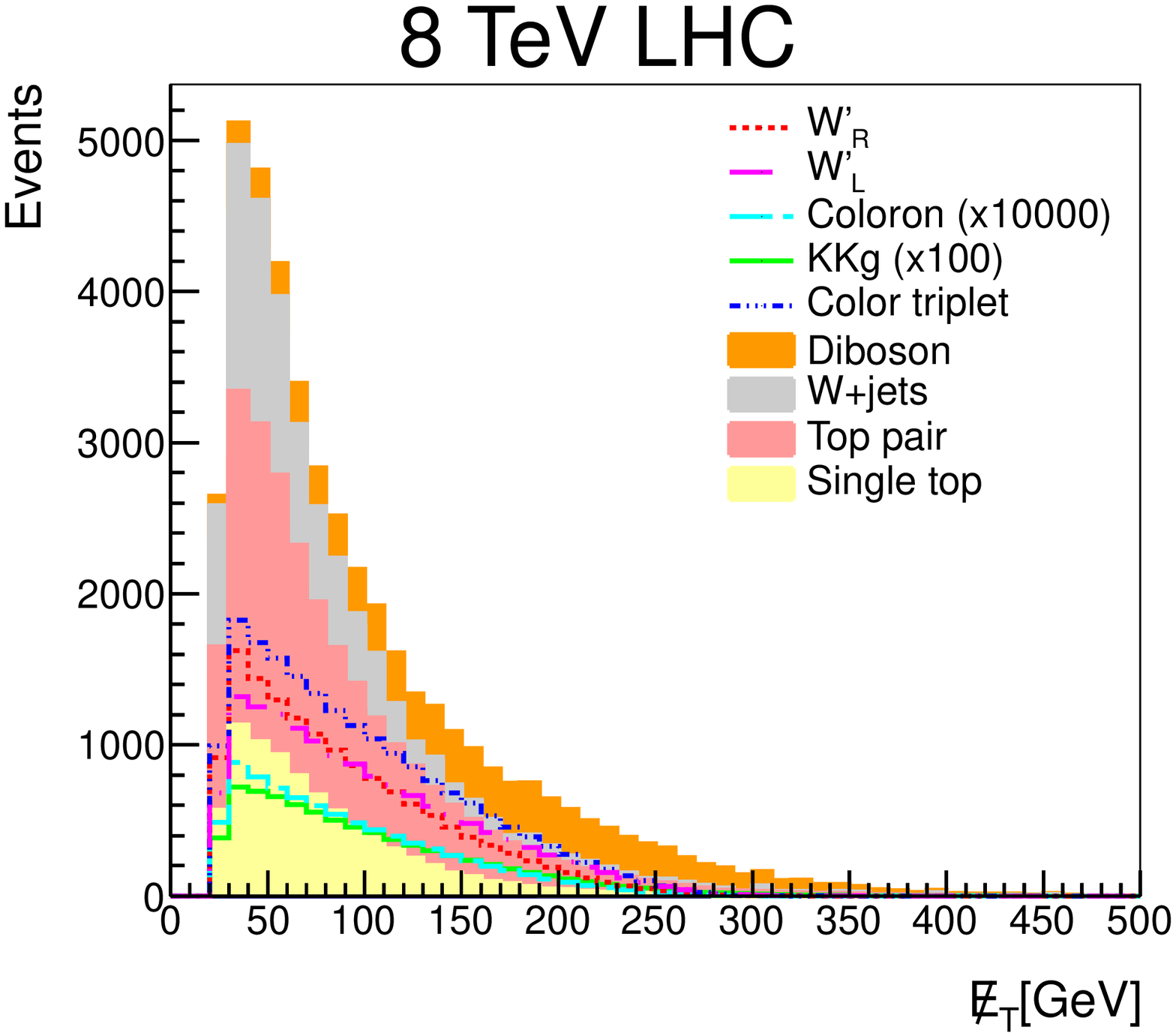}
    \label{fig:lowmet}
  }
  \subfigure[]{
    \includegraphics[width=0.48\textwidth]{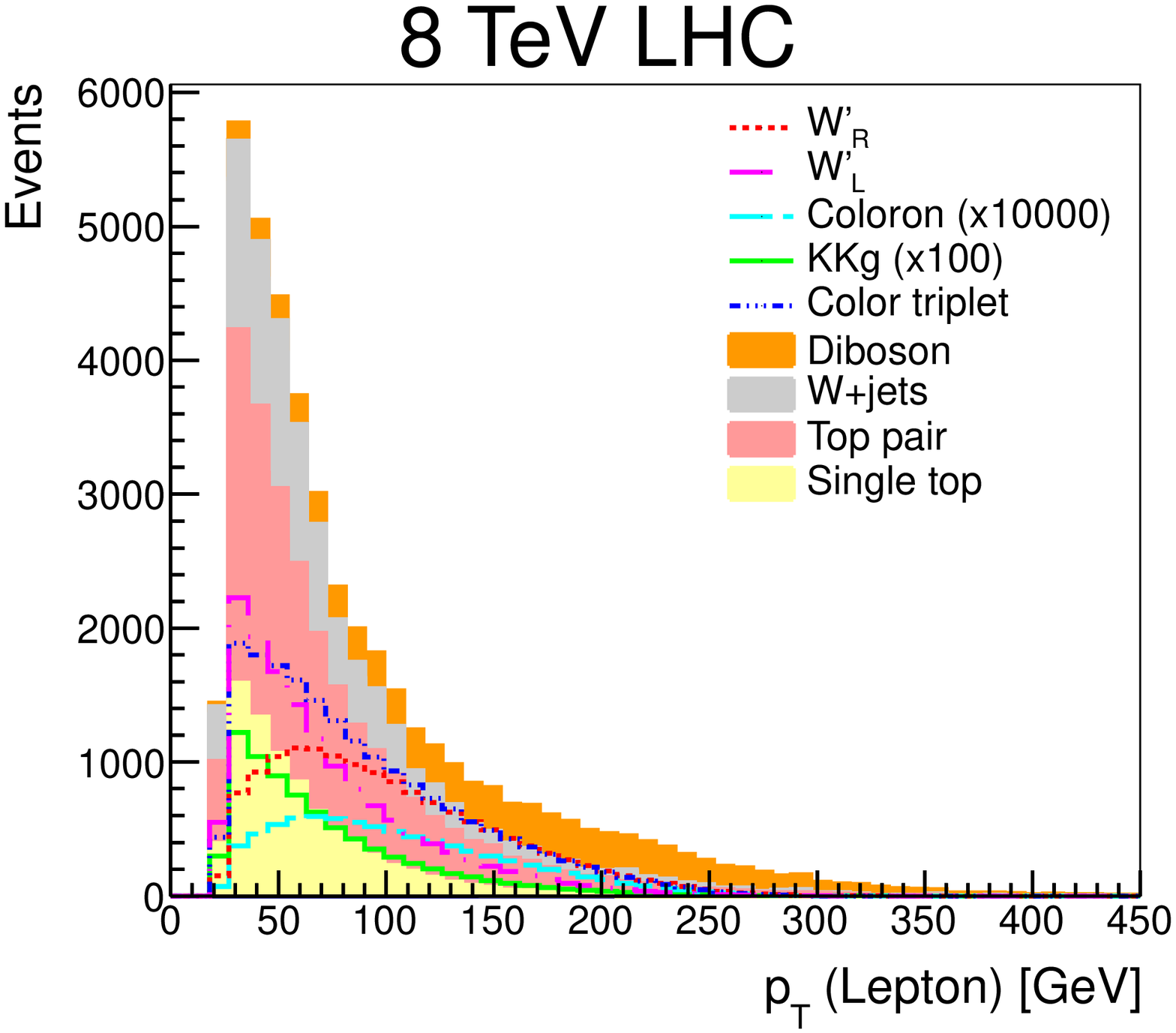}
    \label{fig:lowleppt}
  }
  \caption{The \subref{fig:lowmet} $\etmiss$ and \subref{fig:lowleppt} lepton $p_T$ in 20~fb$^{-1}$ at the 8~TeV LHC, for signal and background processes. (Color online.)}
\label{fig:lowel}
\end{figure}

Fig.~\ref{fig:lowel} shows the $\etmiss$ and lepton $p_T$ distributions. Signals and backgrounds are falling distributions in both cases, but small differences can be seen for the different signals. These differences are related to whether a top quark with left-handed or right-handed helicity decays. For $\Wp_L$ (i.e. SM-like couplings) and $KKg$ production, the $\etmiss$ distribution is broader than the lepton $p_T$ distribution, just as in SM $s$-channel production. For $\Wp_R$ and coloron production, the lepton $p_T$ distribution is broader. Only for color-triplet production are the distributions similar.

\subsection{Event reconstruction}
\label{subsec:loweventreconstruction}

The basic kinematic distributions of objects in the event already show distinguishing features, which will be brought out more in this section where the kinematic properties of the reconstructed $W$~boson, top quark and resonance particle are explored.

\begin{figure}[!h!tbp]
  \centering
    \includegraphics[width=0.48\textwidth]{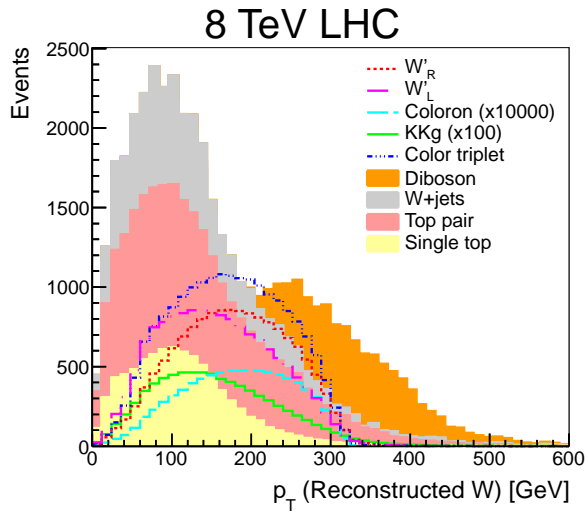}
  \caption{The $p_T$ of the reconstructed $W$~boson in 20~fb$^{-1}$ at the 8~TeV LHC, for signal and background processes. (Color online.)
}
  \label{fig:lowwpt}
\end{figure}

The $W$~boson is reconstructed from the lepton and the missing transverse energy plus the true neutrino $p_Z$. Its $p_T$ distribution shows a shape difference between signal and background, as shown in Fig.~\ref{fig:lowwpt}. The signals all have a distribution that is smooth and almost semi-circular between 0~GeV and the kinematic cutoff at 375~GeV. The backgrounds all have different distributions, with $W$+jets and top backgrounds peaking at lower $p_T$, while the diboson background peaks at higher $p_T$.

\begin{figure}[!h!tbp]
  \centering
  \subfigure[]{
    \includegraphics[width=0.48\textwidth]{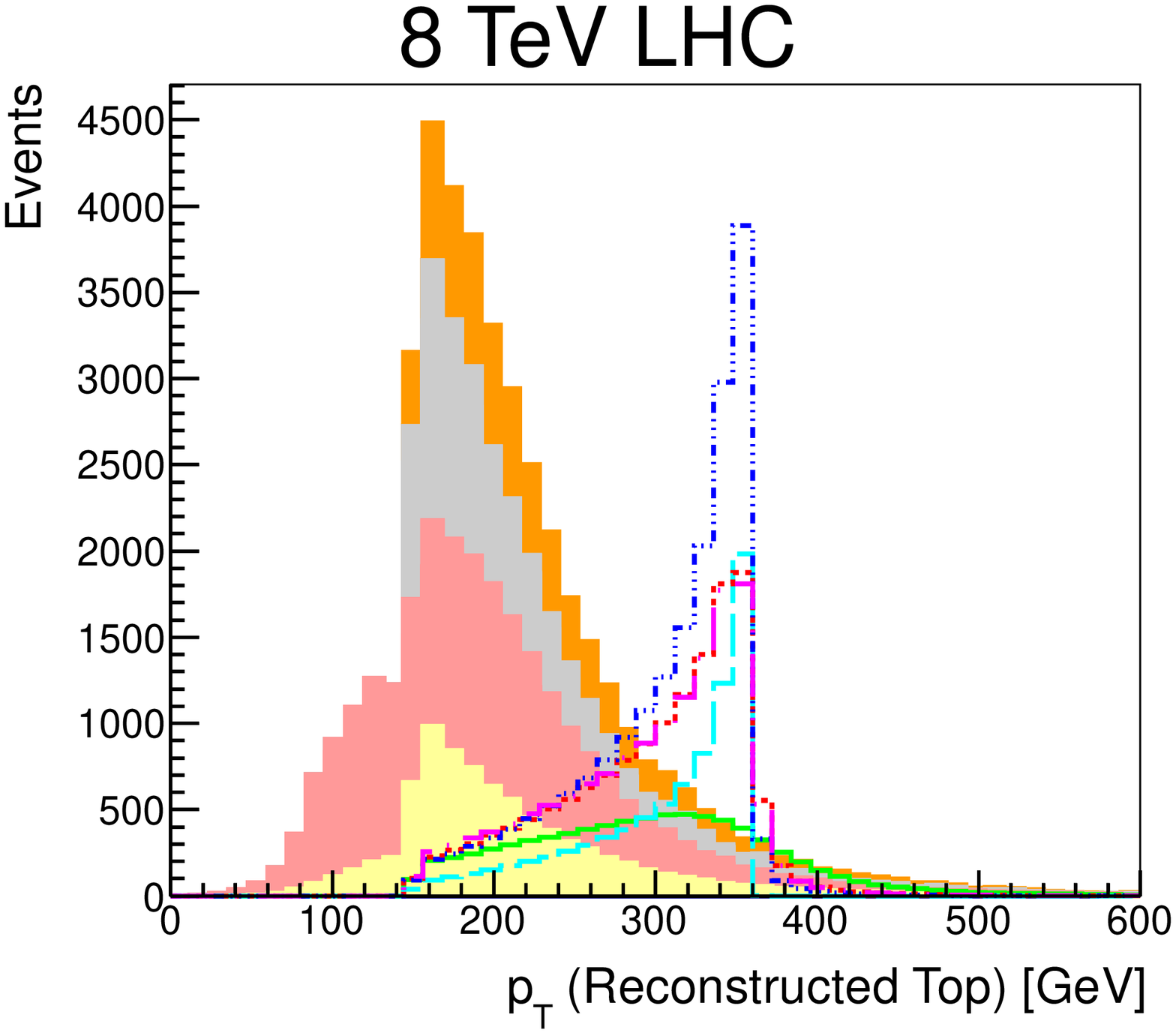}
    \label{fig:lowtoppt}
  }
  \subfigure[]{
    \includegraphics[width=0.48\textwidth]{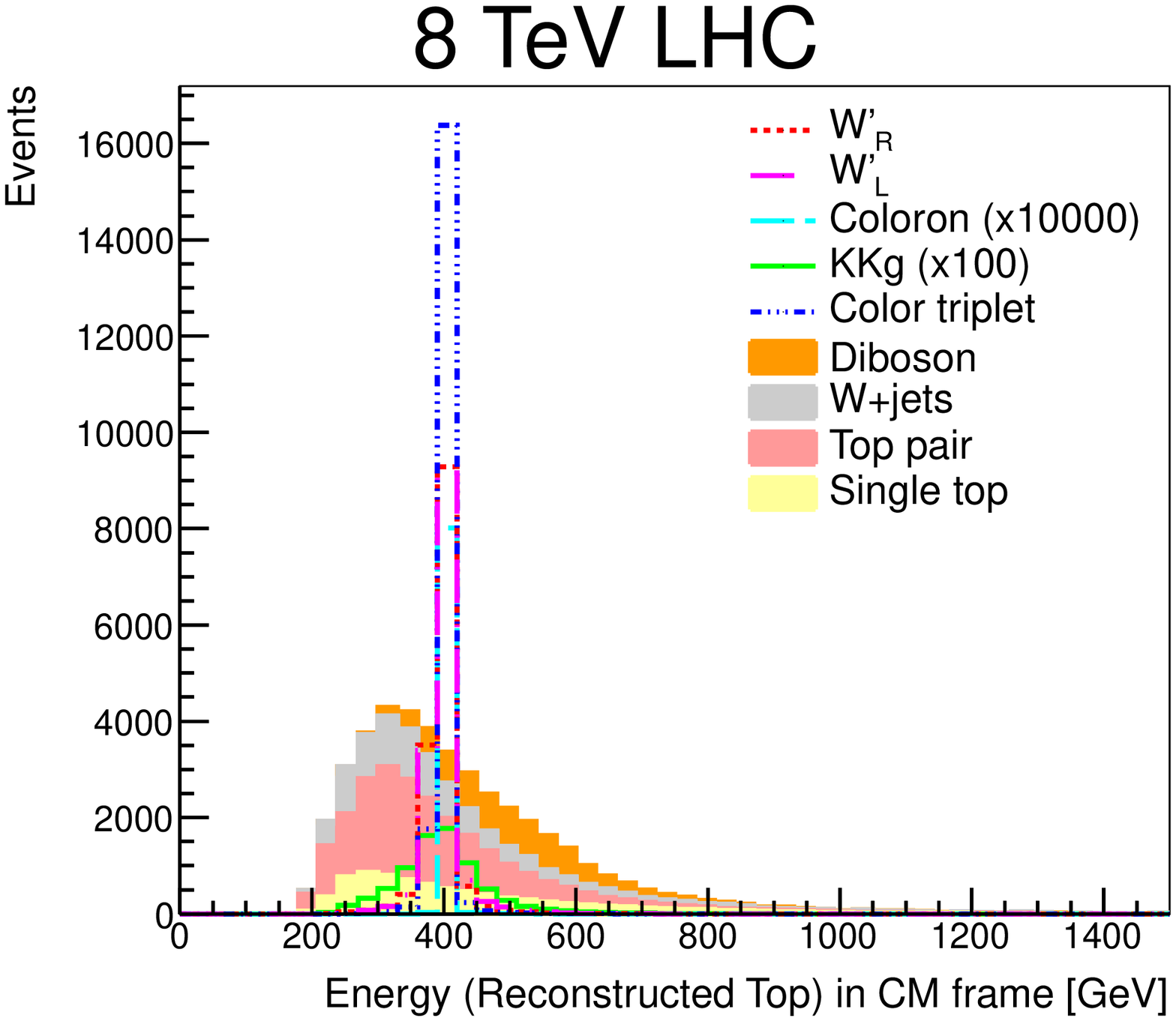}
    \label{fig:lowtopeboost}
  }
  \caption{The \subref{fig:lowtoppt} $p_{T}$ and \subref{fig:lowtopeboost} energy in the CM frame of the reconstructed top quark in 20~fb$^{-1}$ at the 8~TeV LHC, for signal and background processes. (Color online.)
}
  \label{fig:lowtop}
\end{figure}

The top quark is reconstructed from the $W$~boson and the second jet.
The $p_{T}$ of the reconstructed top quark is shown in Fig.~\ref{fig:lowtoppt}. It is clearly a good variable for separating signal from background, with a peak at half of the resonance particle energy. This peak structure is even more pronounced than that of the leading jet (which is back-to-back with the top quark) from Fig.~\ref{fig:lowjet1pt}. The requirement on the $p_T$ of the leading jet is also responsible for the jump at 150~GeV in Fig.~\ref{fig:lowtoppt}.

The energy of the reconstructed top quark, boosted into the CM reference frame, is shown in Fig.~\ref{fig:lowtopeboost}. This shows a very narrow peak, similar to the energy distribution of the leading jet. However, the neutrino $p_Z$ enters twice in the top-quark case, once in the top reconstruction and once in the definition of the CM frame. Hence this distribution will be broadened in experimental measurements.

\begin{figure}[!h!tbp]
  \centering
    \includegraphics[width=0.48\textwidth]{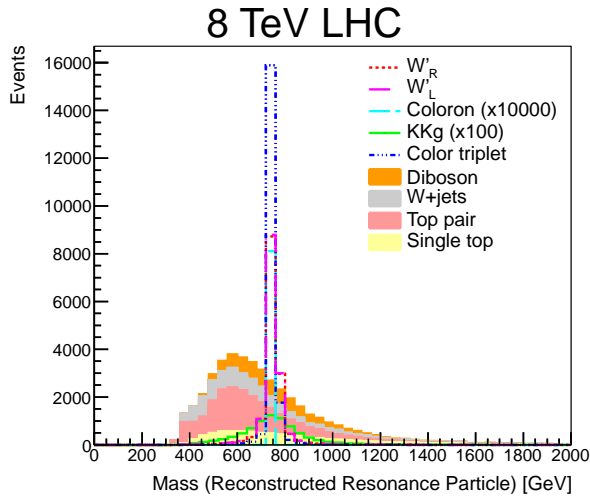}
  \caption{The mass of the reconstructed resonance particle in 20~fb$^{-1}$ at the 8~TeV LHC, for signal and background processes. (Color online.)
}
  \label{fig:lowparticlemass}
\end{figure}

The main variable that experiments have been using so far is the reconstructed particle mass. As this mass increases, all of the backgrounds decrease smoothly, making this an ideal scenario for a narrow-mass resonance search. However, in the low-mass region this advantage no longer holds. Figure~\ref{fig:lowparticlemass} shows that the background peaks around 600~GeV. The location of this background peak will depend on the selection cuts, but generally speaking, for resonance masses between 400~GeV and 800~GeV or so, a search for a narrow mass peak in data is disfavored. And a realistic detector resolutions will broaden out the signal invariant mass peak even more. In this kinematic region, a variable such as the leading jet $p_T$ provides better discrimination (c. f. Fig.~\ref{fig:lowjet1pt}). 

\subsection{Angular correlations}
\label{subsec:lowanglevar}

Angular correlations between final state objects are important in high-mass resonance searches because they provide discrimination from the background that is approximately independent of the resonance mass. They are also important in separating different signals from each other as will be discussed in Sec.~\ref{subsec:highanglevar}.

\begin{figure}[!h!tbp]
  \centering
  \subfigure[]{
    \includegraphics[width=0.48\textwidth]{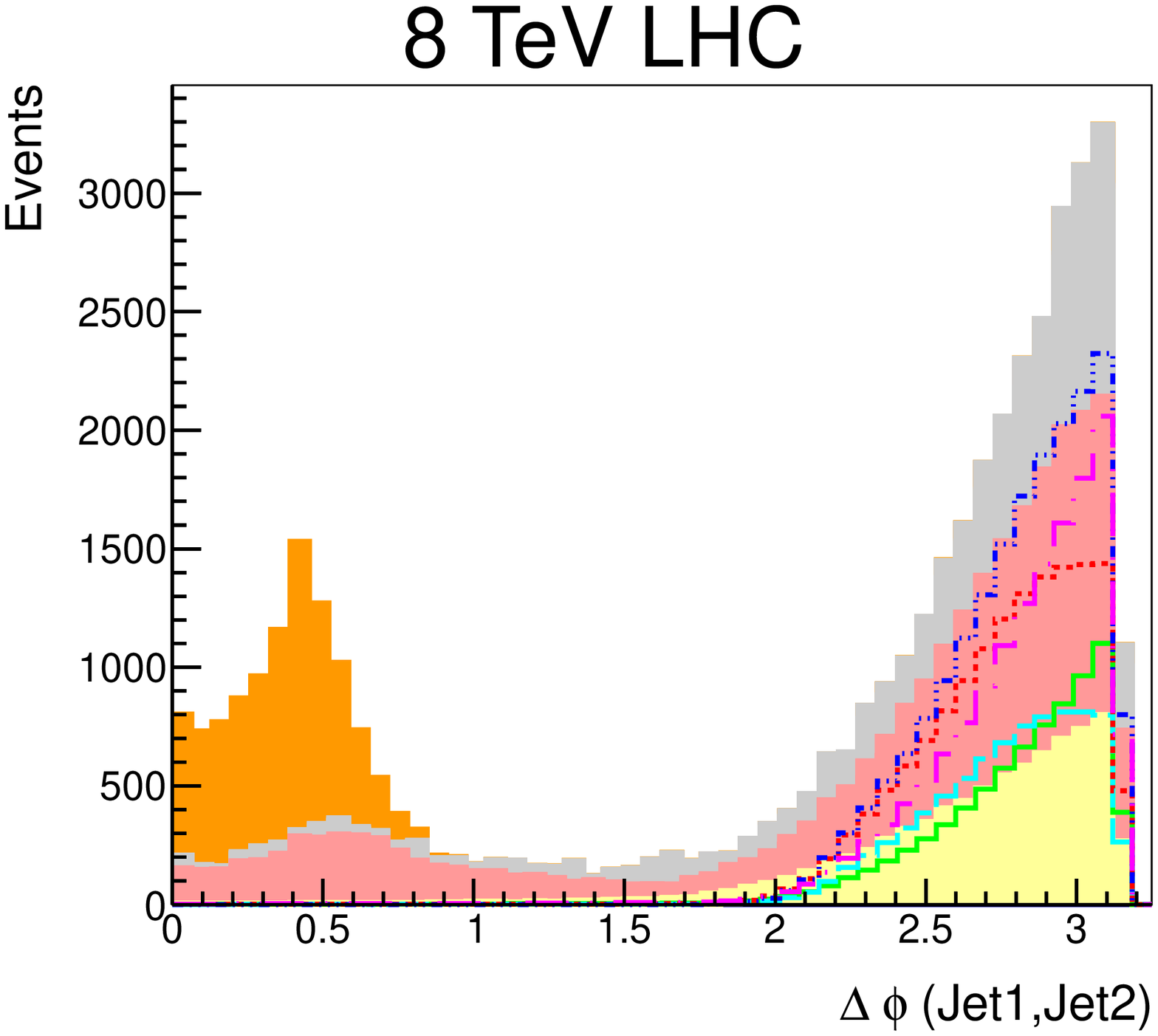}
    \label{fig:lowdeltaphijet12}
}
  \subfigure[]{
    \includegraphics[width=0.48\textwidth]{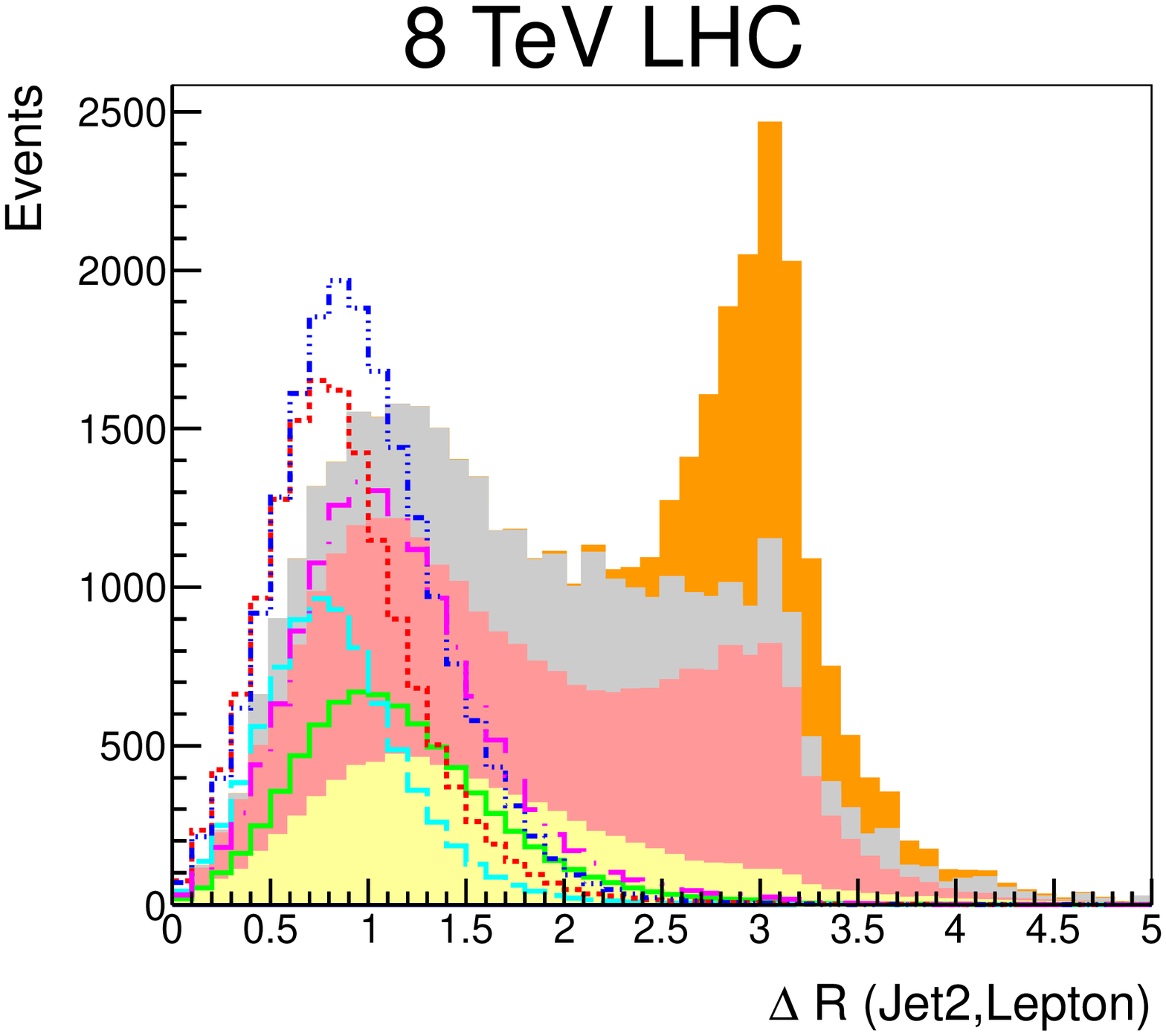}
    \label{fig:lowdeltarjet2lep}
  }
  \subfigure[]{
    \includegraphics[width=0.48\textwidth]{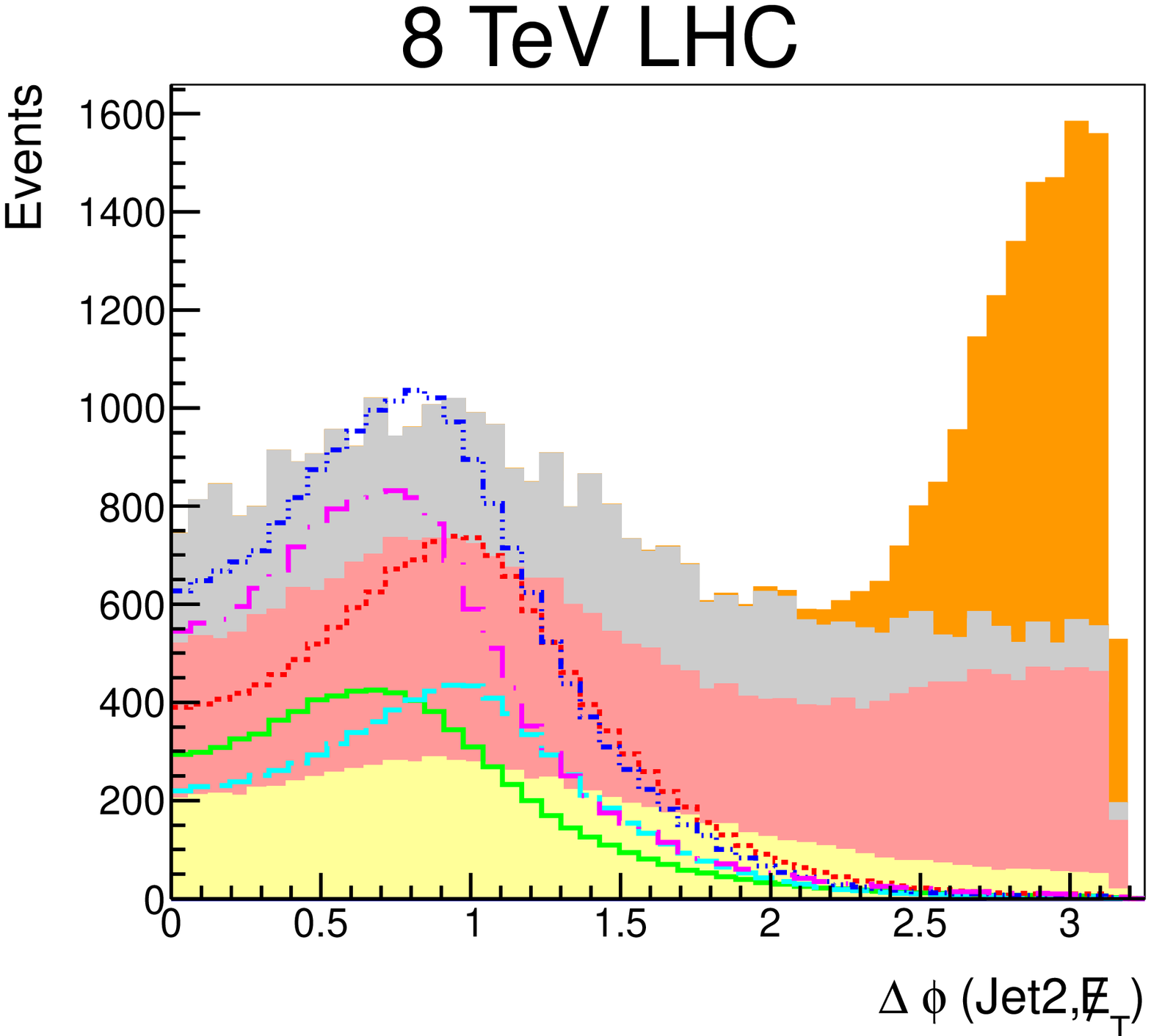}
    \label{fig:lowdeltaphijet2MET}
  }
  \subfigure{
    \includegraphics[width=0.25\textwidth]{figleglow}
  }
  \caption{The \subref{fig:lowdeltaphijet12} $\Delta \phi$ between the leading two jets, \subref{fig:lowdeltarjet2lep} $\Delta R$ between the second jet and the lepton and \subref{fig:lowdeltaphijet2MET} $\Delta \phi$ between the second jet and the $\etmiss$, in 20~fb$^{-1}$ at the 8~TeV LHC, for signal and background processes. (Color online.)
}
\end{figure}

Figure~\ref{fig:lowdeltaphijet12} shows the $\Delta \phi$ between the leading two jets. These jets are back-to-back for the signals and the $W$+jets and top backgrounds. The diboson background by contrast has two jets from a $W$~or $Z$~boson decay, which are typically close together. The $\Delta R$ distribution between the leading two jets (not shown) gives similar discrimination of the diboson background.

\begin{figure}[!h!tbp]
  \centering
  \subfigure[]{
  \includegraphics[width=0.48\textwidth]{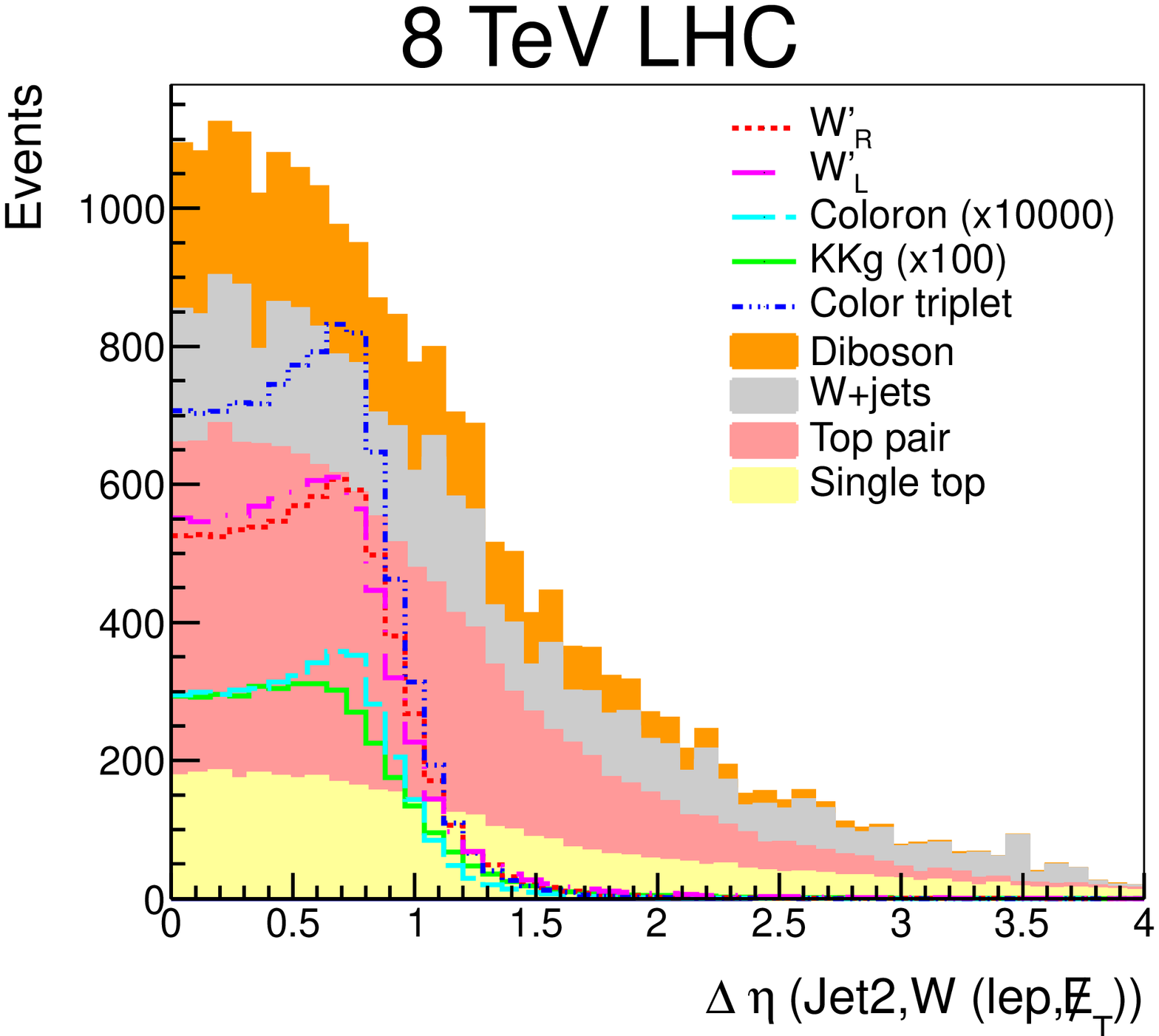}
    \label{fig:lowdeltarjet2W}
  }
  \subfigure[]{
    \includegraphics[width=0.48\textwidth]{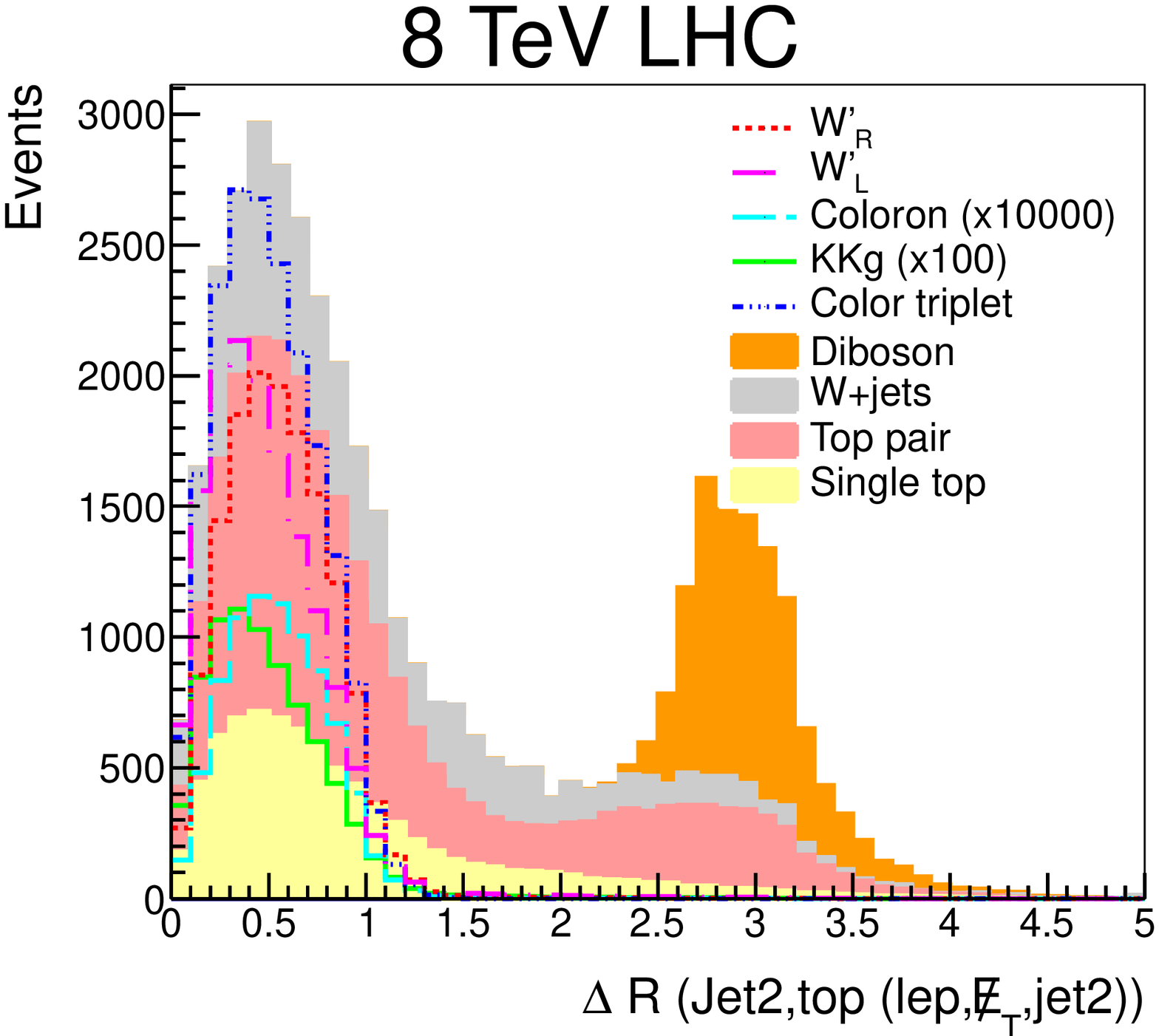}
    \label{fig:lowdeltarjet2top}
  }
  \subfigure{
    \includegraphics[width=0.25\textwidth]{figleglow}
  }
  \caption{The \subref{fig:lowdeltarjet2W} $\Delta \eta$ of the second jet and the reconstructed $W$~boson and \subref{fig:lowdeltarjet2top} $\Delta R$ between the second jet and the reconstructed top quark, in 20~fb$^{-1}$ at the 8~TeV LHC, for signal and background processes. (Color online.)}
  \label{fig:lowdeltaetajet2wtop}
\end{figure}

The $\Delta R$ between the second jet and the lepton, shown in Fig.~\ref{fig:lowdeltarjet2lep}, also separates the signal from the diboson background, for which the lepton and both jets are back-to-back.  Since the second jet comes from the top-quark decay in signal events, this distribution is also a measure of the boost of the top quark. The $W$+jets background has a smoother distribution out to higher $\Delta \phi$. The top pair background shows two peaks, one at low $\Delta \phi$ where the signal peaks, and a second one near $\pi$ from events where the second jet is from the top quark not associated with the lepton. Individually, the $\Delta \phi$ and $\Delta \eta$ distributions between the second jet and the lepton have only minor discrimination power, but when combined into $\Delta R$, the discrimination power is quite high. As can be expected, the distribution of $\Delta \phi$ between the second jet and the $\etmiss$, shown in Fig.~\ref{fig:lowdeltaphijet2MET}, also shows good separation between signal and background. The $\Delta \phi$ and $\Delta R$ between the second jet and the $W$~boson (not shown) have similar distributions. 
The $\Delta \eta$ between the second jet and the reconstructed $W$~boson is shown in Fig.~\ref{fig:lowdeltarjet2W}. The shape of the signal distribution (rising, with a peak below 1) is very different from the backgrounds, which are smoother and have a long tail. This distinction in $\Delta \eta$ also depends on the neutrino $p_Z$ reconstruction.
Finally, the $\Delta R$ between the second jet and the top quark, shown in Fig.~\ref{fig:lowdeltarjet2top}, has similar features of a narrow signal distribution, broader background distributions, and a peak for the diboson around $\pi$.

\begin{figure}[!h!tbp]
  \centering
  \subfigure[selected events]{
    \includegraphics[width=0.48\textwidth]{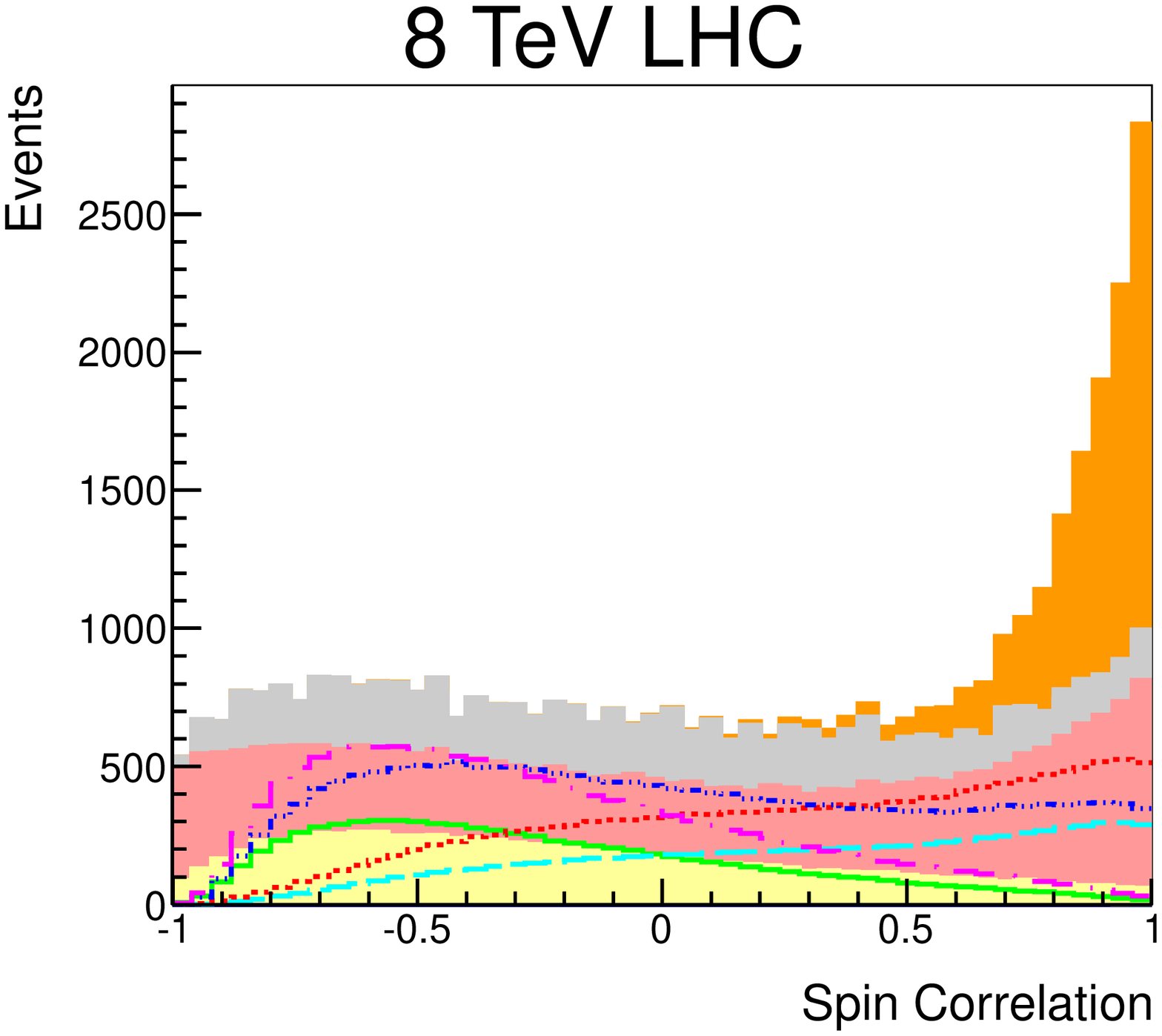}
    \label{fig:lowspincor}
  }
  \subfigure[selected events with 2 $b$-tags]{
    \includegraphics[width=0.48\textwidth]{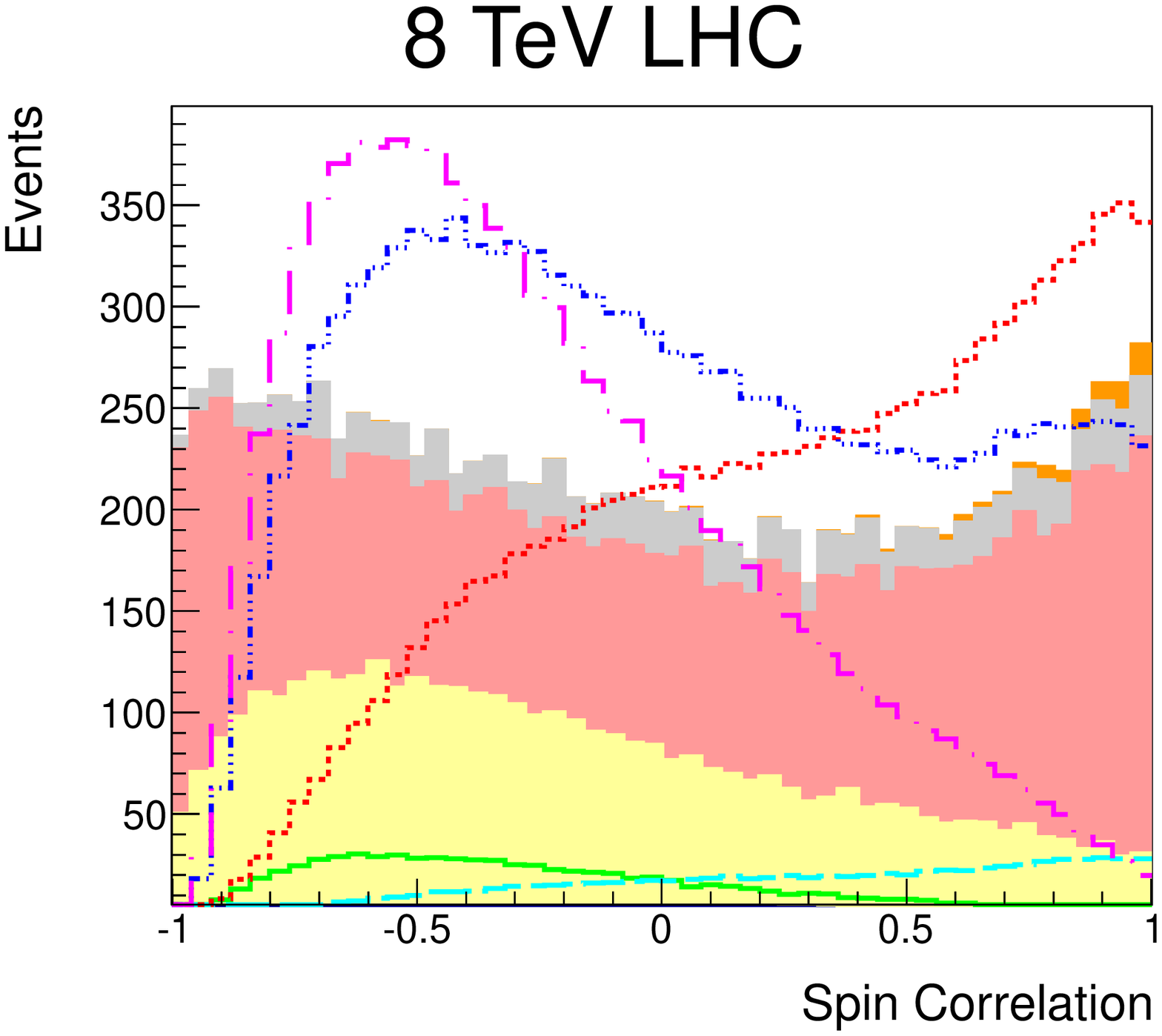}
    \label{fig:lowspincor2tag}
  }
  \subfigure[]{
    \includegraphics[width=0.48\textwidth]{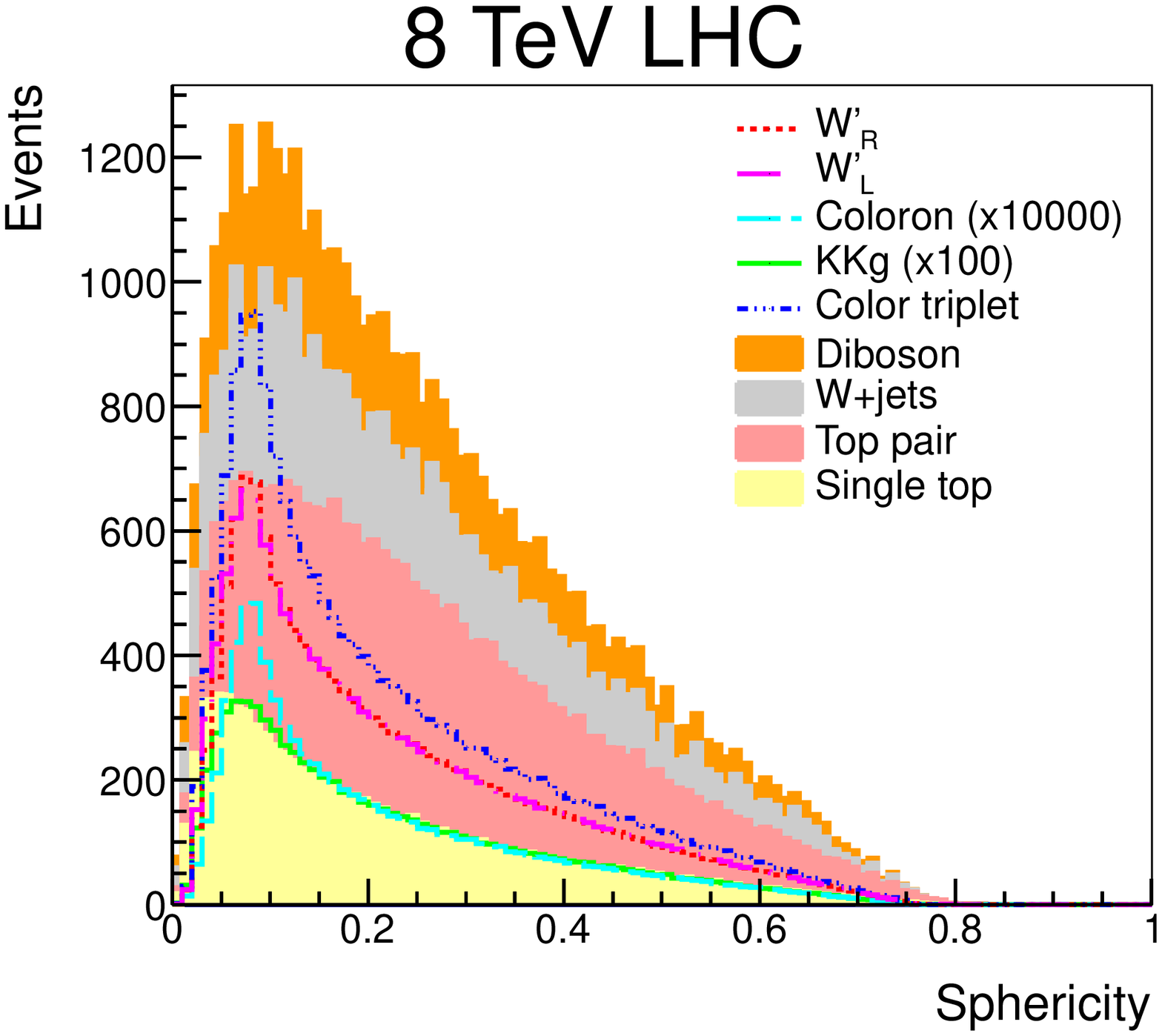}
    \label{fig:lowsphericity}
  }
  \subfigure[]{
    \includegraphics[width=0.48\textwidth]{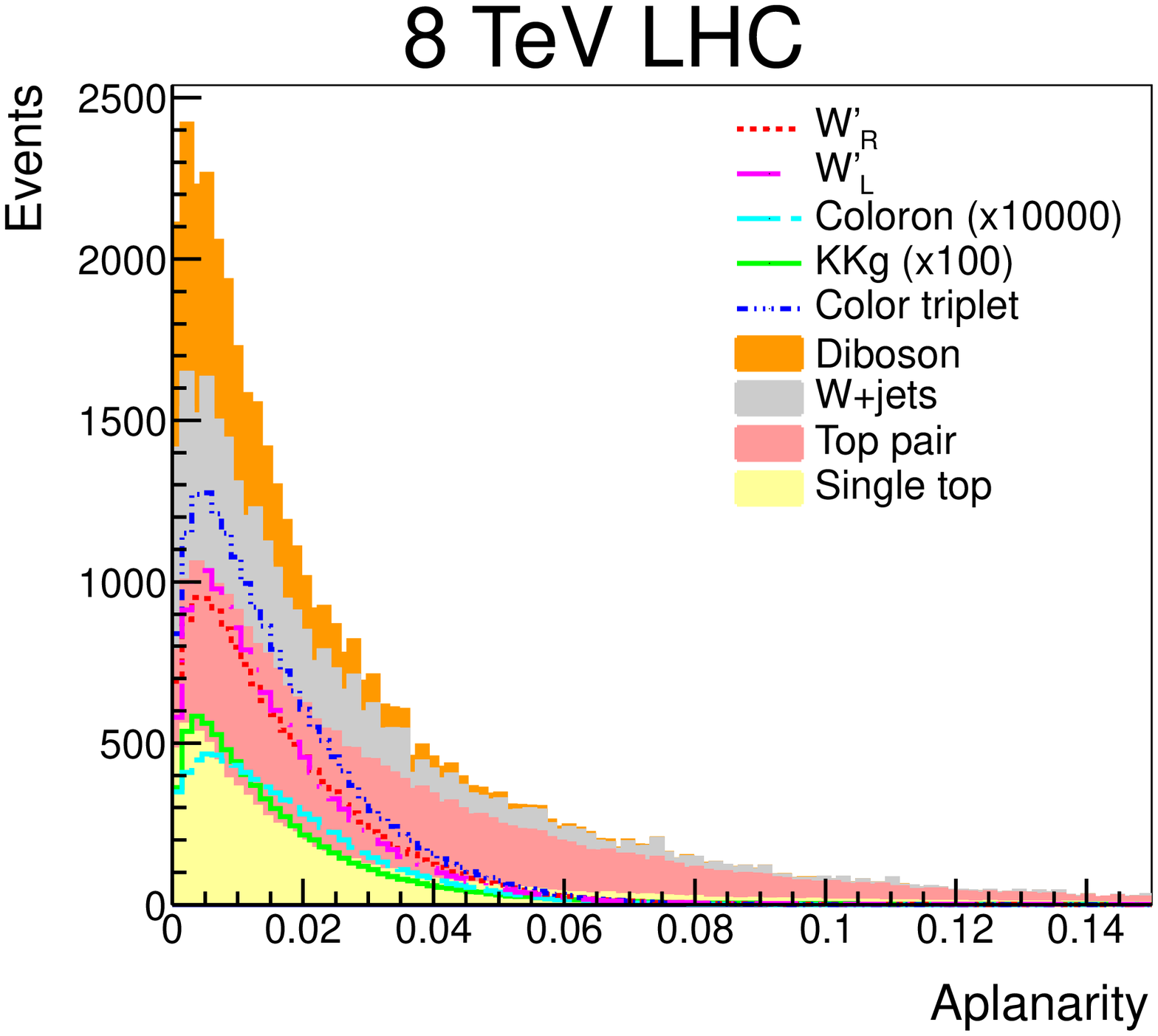}
    \label{fig:lowaplanarity}
  }
  \caption{The spin correlation for \subref{fig:lowspincor} all selected events and \subref{fig:lowspincor2tag} 2-$b$-tag events, as well as the \subref{fig:lowsphericity} sphericity and \subref{fig:lowaplanarity} aplanarity, in 20~fb$^{-1}$ at the 8~TeV LHC, for signal and background processes. (Color online.)
}
\end{figure}

The most direct access to the angular correlation of the decay of the resonance particle is provided by the spin correlation of the top quark, shown in Fig.~\ref{fig:lowspincor}. The spin correlation is calculated in the helicity basis in single top production~\cite{Heim:2009ku} as the angle between the lepton, boosted into CM frame and then the top~quark rest frame, and the top quark moving direction in the CM frame.
Each signal has a unique distribution, and all of them differ from the backgrounds. The features of the spin correlation can be seen even better in two-$b$-tag events. Fig.~\ref{fig:lowspincor2tag} illustrates the dominance of the $\Wp$ and color-triplet signals over the backgrounds. Note that the correct reconstruction of neutrino $p_Z$ is relevant here, otherwise the distinguishing features are broadened.

The sphericity (Fig.~\ref{fig:lowsphericity}) and aplanarity (Fig.~\ref{fig:lowaplanarity}) are two additional angular correlation variables that show separation between signals and backgrounds. They are both calculated from the sphericity tensor
$S^{\alpha \beta} = \sum\limits_{i}p_i^{\alpha}p_i^{\beta} / \sum\limits_i|p_i|^2$, where the sum goes over the momentum components of the lepton and jets. The tensor $S^{\alpha \beta}$ is diagonalized to obtain the three eigenvalues $\lambda_1 \ge \lambda_2 \ge \lambda_3$. Then, 
\begin{itemize}
\item Sphericity = $1.5*(\lambda_2+\lambda_3)$,
\item Aplanarity = $1.5*\lambda_3$.
\end{itemize} 
The signals all have lower sphericity values, while the backgrounds, in particular $\ttbar$ have a higher sphericity. The situation is similar for the aplanarity, though here the signal and background shapes are more similar.

\section{High-mass analysis}
\label{sec:highmass}

The high-mass analysis focuses on identifying the nature of a hypothetical new resonance that might be observed at the LHC at 14~TeV. We investigate kinematic distributions that separate the different signals from each other. We choose a resonance particle mass of 3000~GeV and a collider energy of 14~TeV. The kinematic distributions presented here will be similar at other multi-TeV resonance masses and at other collider energies, for example 13~TeV. Only the shape of distributions is studied here, all signals are normalized to the same area. We focus on those distributions that are the most discriminating between different signals. Some additional separating variables can be found in the low-mass analysis (c.~f. Sec.~\ref{sec:lowmass}).

\begin{figure}[!h!tbp]
  \centering
  \subfigure[]{
    \includegraphics[width=0.48\textwidth]{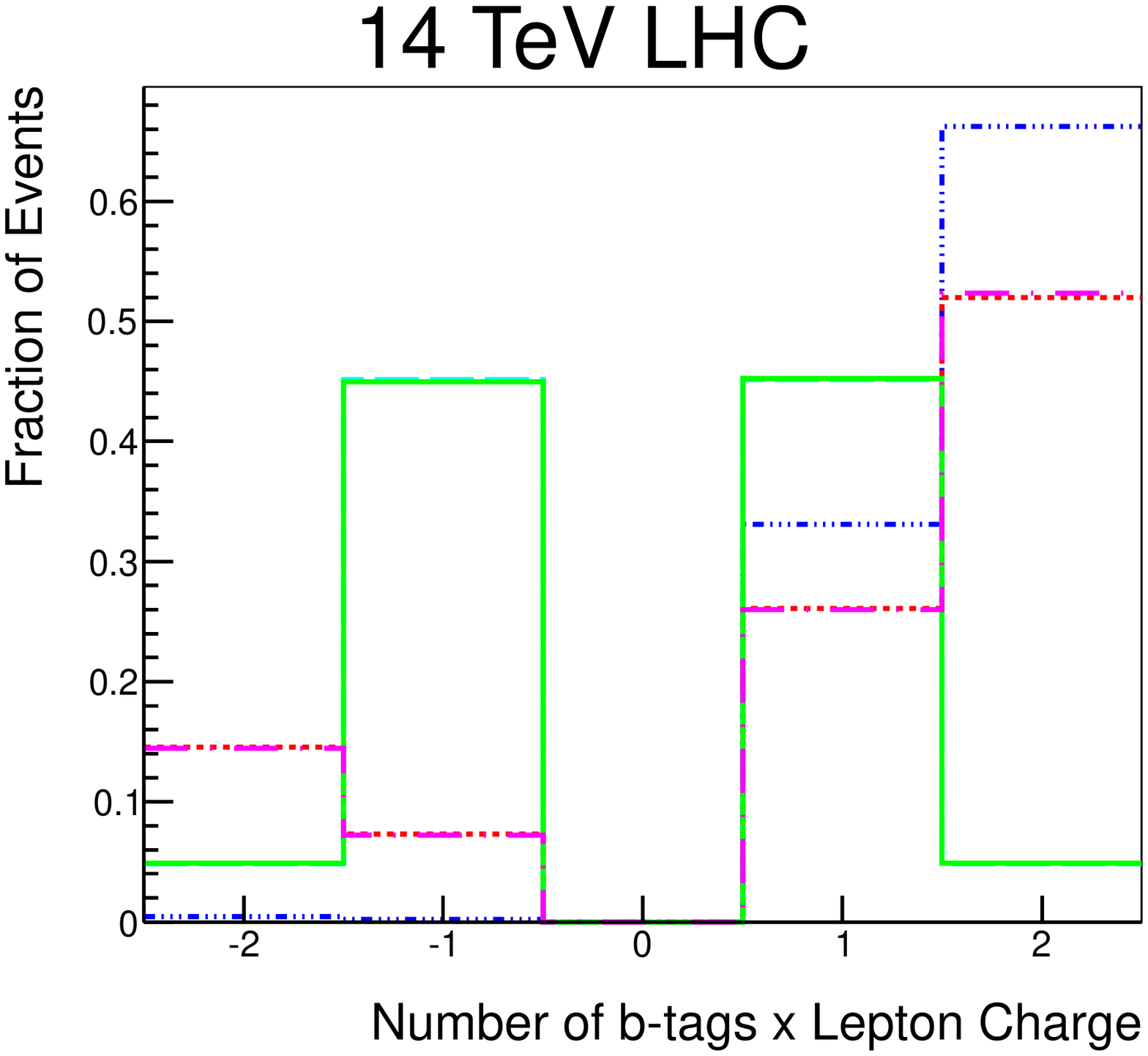}
    \label{fig:highntagslepcharge}
  }
 \subfigure[]{\includegraphics[width=0.48\textwidth]{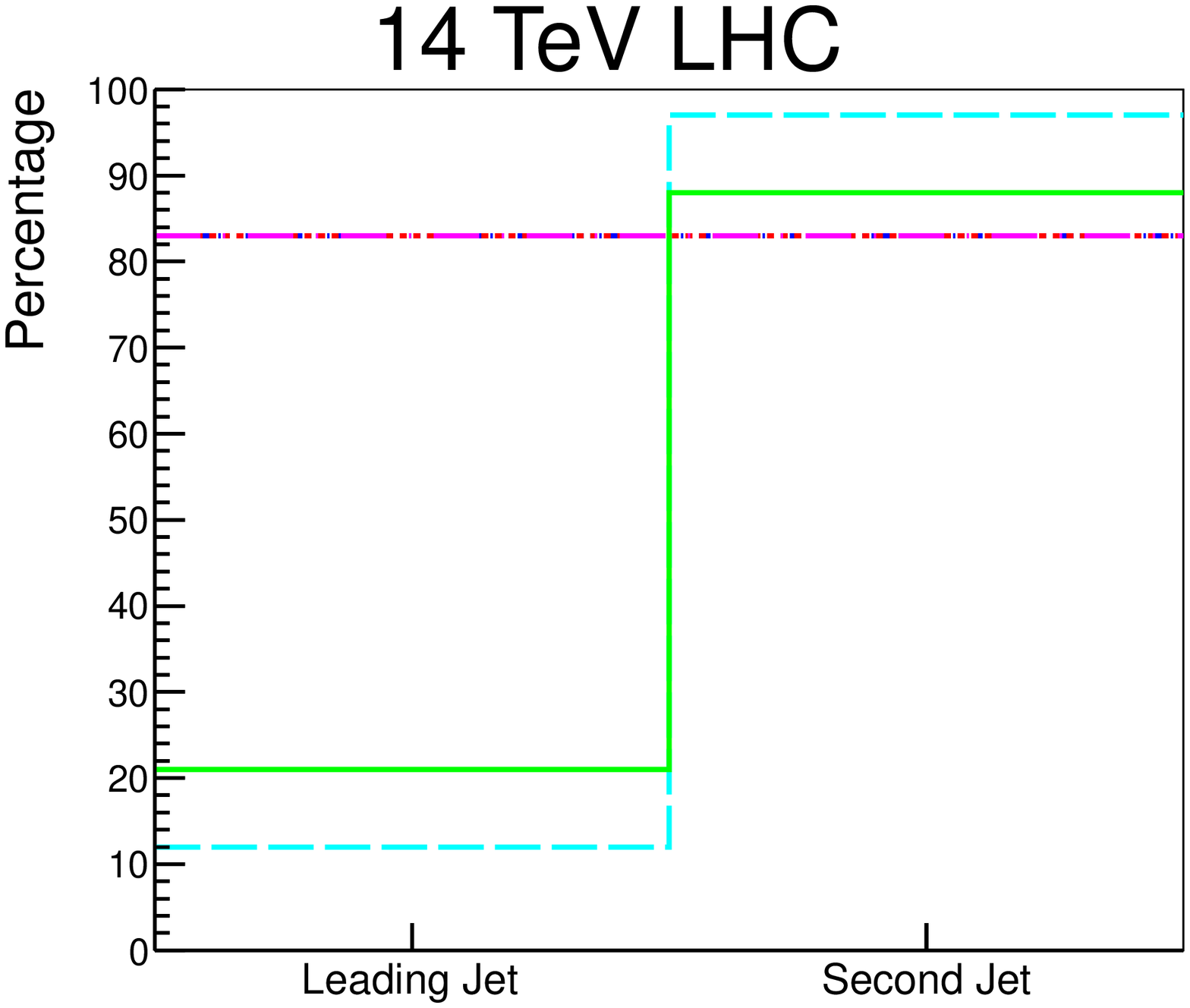}
   \label{fig:isjet1b}  }
  \subfigure{
    \includegraphics[width=0.22\textwidth]{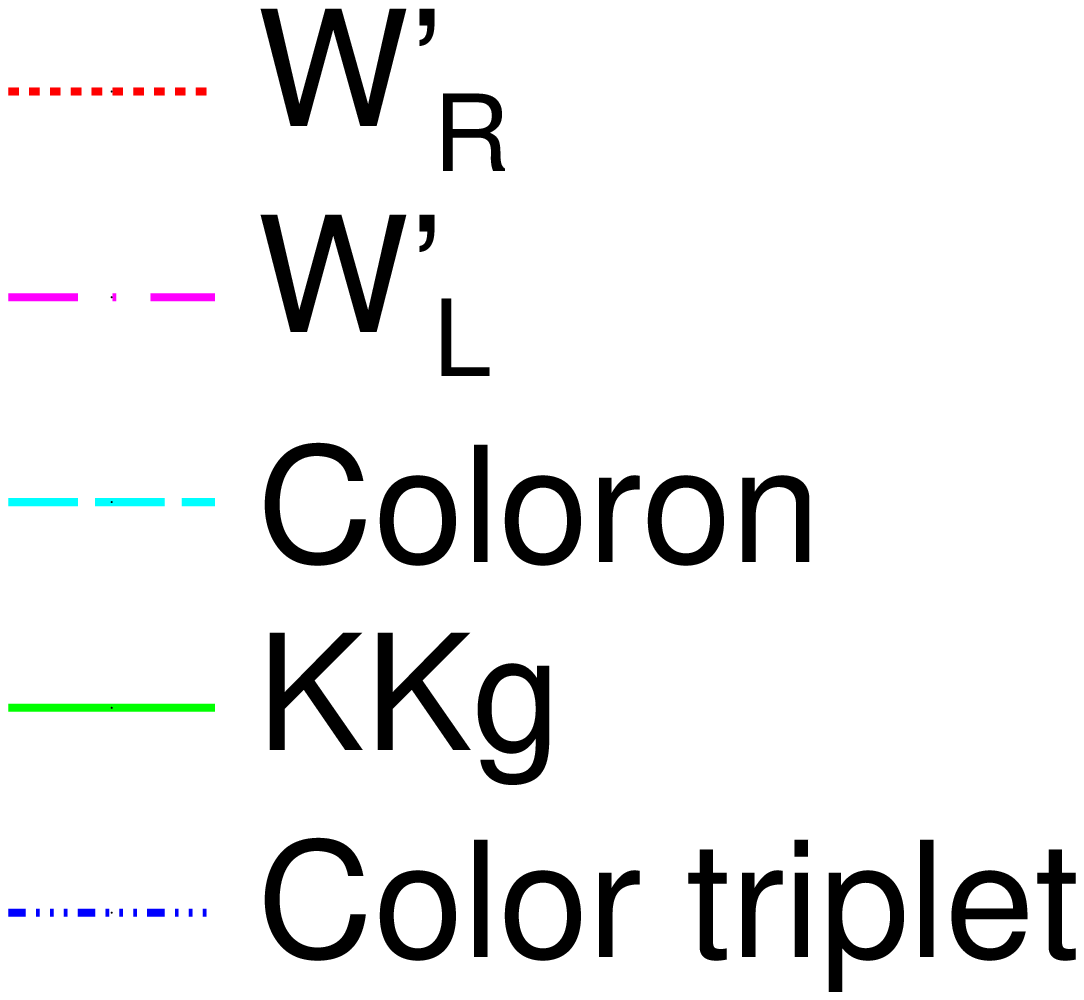}
  }
  \caption[]{\subref{fig:highntagslepcharge} Number of $b$-tagged jets multiplied by the charge of the lepton and \subref{fig:isjet1b} percentage of selected events for which the leading or second jet is  $b$-tagged,
at the 14~TeV LHC. The different signal distributions are  normalized to unit area. (Color online.)}
\end{figure}

Figure~\ref{fig:highntagslepcharge} shows the product of the number of $b$-tagged jets and lepton charge for the different signals. Similar to the 8~TeV distributions shown in Fig.~\ref{fig:lowntagsandlepcharge}, the final states are also unique at 14~TeV. In particular the color-triplet is rarely produced with decay to a negative lepton, as expected from the quark-quark initial state. $\Wp$ production is also asymmetric in lepton charge, but not as much as the color-triplet. The coloron and $KKg$ are symmetric in lepton charge and both mainly have one $b$-tagged jet. There are some two-tag events from the tagging of a $c$~quark jet. 

Though the coloron and $KKg$ distributions look identical in Fig.~\ref{fig:highntagslepcharge}, they differ in which of the jets is $b$-tagged. Figure~\ref{fig:isjet1b} shows the fraction of selected events for which each of the two jets is $b$-tagged. Only events that have at least one $b$-tagged jet enter this distribution, hence the percentage values are higher than the $b$-tag probability for individual jets. For both coloron and $KKg$ production, the second jet is $b$-tagged in most events. But in $KKg$ production, the leading jet is tagged more often than in coloron production. These are events where the leading jet is the $b$~quark from the top-quark decay, which does not happen in coloron events. These are events in the low $p_T$ region of Fig.~\ref{fig:highjet1pt}.

\subsection{Object properties}
\label{subsec:highobjectproperties}

\begin{figure}[!h!tbp]
  \centering
  \subfigure[]{
    \includegraphics[width=0.48\textwidth]{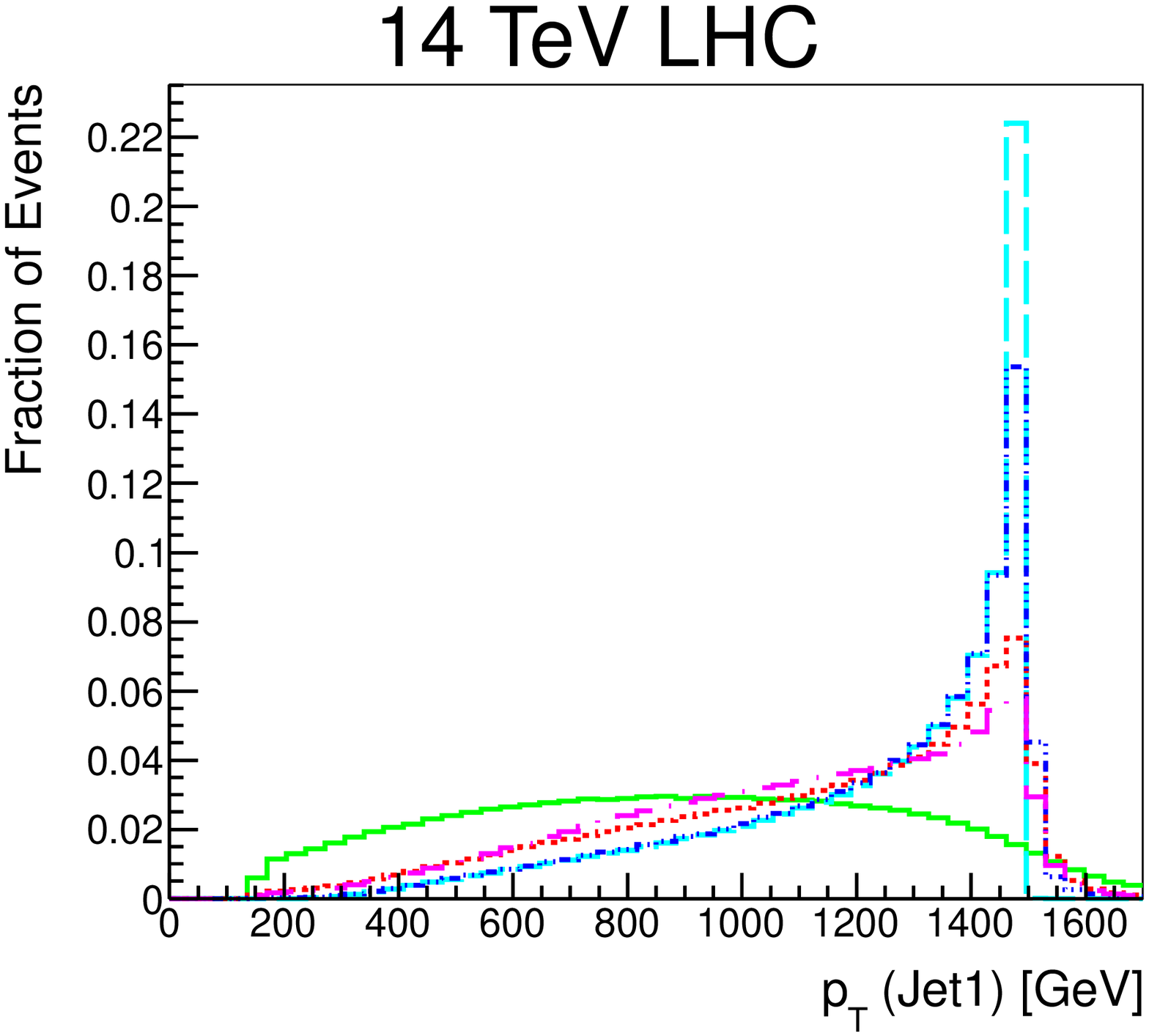}
    \label{fig:highjet1pt}
  }
  \subfigure[]{
    \includegraphics[width=0.48\textwidth]{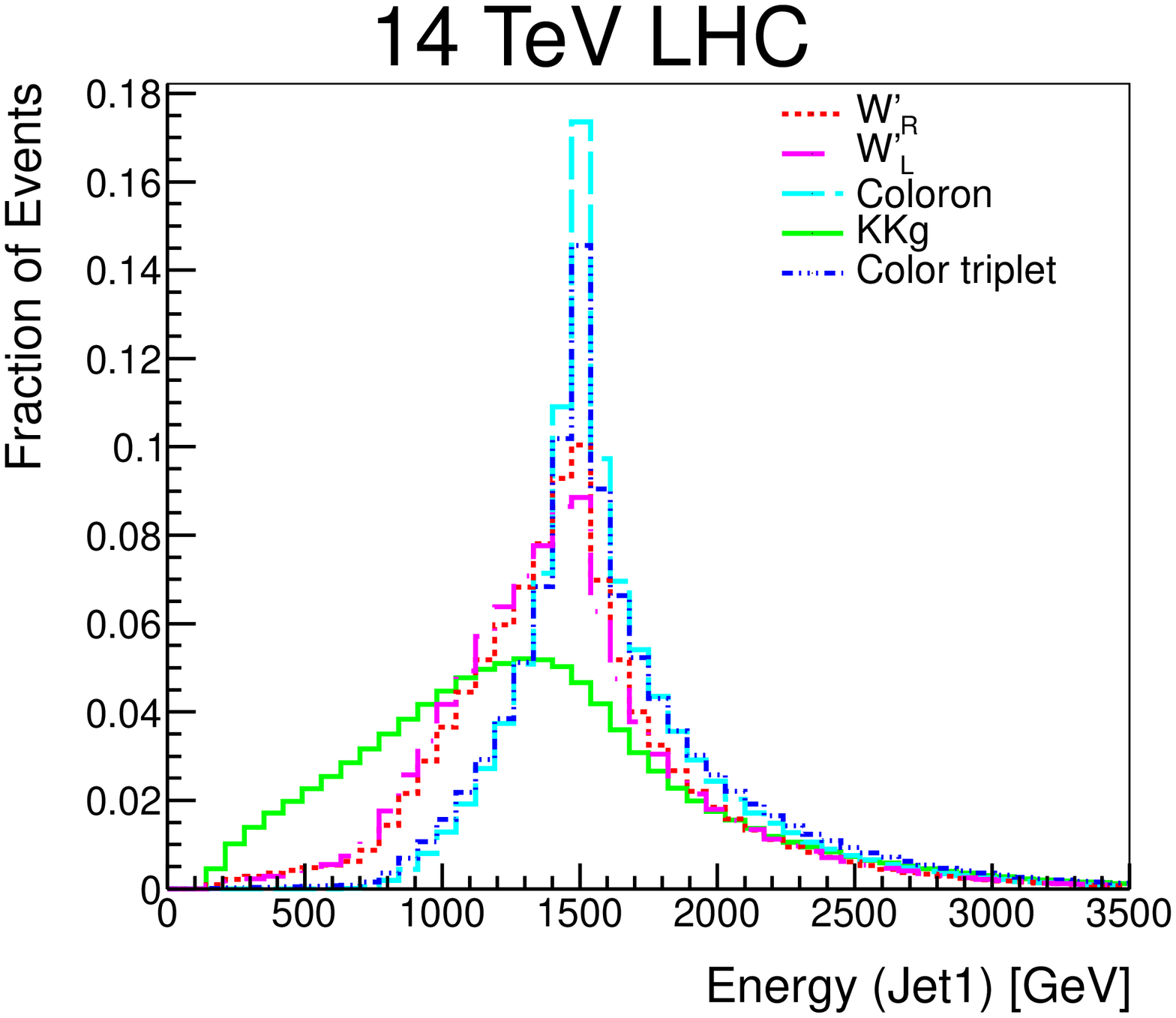}
    \label{fig:highjet1e}
  }
  \caption{The \subref{fig:highjet1pt} $p_{T}$ and \subref{fig:highjet1e} energy (in the lab frame) of the leading jet, at the 14~TeV LHC. The different signal distributions are  normalized to unit area. (Color online.)
}
\label{fig:highjet1}
\end{figure}

The kinematic distributions of the basic reconstructed objects can already distinguish the different signals from each other. The $p_T$ and energy of the leading jet are shown in Fig.~\ref{fig:highjet1}. The $\Wp$, coloron and color-triplet signals all show the expected $p_T$ peak at half of the resonance mass. The $KKg$ distribution shows no such peak structure and is instead broad up to 1500~GeV. The situation is similar for the distribution of the energy of the leading jet, where the coloron and color-triplet have narrow distributions, the $\Wp$ have slightly broader distributions, and the $KKg$ distribution is broader than the others. The distinctive $KKg$ distributions are due to the $KKg$ coupling and width, which also leads to a broader distribution of the resonance mass, see Sec.~\ref{subsec:higheventreconstruction}.

\begin{figure}[!h!tbp]
  \centering
  \subfigure[]{
    \includegraphics[width=0.48\textwidth]{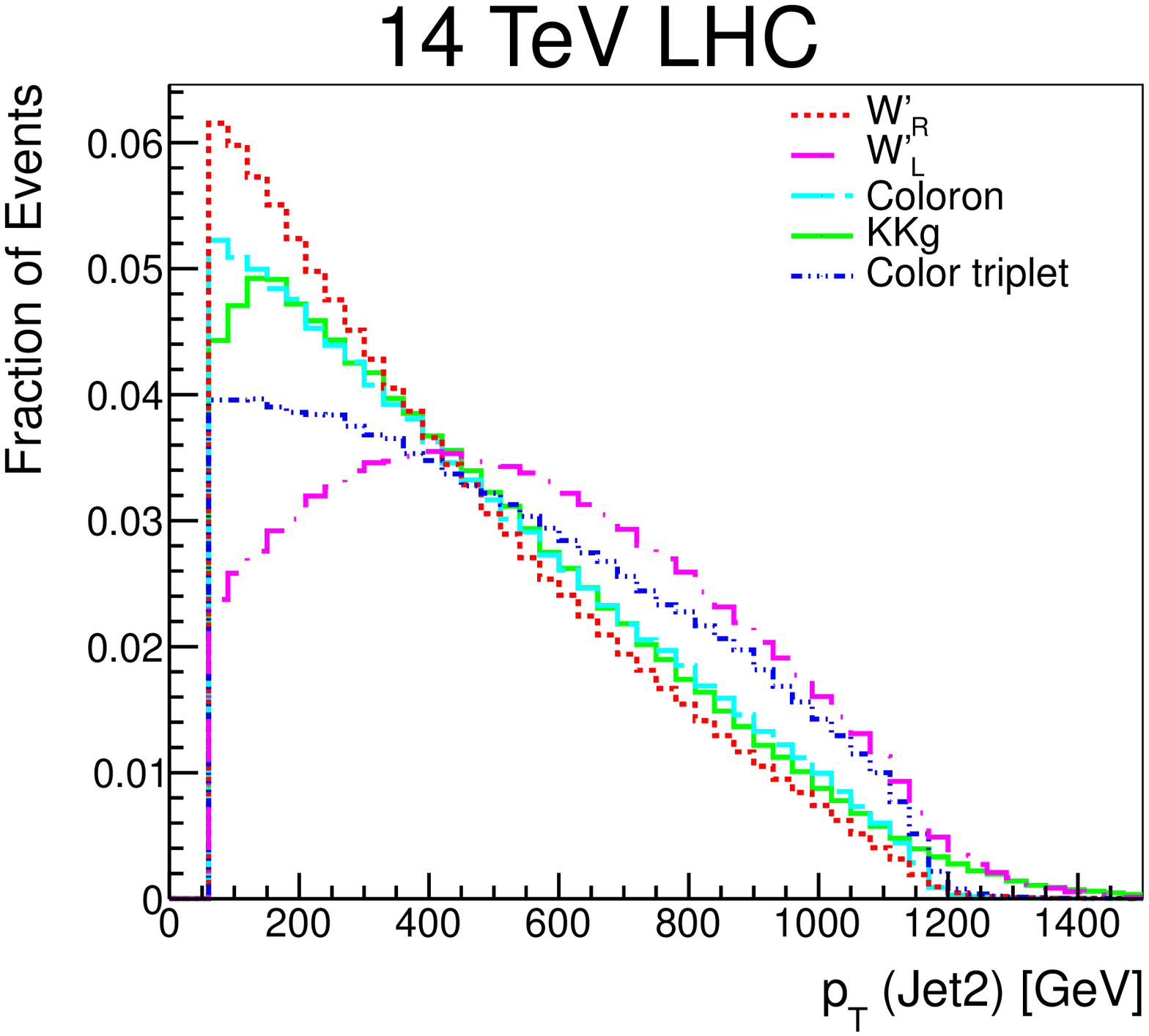}
    \label{fig:highjet2pt}
  }
  \subfigure[]{
    \includegraphics[width=0.48\textwidth]{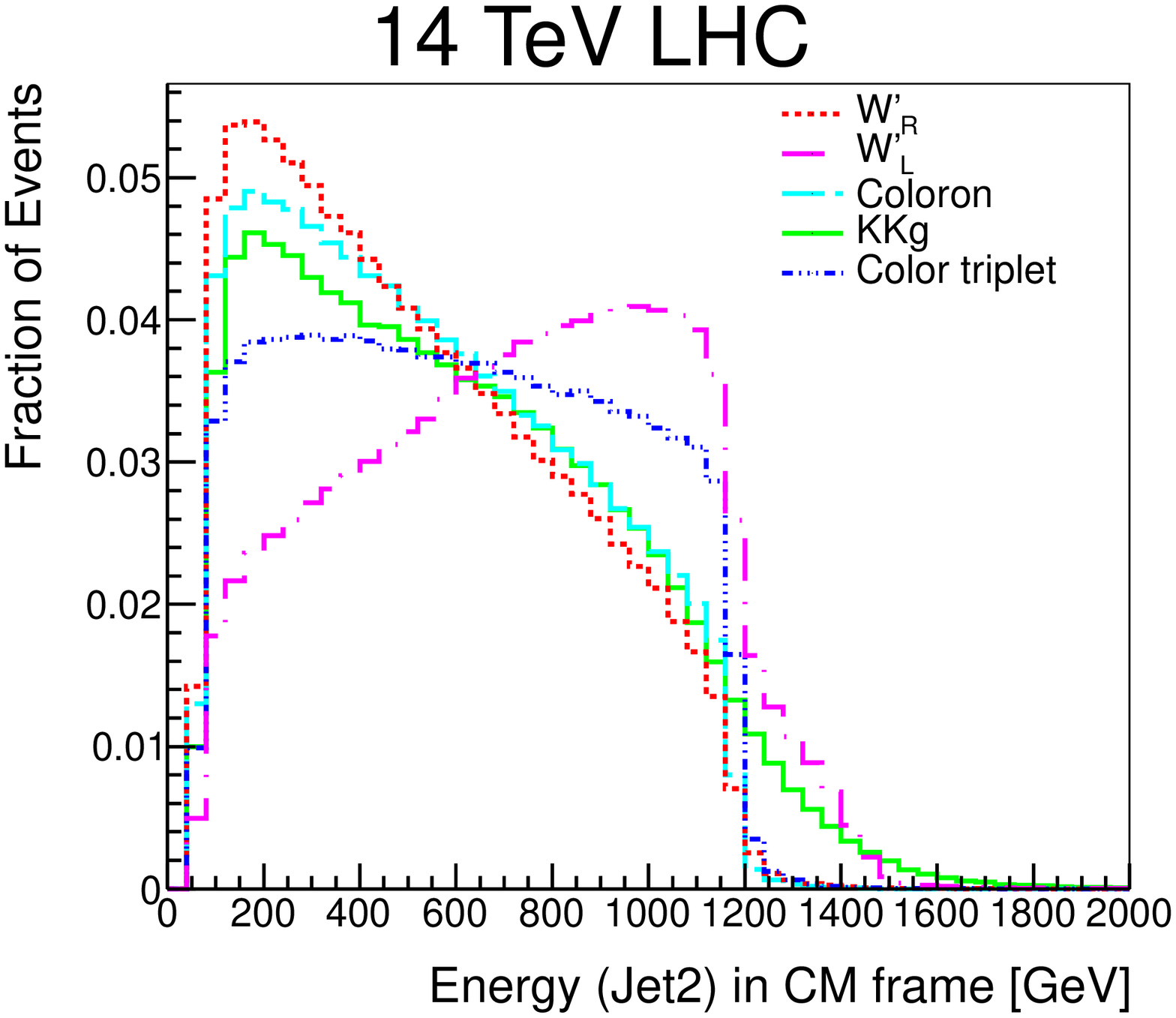}
    \label{fig:highjet2booste}
  }
  \caption{The \subref{fig:highjet2pt} $p_{T}$ and \subref{fig:highjet2booste} energy in the CM frame of the second jet, at the 14~TeV LHC. The different signal distributions are  normalized to unit area. (Color online.)
}
\label{fig:highjet2}
\end{figure}

The $p_T$ and energy distributions of the second jet are shown in Fig.~\ref{fig:highjet2}. Here, all five signals have distinct distributions. Both the initial partons and the helicity of the top quark have an impact. The largest difference is between $\Wp_R$ and $\Wp_L$, which have the same initial state but different helicity top quarks. $\Wp_R$ has lower $p_T$ and peaks at a lower energy than $\Wp_L$, for which the $p_T$ distribution extends to higher momenta and the energy distribution peaks at a higher energy. The energy of the second jet in the CM frame, shown in Fig.~\ref{fig:highjet2booste}, brings out the $\Wp_L$ in particular, with a rising distribution and a kinematic edge at 1200~GeV. This is the upper limit from energy conversation and corresponds to the lower edge of the CM energy distribution of the $W$~boson shown in Fig.~\ref{fig:highwbooste}. The coloron, $KKg$ and color-triplet distributions are in between the two $\Wp$ distributions. The $KKg$ distribution in particular is not as distinct as for the leading jet.

\subsection{Event reconstruction}
\label{subsec:higheventreconstruction}

\begin{figure}[!h!tbp]
  \centering
  \subfigure[]{
    \includegraphics[width=0.48\textwidth]{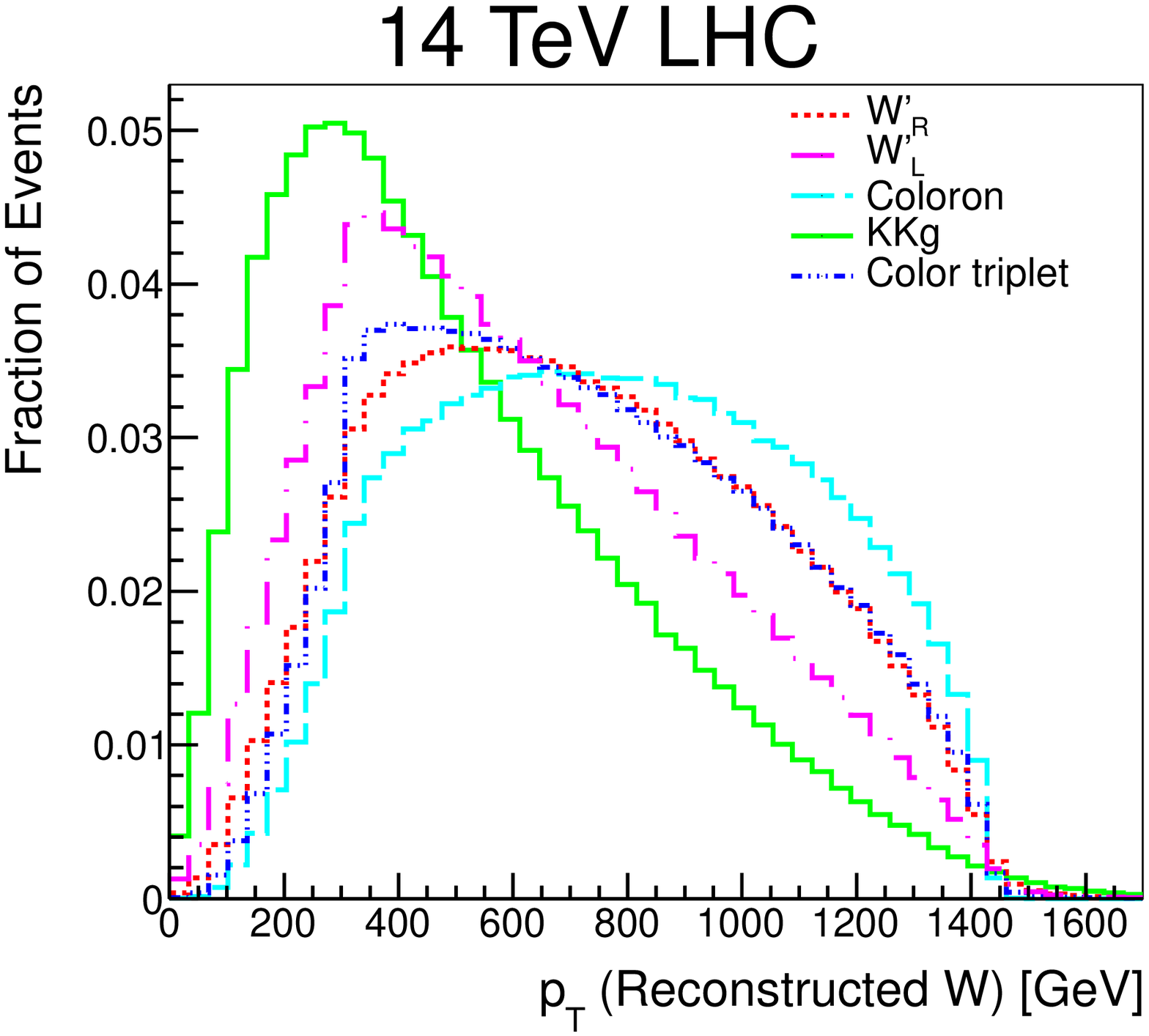}
    \label{fig:highwpt}
  }
  \subfigure[]{
    \includegraphics[width=0.48\textwidth]{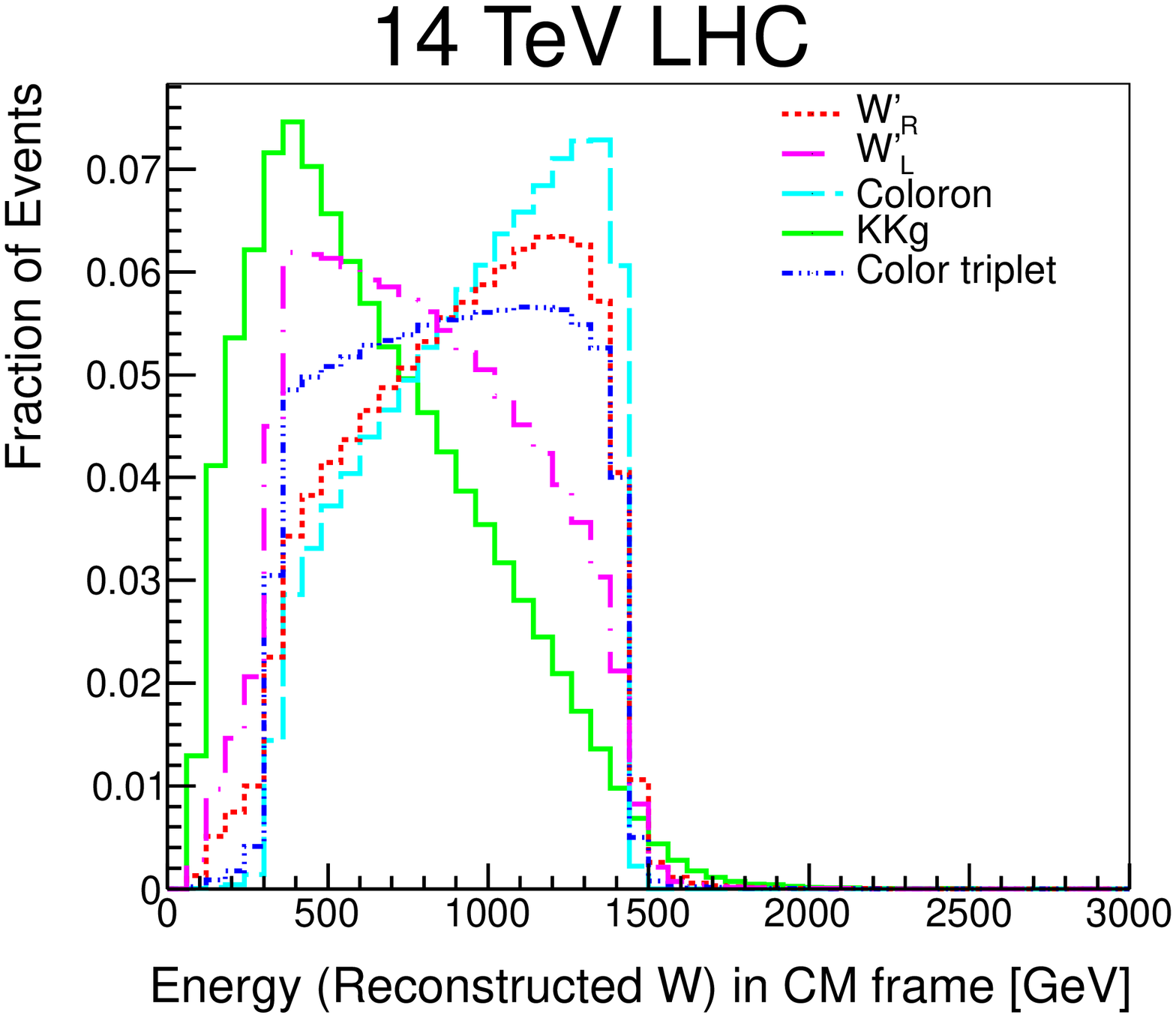}
    \label{fig:highwbooste}
  }
  \caption{The \subref{fig:highwpt} $p_{T}$ and \subref{fig:highwbooste} energy in the CM frame of the reconstructed $W$~boson, at the 14~TeV LHC. The different signal distributions are  normalized to unit area. (Color online.)
}
\label{fig:highw}
\end{figure}

The $W$~boson and top quark are reconstructed as described in Sec.~\ref{sec:analysis}. 
The $p_T$ and energy (in the CM frame) of the reconstructed $W$~boson are shown in Fig.~\ref{fig:highw}. For $\Wp$ and color-triplet, the distributions show the reverse pattern from the second jet, c.~f. Fig.~\ref{fig:highjet2}. In particular, the lower kinematic edge for $\Wp_L$ and the color-triplet visible in Fig.\ref{fig:highwbooste} corresponds to the upper kinematic edge for the same two signals in Fig.\ref{fig:highjet2booste}.

\begin{figure}[!h!tbp]
  \centering
  \subfigure[]{
    \includegraphics[width=0.48\textwidth]{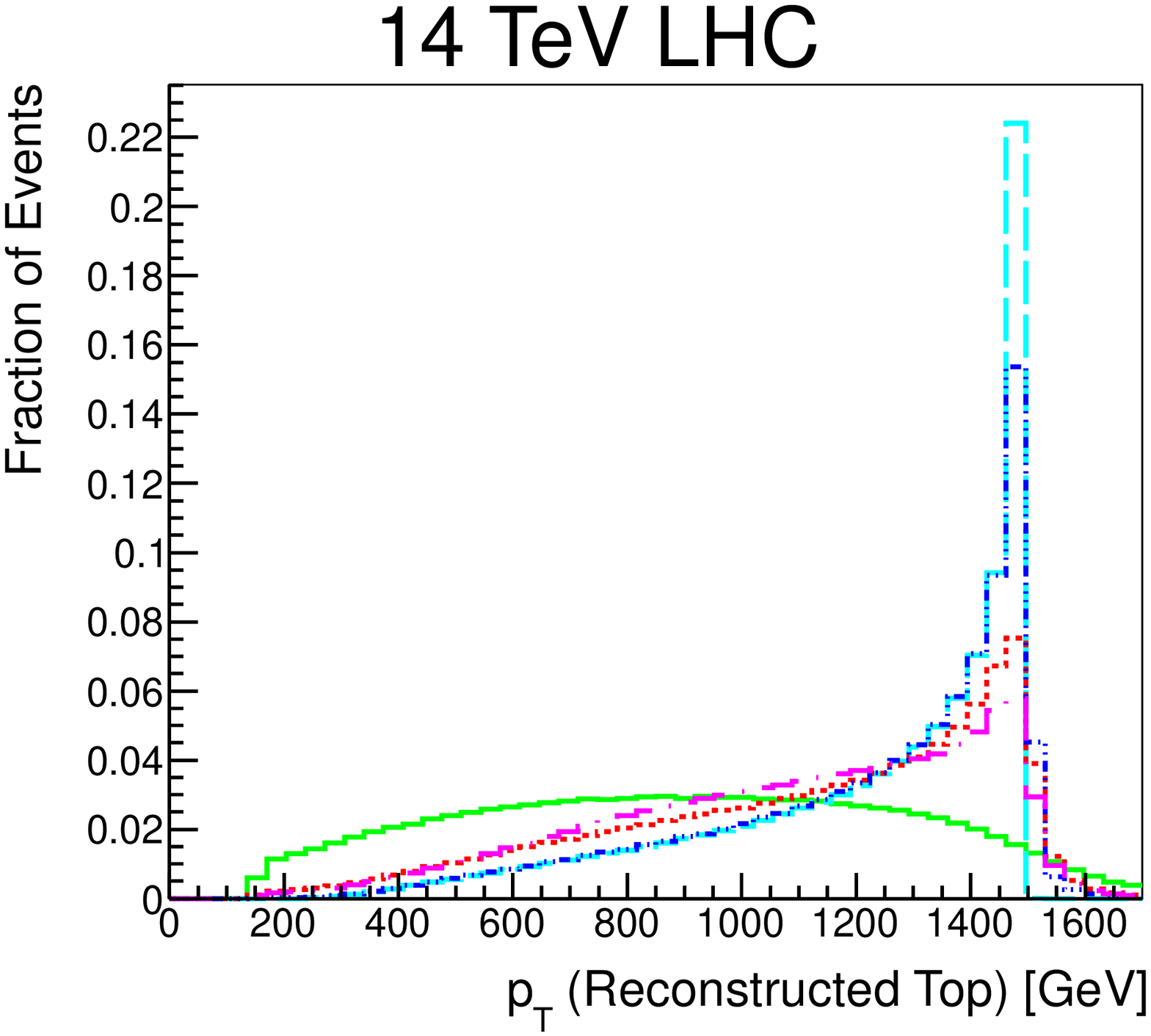}
    \label{fig:hightoppt}
  }
  \subfigure[]{
    \includegraphics[width=0.48\textwidth]{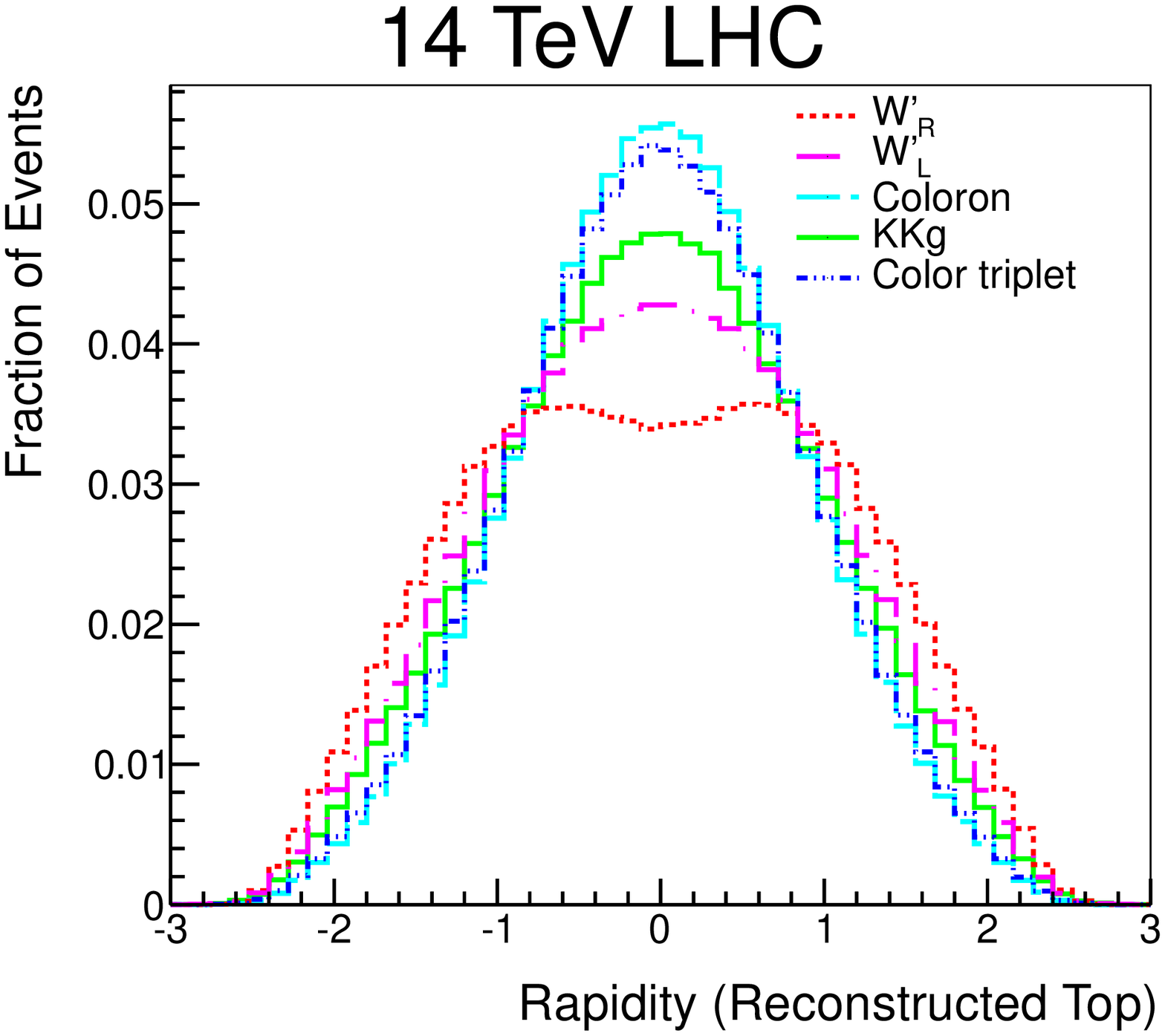}
    \label{fig:hightoprapidity}
  }
  \caption{The \subref{fig:hightoppt} $p_{T}$ and \subref{fig:hightoprapidity} rapidity of the reconstructed top quark, at the 14~TeV LHC. The different signal distributions are  normalized to unit area. (Color online.)
}
\end{figure}

The reconstructed top-quark mass is shown in Fig.~\ref{fig:wpinttop} for the different $\Wp$ signals. The distributions of the coloron and $KKg$ (not shown) are similar to $\Wp_L$, while the color-triplet distribution (not shown) is similar to $\Wp_R$. The $p_{T}$ of the reconstructed top quark is shown in Fig.~\ref{fig:hightoppt}. It separates $KKg$, which has a very broad distribution, from the other signals, which show a Jacobian peak at the resonance particle mass. The energy of the top quark, boosted into the CM frame, is not as good at separating signals from each other but better at separating all of them from the background, see Fig.~\ref{fig:lowtopeboost}. 
The distribution of the rapidity of the top quark is shown in Fig.~\ref{fig:hightoprapidity}. 
This variable separates right-handed $\Wp$ production from the other signals. While the distributions themselves depend somewhat on the neutrino $p_Z$, setting it to zero only enhances the visible difference between the different signals.

\begin{figure}[!h!tbp]
  \centering
  \subfigure[]{
    \includegraphics[width=0.48\textwidth]{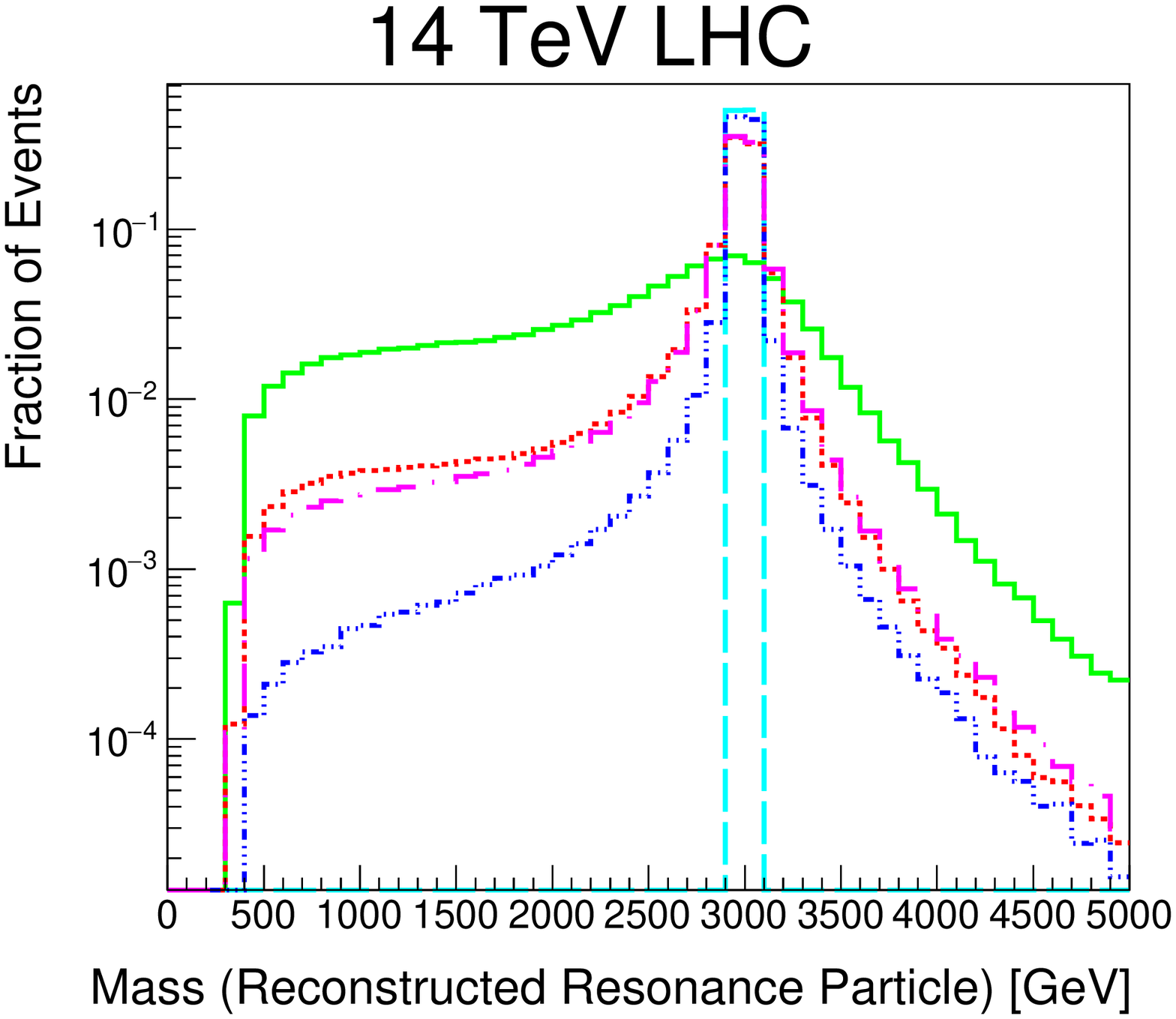}
  }
  \subfigure{
    \includegraphics[width=0.22\textwidth]{figleghigh}
  }
  \caption{The mass of the reconstructed resonance particle at the 14~TeV LHC. The different signal distributions are  normalized to unit area. (Color online.) 
}
  \label{fig:highparticlemass}
\end{figure}

The mass of the reconstructed resonance particle is shown in Fig.~\ref{fig:highparticlemass}. It is sharply peaked for all signals except $KKg$ because effects of limited detector resolution are not included in this parton-level analysis. It is much wider for $KKg$ and extends to lower masses, thus at least in principle this is a potential variable to isolate $KKg$.

\subsection{Angular correlations}
\label{subsec:highanglevar}

\begin{figure}[!h!tbp]
  \centering
  \subfigure[]{
    \includegraphics[width=0.48\textwidth]{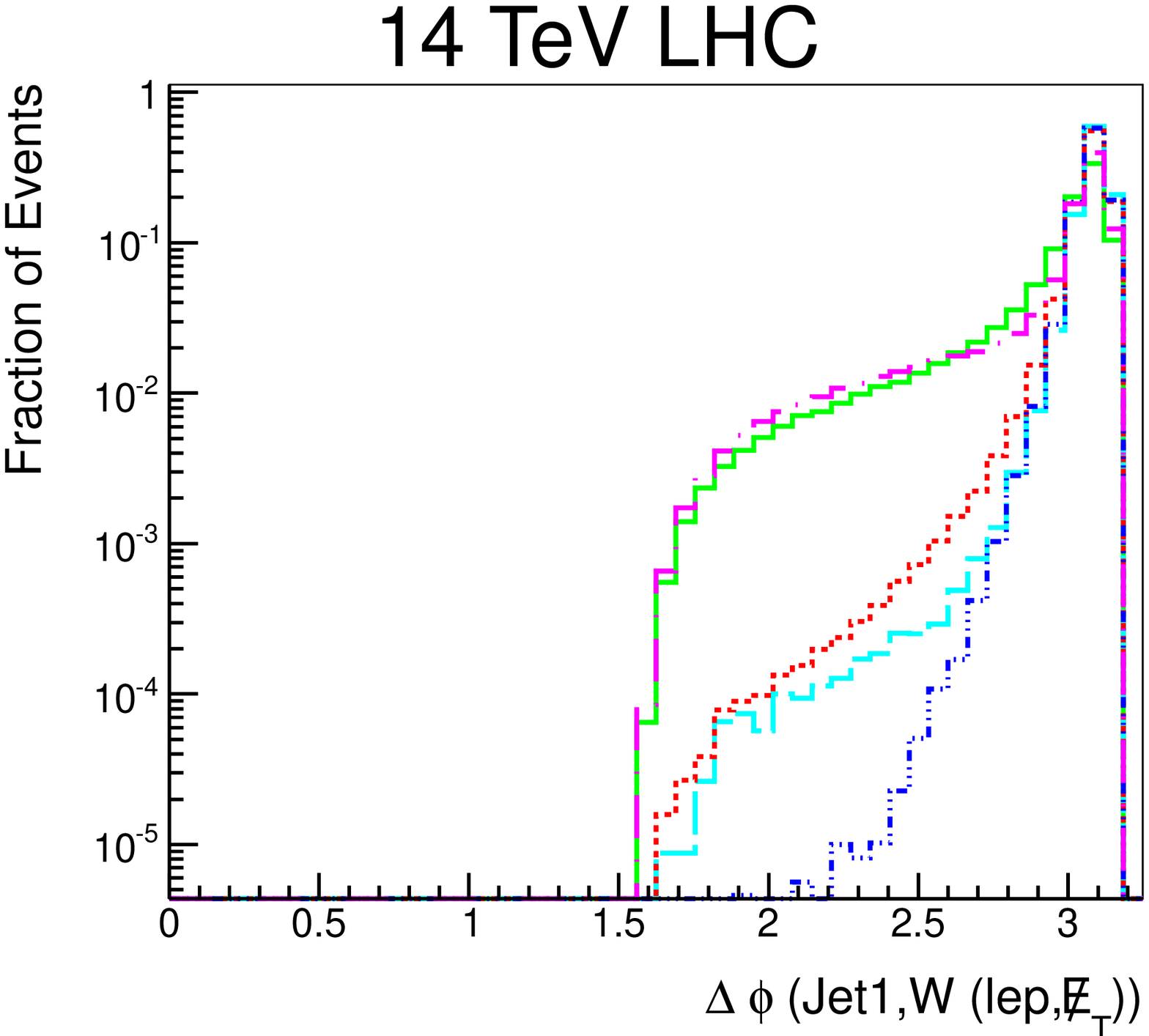}
  \label{fig:highdeltaphijet1w}
  }
  \subfigure[]{
    \includegraphics[width=0.48\textwidth]{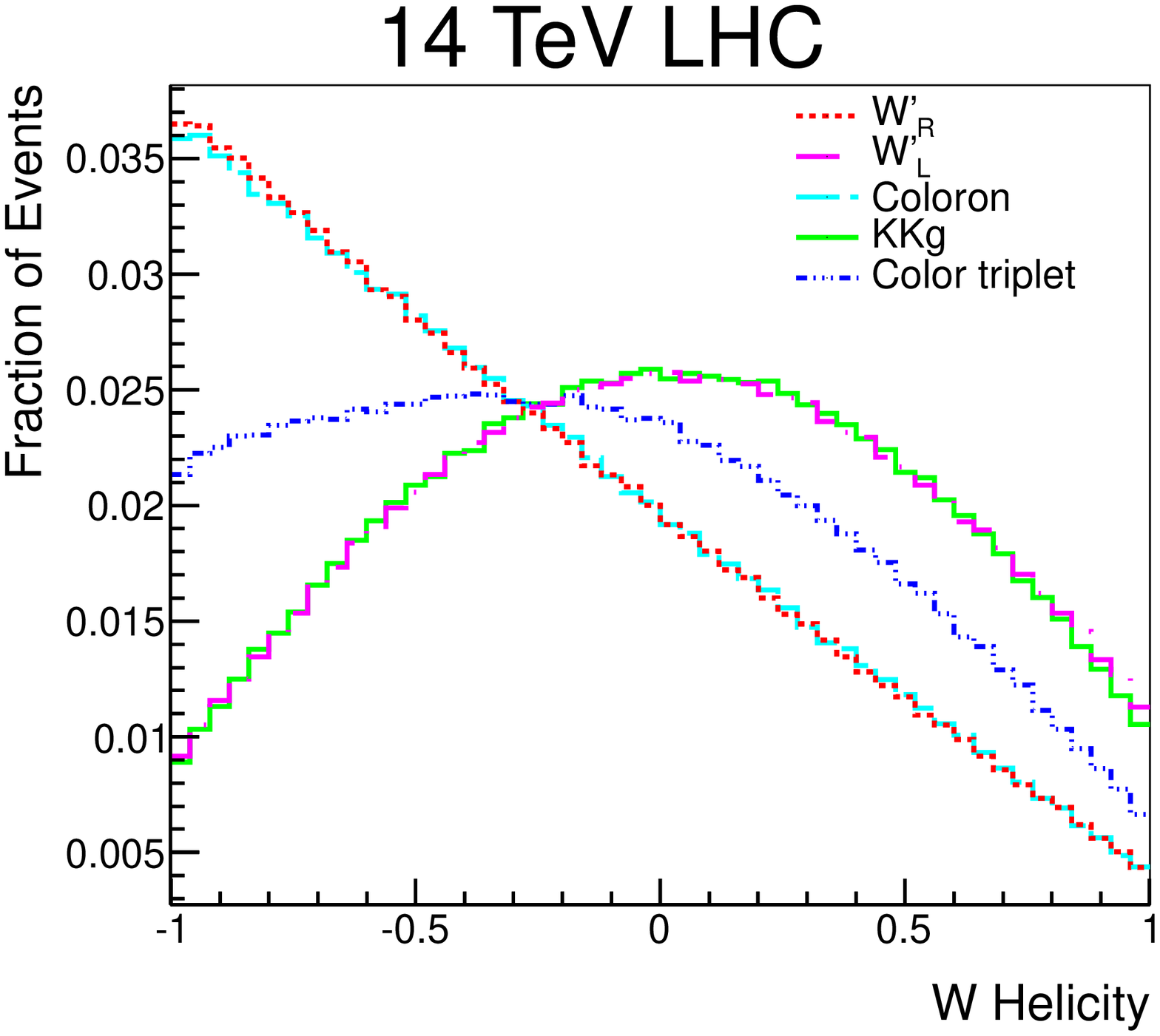}
  \label{fig:highwhelicity}
  }
  \subfigure[]{
    \includegraphics[width=0.48\textwidth]{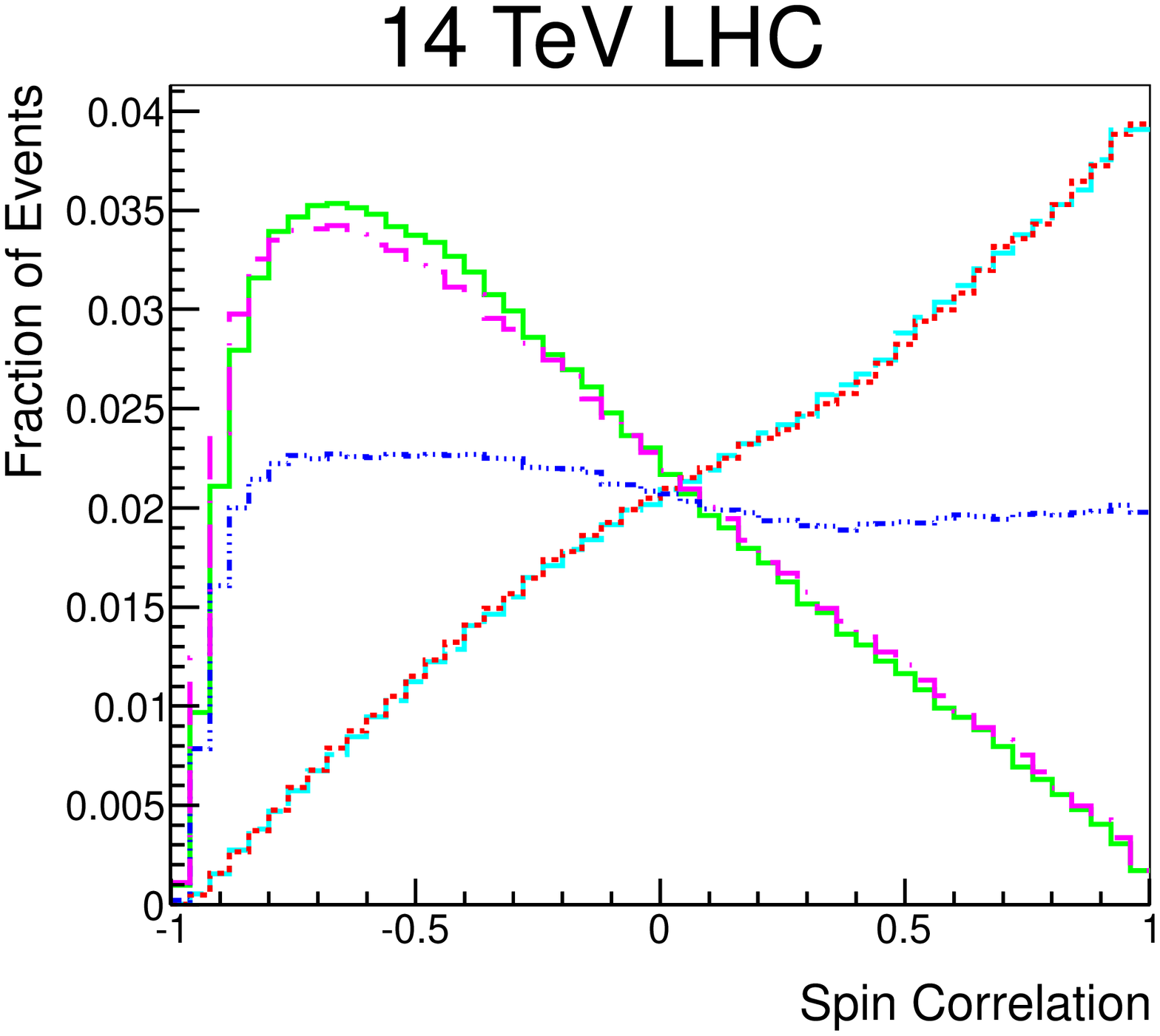}
  \label{fig:highspincor}
  }
  \caption{The \subref{fig:highdeltaphijet1w} $\Delta \phi$ between the leading jet and the $W$~boson, 
\subref{fig:highwhelicity} helicity of the $W$~boson and
\subref{fig:highspincor} top-quark spin correlation in the helicity basis, 
at the 14~TeV LHC. The different signal distributions are  normalized to unit area. (Color online.)
}
  \label{fig:highang}
\end{figure}

The angular correlation variables discussed in Sec.~\ref{subsec:lowanglevar} also discriminate some signals from each other. Here we focus on those that show the most discrimination at 14~TeV.
The distribution of the $\Delta \phi$ between the leading jet and the reconstructed $W$~boson is shown in Fig.~\ref{fig:highdeltaphijet1w} and shows a clear distinction between $\Wp_L$ and $KKg$ (broader distribution) on one side and $\Wp_R$, the coloron and the color-triplet on the other side. All distributions show that the $W$~boson and leading jet are back-to-back as is expected, but the details are different due to spin correlations. This becomes clear from the helicity distribution of the $W$~boson, shown in Fig.~\ref{fig:highwhelicity} and the spin correlation in the helicity basis, shown in Fig.~\ref{fig:highspincor}. 

The coloron exhibits the same $W$~helicity and spin correlation as the right-handed $\Wp$, while the $KKg$ has the same distribution as the left-handed $\Wp$. The color-triplet is unique in these distributions. Note that as in the 8~TeV analysis, the neutrino $p_Z$ plays a very important role in the $W$~helicity and spin correlation calculations. If neutrino $p_Z$ is set to zero, all differences between signals disappear.


\section{Conclusions}
\label{sec:concl}

We have presented an overview of resonance production in the single top plus jet final state at a hadron collider, including a phenomenological analysis at the 8~TeV and 14~TeV LHC. So far, LHC searches have only considered $\Wp$ production. We have demonstrated that the production of $\Wp$ with left-handed couplings and its decay to $t\bar{b}$ receives a significant boost in cross section from the production of off-shell top quarks. The searches can easily be expanded to also look for colored resonances such as scalar color octets, Kaluza-Klein gluons and color-triplet scalars.
With the 8~TeV dataset, the LHC is already sensitive to several of these models over a wide range of resonance masses and couplings. The background to the single top plus jet signature is large, and we have presented several variables that are able to isolate one or all of the signals from the backgrounds. We have focused on the low-resonance-mass region for the 8~TeV LHC, where the background is largest and the isolation of the signal most difficult. Care must be taken in separating the signals from the background because the kinematic distributions are different for the different signals. In particular the application of multivariate analysis techniques will only be optimal for one of the signals and will not be sensitive to others. 
We have demonstrated for the example of the 14~TeV LHC that each of the signals has its own unique signature in the detector. If an excess should be observed in the single top plus jet final state in Run~2 at the LHC, then the distributions presented here are able to identify the nature of that new signal. The charge of the lepton as well as the $b$-tag multiplicity in the event are two simple yet powerful variables that distinguish signals from each other and from the backgrounds.

\section*{Acknowledgements}
We thank K.~Agashe, B.~Alvarez-Gonzalez, S.~Chivukula, M.~Peskin, B.~Schoenrock, E.~Simmons and C.-P. Yuan for useful discussions.
The work of E.D., J.N. and R.S. is supported in part by the US National Science Foundation under Grant No. PHY-0952729.  N.V is supported by U.S. National Science Foundation under Grant No. PHY-0854889. D.W. is supported by Department of Energy under Grants No. DE-AC02-76SF00515 and in part by a grant from the Ford Foundation via the National Academies of the Sciences as well as the National Science Foundation under Grants No. NSF-PHY-0705682, the LHC Theory Initiative. 
The research of J.H.Y. was supported by the National Science Foundation under Grant Numbers PHY-1315983 and PHY-1316033. 

\bibliography{top_colored}

\end{document}